\documentclass[11pt]{article}
\usepackage{amssymb,amsmath}
\usepackage{bm} 
\usepackage{booktabs} 
\usepackage{array}
\usepackage{latexsym}
\usepackage{graphicx}
\usepackage{color}
\usepackage{slashed}
\usepackage{datetime}
\usepackage[nosort]{cite}
\usepackage{verbatim}
\usepackage{enumerate}
\usepackage{chngpage} 
\allowdisplaybreaks
\usepackage{psfrag}

\usepackage{mciteplus}

\usepackage[colorlinks=true,      linkcolor=blue,      urlcolor=blue,      
            filecolor=blue,      citecolor=blue,       pdfstartview=FitH,     
						pdfpagemode=UseNone,      bookmarksopen=true]{hyperref}  
\usepackage[all]{hypcap}     

\newcommand{\comm}[1]{} 

\renewcommand{\arraystretch}{1.3}

\newcommand{\ket}[1]{|#1 \rangle}

\def\({\left(}
\def\){\right)}
\def\[{\left[}
\def\]{\right]}

\def\half{\frac12}

\def\One{{\hbox{ 1\kern-.8mm l}}}

\def\barray{\begin{array}}
\def\earray{\end{array}}
\def\be{\begin{equation}}
\def\ee{\end{equation}}
\def\bea{\begin{eqnarray}}
\def\eea{\end{eqnarray}}
\def\bal{\begin{align}}
\def\eal{\end{align}}

\def\Et{{\widetilde{E}}}

\def\Gt{{\widetilde{G}}}

\def\Jt{{\widetilde{J}}}
\def\Kt{{\widetilde{K}}}

\def\Mt{{\widetilde{M}}}

\def\chit{{\widetilde{\chi}}}

\def\phit{{\widetilde{\phi}}}
\def\psit{{\widetilde{\psi}}}

\def\xit{{\widetilde{\xi}}}
\def\zetat{{\widetilde{\zeta}}}

\def\etat{{\widetilde{\eta}}}
\def\kappat{{\widetilde{\kappa}}}

\def\uu{{\underline{u}}}
\def\vu{{\underline{v}}}

\def\psitu{{\underline{\widetilde{\psi}}}}
\def\phitu{{\underline{\widetilde{\phi}}}}
\def\thet{{\underline{\theta}}}
\def\ru{{\underline{r}}}

\numberwithin{equation}{section} 
\allowdisplaybreaks  

\makeatletter
\g@addto@macro\bfseries{\boldmath}
\makeatother

\definecolor{cardinal}{rgb}{0.6,0,0}
\definecolor{darkgreen}{rgb}{0,0.4,0}
\definecolor{golden}{rgb}{0.92, 0.7, 0}
\definecolor{midnight}{rgb}{0, 0, 0.5}
\definecolor{darkblue}{rgb}{0, 0, 0.7}

\def\cA{{\cal A}}

\def\cC{{\cal C}}
\def\cD{{\cal D}}
\def\cF{{\cal F}}

\def\cL{{\cal L}}

\def\cN{{\cal N}}
\def\cP{{\cal P}}

\newcommand{\Gammau}[1]{\Gamma^{\underline{#1}}}
\newcommand{\epsilonu}[1]{\epsilon^{\underline{#1}}}
\newcommand{\theangle}{\gamma}

\def\p{\partial}
\def\cO{{\cal O}}

\def\Jt{\tilde{J}}

\def\psit{\widetilde{\psi}}

\def\vh{\hat{v}}

\def\bbR{\mathbb{R}}

\def\ket#1{|{#1}\rangle}
\def\half{\frac{1}{2}}

\newcommand{\flap}{\widehat \cL}

\newcommand{\MM}{\mathcal{M}}

\newcommand{\muh}{\widehat{\mu}}

\newcommand{\xh}{\widehat{x}}
\newcommand{\bhat}{\widehat{b}}

\newcommand{\gu}[1]{\Gamma^{\underline{#1}}}
\newcommand{\guh}[1]{\widehat{\Gamma}^{\underline{#1}}}

\begin{document}


\begin{flushright}
QMUL-PH-18-32\\
YITP-18-134\\
\end{flushright}

\vspace{3mm}

\begin{center}

{\huge {\bf Supercharging Superstrata}}
\vspace{14mm}

{\large
\textsc{Nejc \v{C}eplak$^1$,~ Rodolfo Russo$^{1}$, Masaki Shigemori$^{2,3}$}}

\vspace{12mm}

$^1$ Centre for Research in String Theory, School of Physics and Astronomy,\\
Queen Mary University of London, Mile End Road, London, E1 4NS, United Kingdom\\
\medskip
$^2$
Department of Physics, Nagoya University,\\
Furo-cho, Chikusa-ku, Nagoya 464-8601, Japan\\
\medskip
$^3$ Center for Gravitational Physics,
Yukawa Institute for Theoretical Physics, \\ Kyoto University,
Kitashirakawa-Oiwakecho, Sakyo-ku, Kyoto 606-8502 Japan\\
\medskip

\vspace{4mm} 
%

\vspace{13mm}
 
\textsc{Abstract}

\end{center}

\begin{adjustwidth}{13.5mm}{13.5mm} 
 
\vspace{1mm}
\noindent
We construct a new class of smooth horizonless microstate geometries of
the supersymmetric D1-D5-P black hole in type IIB supergravity.
We first work in the AdS$_3 \times S^3$ decoupling limit and use the
fermionic symmetries of the theory to generate new momentum carrying
perturbations in the bulk that have an explicit CFT dual description.
We then use the supergravity equations to calculate the backreaction of
these perturbations and find the full non-linear solutions both in the
asymptotically AdS and asymptotically flat case. These new geometries
have a simpler structure than the previously known superstrata solutions.

\end{adjustwidth}

\thispagestyle{empty}
\newpage



\baselineskip=14.5pt
\parskip=3pt

\tableofcontents

\baselineskip=15pt
\parskip=3pt

\section{Introduction and summary}
\label{Sect:introduction}

The thermodynamic entropy of certain extremal black
holes was successfully explained by enumerating the microstates
predicted by string theory \cite{Strominger:1996sh, Breckenridge:1996is,
Maldacena:1997de}.  In particular, the D1-D5-P black hole, which is a
supersymmetric, 3-charge solution in type IIB string theory, has been
an important arena for studying microscopic physics of black holes.
This is partly because it allows a holographic description in terms of a
two-dimensional CFT, called the D1-D5 CFT, which is under good
theoretical control.

Although by now we understand the counting of black-hole microstates very
well, their physical nature remains mysterious.
The fuzzball conjecture \cite{Mathur:2005zp, Bena:2007kg,Skenderis:2008qn,Mathur:2009hf} claims that
black-hole microstates are made of stringy fuzz that is free of a
horizon and singularities, and spreads over the size of the would-be
horizon.
The typical microstates of generic black holes such as the Schwarzschild
black hole are expected to involve stringy modes and cannot be described within supergravity, the massless-mode truncation of
the full string theory.  However, constructing microstates in the full
string theory is beyond the reach of our current technology and
understanding.

The microstate geometry programme (see \cite{Bena:2013dka} and references therein) aims to construct as many
black-hole microstates as possible within supergravity as smooth,
horizonless geometries.
The programme has been particularly successful for supersymmetric black
holes, where a large number of microstate geometries have been
explicitly constructed \cite{Bena:2005va, Berglund:2005vb}.\footnote{See
\cite{Mathur:2003hj, Lunin:2004uu, Giusto:2004id, Giusto:2004ip,
Giusto:2006zi, Ford:2006yb, Mathur:2011gz, Mathur:2012tj, Lunin:2012gp,
Giusto:2012yz} for explicit constructions of microstate geometries of
the D1-D5-P system, before the superstratum technology was developed.  }
It is still unclear how large a subset of all microstates is describable
within supergravity but, even if not all microstates allow a
supergravity description, explicit microstate geometries are important
because they provide the only top-down, direct tool available for
studying and understanding the microstate structure of black holes.

Superstrata \cite{Bena:2011uw,Giusto:2012jx,Bena:2015bea, Bena:2016ypk, Bena:2017xbt} (see also
\cite{Bena:2017geu}) represent the largest family of microstate
geometries constructed thus far for the D1-D5-P black hole with known
CFT duals and have various interesting features.  The D1-D5-P black hole
contains momentum (P) charge along an $S^1$ coordinatised by~$v$, and
superstrata contain $v$-dependent travelling waves corresponding to the P
charge.
As will be detailed below, superstrata are constructed based on a
solution-generating technique whose holographic meaning is
well-understood.  Consequently, the CFT states dual to superstrata are
explicitly known, which makes them an ideal setup for
studying precision holography \cite{Kanitscheider:2006zf,Kanitscheider:2007wq,Giusto:2015dfa,Tormo:2018fnt}.
Superstrata also give interesting clues for the physical nature of
typical microstates.  Although the superstrata written down thus far are
not typical microstates of the black-hole ensemble, they are
expected to evolve into more typical states when perturbed, and the
endpoint of such a process is a subject of much physical interest
\cite{Eperon:2016cdd, Keir:2016azt, Marolf:2016nwu, Eperon:2017bwq,
Bena:2017upb, Tyukov:2017uig, Bena:2018mpb}.

In this paper, we present an explicit construction of an entirely new
class of superstrata.  They share the same features as the original
superstrata \cite{Bena:2015bea, Bena:2016ypk, Bena:2017xbt}, such as
representing microstates of the D1-D5-P black hole and having dual CFT
states.  However, they are simpler than the original ones in that they
involve a smaller number of non-trivial fields.

The original superstrata were constructed using the solution-generating
technique as follows.  First we take, as a seed, some 2-charge solution
of linear supergravity around AdS$_3\times S^3$, for which the dual CFT
state $\ket{\psi}$ is known\footnote{In our convention we take $\ket{\psi}$ to be an anti-chiral primary state.}.  Next, we act on it with the generators of
the $SU(1,1|2)_L\times SU(1,1|2)_R$ (super)isometry group of
AdS$_3\times S^3$ \cite{Mathur:2003hj}. Specifically, we apply $L_{-1}$ and $J^+_{0}$
generators\footnote{These are generators in the NS-NS sector. } of the
bosonic subgroup $SL(2,\bbR)_L \times SU(2)_L\subset SU(1,1|2)_L$
\cite{Bena:2015bea, Bena:2016ypk, Bena:2017xbt}. This process generates
a new linear solution with non-vanishing third (P) charge.  Acting on
the seed $m$ times with $J_0^+$ and $n$ times with $L_1$ generates the
solution dual to the CFT state $(J^+_0)^m (L_{-1})^n \ket{\psi}$.
Finally, we use the structure of the BPS equations to complete the
linear solution to a fully backreacted non-linear solution. In CFT, this
final process corresponds to having the same excitation many times, namely,
$[(J^+_0)^m (L_{-1})^n \ket{\psi}]^{N_{mn}}$, with $N_{mn}\gg 1$.

However, the supergroup $SU(1,1|2)_L$ also includes fermionic generators
$G^{+A}_{-1/2}$, where $A=1,2$ is the index for an $SU(2)_B$ group
related to the internal manifold.  Therefore, alternatively, we can act
with these fermionic generators on the seed (namely, ``supercharge'' it)
to generate a completely new class of linear solutions.  In order to get
a bosonic solution, we need to act twice with fermionic generators.  In
addition to this, one can act on the newly obtained state with the
bosonic symmetry generators to obtain a new state $(J^+_0)^m (L_{-1})^n
G^{+1}_{-1/2} G^{+2}_{-1/2}\ket{\psi}$.\footnote{More precisely, this
state is a linear superposition of an old superstratum and a new one,
and the former must be subtracted; see \eqref{eq:stateOkmnq}.}  The
non-linear completion goes much the same way as before, and produces a
new set of superstrata.

The supersymmetric solutions in supergravity are parametrised by a
number of scalars and forms \cite{Gutowski:2003rg, Cariglia:2004kk,
Giusto:2013rxa, Bena:2015bea}.  In microstate geometries, these
quantities get excited in non-trivial ways, representing the structure
of the microstate.  In the original superstrata, a scalar and a 2-form
(which is related to the NS-NS $B$-field) get excited at linear order
and, at quadratic order, more scalars and forms are turned on in a very
specific way so that the combination that enters the metric is
$v$-independent.  This mechanism was crucial for the explicit
construction of the geometry and was called ``coiffuring''
\cite{Bena:2014rea}. In contrast, in our new superstrata, at linear
order, only the 2-form is excited and there is no scalar excited that
must be cancelled by other scalars excited at higher order. Thus we will
see that coiffuring is not necessary for the new superstrata and,
consequently, they are simpler than the ones generated just by using bosonic symmetries.
In hindsight, the existence of such a simple branch of superstrata could
have been expected from the analysis of the excitation spectrum of
linear supergravity around AdS$_3\times S^3$ \cite{Deger:1998nm,
Maldacena:1998bw, Larsen:1998xm, deBoer:1998kjm}; see Appendix
\ref{app:DegerEtal} for details.  However, we go beyond such linear
analysis and construct fully non-linear solutions using the structure of
the BPS equations.

Due to their simplicity, studying the structure of the solution is
easier for the new superstrata than for the original ones.
For example, although the solution-generating technique can only produce
asymptotically AdS$_3$ solutions, we can trivially extend the new
superstrata to asymptotically flat solutions, as we do in section \ref{sect:AsyFlat}.
In contrast, extending the original superstrata to asymptotically flat
ones required a non-trivial step of solving differential equations
\cite{Bena:2017xbt}.
Also, in the original superstrata, there was technical difficulty in
constructing solutions that involves two modes with completely different
quantum numbers $(k_1,m_1,n_1)$, $(k_2,m_2,n_2)$ \cite{Bena:2017xbt}.
The simple setup of the new superstrata may shed light on this technical
point.

Investigating physical aspects of this new class of superstrata, such
as the integrability of the geometry \cite{Bena:2017upb} and their
precision holography
\cite{Kanitscheider:2006zf,Kanitscheider:2007wq,Giusto:2015dfa,Tormo:2018fnt}
would be very interesting.
As mentioned above, possible instabilities of microstate geometries have
attracted much interest lately \cite{Eperon:2016cdd, Keir:2016azt,
Marolf:2016nwu, Eperon:2017bwq, Bena:2017upb, Tyukov:2017uig,  Bena:2018mpb}.  In
particular, it has been argued \cite{Eperon:2016cdd} that supersymmetric
microstate geometries are non-linearly unstable when a small amount of
energy is added, leading to a formation of a near-extremal black hole.
The metric of the original superstrata in the asymptotically-flat
setting \cite{Bena:2017xbt} has no isometry in the $v$ direction, which
violates one of the assumptions in the analysis of \cite{Eperon:2016cdd}.
However, the asymptotically-flat version of the new single-mode superstrata \eqref{eq:asy-flatsinglemodesuperstrata} is
$v$-independent metrically, and it would be interesting to examine their
possible instability and its endpoint.

The structure of the current paper is as follows. 
In section \ref{Sect:CFT} we give a brief description of the D1-D5 CFT, which is dual to the supergravity picture, and present the family of CFT states whose dual gravitational geometries we want to construct. 
In section \ref{Sect:sugra} we then go to the supergravity side and introduce the setup of BPS equations governing the ansatz quantities that define the fields of our solutions. We then introduce the two-charge geometry which we will use as a seed in our solution-generating technique and present the superstrata that were previously constructed in \cite{Bena:2015bea, Bena:2017xbt}. 
In section \ref{Sect:KillingSpinors} we construct the Killing spinors of empty global AdS$_3 \times S^3 \times T^4$ and use them to generate the fermionic variations of the supergravity fields, which are given in section  \ref{Sect:QQsol}. There we construct the new superstrata solutions to linear order in the perturbation parameter, corresponding to an infinitesimal deformation of AdS$_3 \times S^3 \times T^4$. We then derive the fully backreacted, non-linear solution in section \ref{sect:NonLin}, where we also discuss the asymptotically flat extension, calculate the conserved charges obtained from the geometry, and compare them to the conserved charges calculated on the CFT side.
In section~\ref{Sect:Summary}, we collect formulas for the new
 superstrata and, in addition, present two families of
solutions for which all the excited scalars and forms can be written down in
an explicit way.  The reader who is interested in the explicit form of the
superstrata geometry may find this section useful.
Appendix~\ref{app:susyv} summarises our convention for type IIB
supergravity, in particular, supersymmetry variations, spinors, and
gamma matrices.
In Appendix~\ref{app:ExplRes}, we discuss some technical aspects of the
supersymmetry variations that are not fully covered in the main text.
In Appendix~\ref{app:DegerEtal}, we summarise the spectrum of linear
supergravity around AdS$_3\times S^3$ worked out in \cite{Deger:1998nm}.
This predicts the structure of excited fields in the new superstrata
that we found in this paper, and further suggests other simple kinds of
superstratum that would be interesting to investigate.
In Appendix~\ref{app:RRspinors}, we work out the map between the
Killing spinors in the NS-NS coordinates presented in the main text and
the ones in the RR coordinates.

\section{A CFT starter}
\label{Sect:CFT}

According to the AdS/CFT duality, type IIB string theory on AdS$_3
\times S^3 \times \MM$ (where $\MM$ can be either $T^4$ or $K3$) is
equivalent to a 2D SCFT with $\cN=(4,4)$ supersymmetry, called the D1-D5
CFT\@.\footnote{See {\it e.g.}~\cite{David:2002wn, Avery:2010qw} for
reviews of the D1-D5 CFT.} Besides Virasoro's symmetry, this SCFT has two sets of
fermionic generators $G^{\alpha A}$, $\tilde{G}^{\dot\alpha A}$ and of
bosonic currents $J^{i}$, $\tilde{J}^{\bar{i}}$ that together form a
holomorphic and an anti-holomorphic copy of the small ${\cal N}=4$
superconformal algebra. The Greek indices $\alpha$ and $\dot\alpha$ are
in the fundamental representation of $SU(2)_L$ and $SU(2)_R$
respectively: this $SU(2)_L \times SU(2)_R$ is the R-symmetry of the
theory.  The indices $i$ and $\bar{i}$ are the triplet indices of
$SU(2)_L$ and $SU(2)_R$, respectively.  The index $A$ is in the
fundamental representation of another $SU(2)_B$, which acts as an outer
automorphism on the superalgebra. Similar to ${\cal N}=4$
Super-Yang-Mills theory, this holographic SCFT also has a free locus in
its moduli space where it can be described by a collection of free
fields\footnote{In order to have a free theory description of the $K3$
case one can as usual consider the orbifold limit $K3 =
T^4/\mathbb{Z}_2$, where the fields with an odd number of $\dot{A}$
indices are odd under $\mathbb{Z}_2$.  See {\it e.g.}~\cite{Bena:2017xbt} for a discussion on the moduli space and the
position of the free orbifold point in it.}: $\partial
X_{A\dot{A}\,(r)},\; \psi^{\alpha\dot{A}}_{(r)}$ in the holomorphic and
similarly $\bar\partial X_{A\dot{A}\,(r)},\;
\tilde\psi^{\dot\alpha\dot{A}}_{(r)}$ in the anti-holomorphic sector,
where $r=1,\ldots, N$, so the total central charge is $c=6N$. Notice
that the new type of indices $\dot{A}$, appearing on the free fields,
belongs to a $SU(2)_C$ that is not part of the symmetry group of the
theory. Since the bosonic fields are in the $({\bf 2},{\bf 2})$ representation of
$SU(2)_C \times SU(2)_B$, one can think of this group as acting on the
tangent space of the target space $\MM$ of one boson. The free locus is
described by $N$ copies of the elementary fields which need to be
treated as identical, so the full target space is the orbifold
$\MM^N/{\cal S}_N$, where ${\cal S}_N$ is the symmetric group acting on
the index $(r)$ labelling the copies.

In the following we will use the anomaly-free part of the small ${\cal N}=4$ superconformal algebra in the NS-NS sector,
\begin{equation}
  \label{n4algc0}
\begin{aligned}
  & [L_m,L_n]=(m-n) L_{m+n}\;,~~~ [J^j_0,J^k_0] = i \epsilon^{jkl}J^l_0\;,~~~[L_n,J^{\alpha\beta}_0] = 0 \;,\\
  & \{G^{\alpha A}_r,G^{\beta B}_s\} = \epsilon^{\alpha\beta} \epsilon^{AB} L_{r+s} + (r-s) \epsilon^{AB}  (\sigma^{i\, T})^\alpha_{~\gamma} \epsilon^{\gamma\beta} J^{i}_{r+s}\;,\\
  & [J^j_0,G_s^{\alpha A}] = \frac{1}{2} G^{\beta A}_s (\sigma^j)^{~\alpha}_{\beta} \;,~~~ [L_m,  G^{\alpha A}_{s}] = \left(\frac{m}{2} - s\right) G^{\alpha A}_{m+s}\;,
\end{aligned}
\end{equation}
with $n,m=-1,0,1$ and $s,r=\pm \frac 12$, while $(\sigma^j)^{~\alpha}_{\beta}$ are the Pauli matrices and all $SU(2)$ indices are lowered/raised with the $\epsilon$ satisfying\footnote{We will use $\alpha=+,-$ and $A=1,2$ to highlight the difference between the R-symmetry and the outer automorphism indices.} $\epsilon_{12}=\epsilon_{+-}=\epsilon^{21}=\epsilon^{-+}=1$. When $\MM$ is $T^4$, there are additional $U(1)$ currents that in the free theory description are simply $\sum_{r=1}^N \partial X_{A\dot{A}\,(r)}$ and $\sum_{r=1}^N \bar\partial X_{A\dot{A}\,(r)}$. They will play no role in our discussion which is valid for both the $T^4$ and $K3$ cases.

In the NS-NS sector, $SL(2,\mathbb{C})$-invariant vacuum $|0\rangle$ satisfies $L_n |0\rangle =\tilde{L}_n |0\rangle =0$ for any $n \geq -1$ and  $G_{r}^{\alpha A} |0\rangle = \tilde{G}_{r}^{\dot\alpha A} |0\rangle =0$ for $r\geq - 1/2$ which implies $J^{i}_n |0\rangle =\tilde{J}^{\bar{i}}_n |0\rangle =0$ for $n\geq -1$. In what follows a particular kind of state, called an anti-chiral primary state, will play the central role in the construction of new superstrata solutions. An anti-chiral state  $|s\rangle$ satisfies
\begin{equation}
  \label{eq:aCPO}
  \begin{aligned}
   L_n |s\rangle =0\;,~~n\geq 1\;;~~~ G^{\alpha A}_r |s\rangle = 0 \;,&~~r\geq\frac{1}{2}\;;~~~J^-_n |s\rangle =0\;,~~n\geq 0\;; \\
   J^+_n |s\rangle =  J^3_n |s\rangle =0\;,~~n\geq 1\;;&~~~L_0 |s\rangle = - J^{3}_0 |s\rangle = h |s\rangle\;,
  \end{aligned}
\end{equation}
where, as usual,
$J^\pm= J^1\pm i J^2
$. A simple example of such a state  in the free theory is $\left|O^{--} \right\rangle \equiv O^{--} \left|0\right\rangle $ where\footnote{In our conventions the OPEs between the elementary fields are
  \begin{equation}
    \partial X_{A\dot{A}}(z_1) \partial X_{B\dot{B}}(z_2) \sim \frac{\epsilon_{AB}\epsilon_{\dot{A}\dot{B}}}{(z_1-z_2)^2}\;,~~~\psi^{\alpha \dot{A}}(z_1) \psi^{\beta \dot{B}} \sim -\frac{{\epsilon^{\alpha \beta}}\epsilon^{\dot{A}\dot{B}}}{z_1-z_2}\;
    \end{equation}
    and $G^{\alpha A} =\sum_{r=1}^N \psi^{\alpha \dot{A}}_{(r)} \partial X_{A \dot{B}\,(r)}$. Similar relations hold for the anti-holomorphic sector.}
\begin{equation}
  \label{eq:Oab}
  O^{\alpha \dot\alpha} = -\frac{i}{\sqrt{2 N}} \sum_{r=1}^N \psi^{\alpha \dot{A}}_{(r)}\, \tilde{\psi}^{\dot\alpha \dot{B}}_{(r)}\, \epsilon_{\dot{A}\dot{B}}\;,
\end{equation}
which has $h=\bar{h}=1/2$. There is a family of anti-chiral primary operators\footnote{The subscript $[k]$ in the square parenthesis refers to the order of the twisted sector where the operator under discussion lives.} $\Sigma_{[k]}^{--}$ with $h=\bar{h}=\frac{k-1}{2}$ that, at the free locus, live in the twisted sectors of the orbifold ${\cal S}_N$ mentioned above. These operators change the boundary conditions of the elementary fields. For instance when acting on the vacuum they link together $k$ copies of the elementary fields into a single object that we call a ``strand''. In the following, we will use the anti-chiral primary $O_{[k]}^{--}$ with $h=\bar{h}=\frac{k}{2}$, where a strand of length $k$ is further excited by a holomorphic and an anti-holomorphic elementary fermionic field and, for the corresponding states, we introduce the notation
\begin{equation}
  \label{eq:stateOk}
   |O^{--}\rangle_k = \lim_{z,\bar{z}\to 0} O^{--}_{[k]} |0\rangle\,.
\end{equation}
Acting with $J^+_0$, $L_{-1}$ and $G^{+A}_{-\frac 12}$ on $ |O^{--}\rangle_k$ we can obtain new bosonic states in the same multiplet
\begin{equation}
  \label{eq:stateOkmnq}
   |k,m,n,q\rangle^{\rm NS} = (J^+_0)^m (L_{-1})^n  \left(G_{-\frac12}^{+1}G_{-\frac12}^{+2} + \frac{1}{k} J^+_0 L_{-1}\right)^q |O^{--}\rangle_k \;,
\end{equation}
where $m\leq k -2q$, $q=0,1$, otherwise the state is trivially zero, while $n=0,1,2 \ldots$ can be any non-negative integer. The eigenvalues $(h,j)$ of $L_0$ and $J^3_0$ are $h=\frac{k}{2}+n+q$ and $j=-\frac{k}{2}+m+q$, while $(\bar{h},\bar{j})$ are unchanged. Notice that, due to the commutation relations~\eqref{n4algc0}, the order of the operators in~\eqref{eq:stateOkmnq} is immaterial. The combination in the parenthesis (weighted by $q$) is chosen so as to make the states $|k,m+1,n+1,0\rangle^{\rm NS}$ and $|k,m,n,1\rangle^{\rm NS}$ orthogonal, which means that, under the AdS/CFT dictionary, they will correspond to two independent supergravity perturbations. It is straightforward to check this by using the commutation relations~\eqref{n4algc0}. It is easier to start with the $n=0$ case:
\begin{equation}
  \label{eq:checkor}
  \begin{aligned}
    {}^{\rm NS}\langle k,m+1,1,0|k,m,0,1\rangle^{\rm NS} = &~ {}^{\rm NS}\langle k,m+1,0,0|\, L_1 \left(G_{-\frac12}^{+1}G_{-\frac12}^{+2} + \frac{1}{k} J^+_0 L_{-1}\right) |k,m,0,0\rangle^{\rm NS} \\ = &~  {}^{\rm NS}\langle k,m+1,0,0| \left(- J_0^+ + \frac{2}{k} L_0 J_0^+\right)|k,m,0,0\rangle^{\rm NS} = 0\;.
    \end{aligned}
  \end{equation}
This shows that $L_1 |k,m,n,q\rangle^{\rm NS} \sim |k,m,n-1,q\rangle^{\rm NS}$, so we can recursively prove the orthogonality of  $|k,m+1,n+1,0\rangle^{\rm NS}$ and $|k,m,n,1\rangle^{\rm NS}$ for general $n$. Finally notice that the state $|1,0,0,1\rangle^{\rm NS}$ is trivial\footnote{This follows form $G^{- A}_{-\frac 12} |O^{--}\rangle_1 = 0$, while the same does not hold for states of winding $k>1$.} since it has zero norm and so all the states $|1,0,n,1\rangle^{\rm NS}$ are zero since they are constructed from $|1,0,0,1\rangle^{\rm NS}$ by dressing it with powers of $L_{-1}$.

The final CFT ingredient we need is the possibility of realising the superalgebra in equivalent ways by taking the spectral flow of the generators (here we follow the conventions of~\cite{Giusto:2013bda})
\begin{equation}
  \label{eq:spfl}
  T_\nu = T - \frac{2 \nu}{z} J^3 + \frac{c \nu^2}{6 z^2} ~,~~~ J_\nu^3 = J^3 - \frac{c \nu}{6 z}~,~~~ J^\pm_\nu(e^{2\pi i}z) = e^{\mp 2\pi i 2\nu} J_\nu^\pm(z)\;,
\end{equation}
and similarly for the anti-holomorphic sector. When the spectral flow parameter $\nu$ takes half-integer values, the theory is in the RR sector, and in particular, for $\nu=\bar\nu=-1/2$, an anti-chiral primary state flows to RR ground state with $h_{\rm R}=c/24$ and $j_{\rm R}=-h_{\rm NS} + c/12$ (since $j_{\rm NS}=-h_{\rm NS}$ for anti-chiral primary states). Thus, after this spectral flow, the states in~\eqref{eq:stateOkmnq} become excited RR states that we denote as
\begin{equation}
  \label{eq:stateOkmnqeigen} 
  \left|++\right\rangle^{N-k}\, |k,m,n,q\rangle^{\rm R}~,~~~{\rm with}~~~
  h_{\rm R} = \frac{N}{4} + m + n + 2 q\;,~~ j_{\rm R} = \frac{N-k}{2}+m+q ~,
\end{equation}
where we used $c=6N$ and used that the NS-NS vacuum state $\left| 0\right\rangle^{\rm NS}$ goes into the RR ground state $\left| + + \right\rangle$. 

In this work we focus on protected RR states that are dual to smooth supergravity solutions. A nice class of such states is obtained by starting from a NS-NS multi-particle state which is the product of $N_b$ copies of~\eqref{eq:stateOkmnq} and then by performing the spectral flow to the RR sector mentioned above. In this paper we consider states involving just one type of excitation, dual to a single-mode superstrata solution, which in the RR sector then take the following form:
\begin{equation}
  \label{eq:RRsmQ}
  \left|++\right\rangle^{N_a}\, \left(|k,m,n,q\rangle^{\rm R}\right)^{N_b}\;,~~~\mbox{with}~~~  N_a+k N_b = N \,.
\end{equation}
To be precise, the CFT states of this type are written as a coherent sum of terms involving a different number elementary excitations~\eqref{eq:stateOkmnq}, but in the large $N$ limit the sum is sharply peaked ~\cite{Skenderis:2006ah} and~\eqref{eq:RRsmQ} represents the dominant term in the sum. For a discussion of this point in the context of three-charge states, see~\cite{Bena:2017xbt} and reference therein. Here it is sufficient to say that the coherent sum with a peak at~\eqref{eq:RRsmQ} is defined in terms of two continuous parameters $a^2$ and $b^2$, related to $N_a$ and $N_b$ respectively which determine also the dual gravity solution. Such a coherent sum of CFT states will have charges equal to the those of the dominant term~\eqref{eq:RRsmQ}. These charges are given by the eigenvalues of the operators $L_0$, $\bar L_0$, $J_0^3$ and $\bar J_0^3$ which are equal to 
\begin{equation}
  \label{eq:QQchcft}
  h_{\rm R} = \frac{N}{4} + \left(m + n + 2 q\right) N_b\;,~~ \bar{h}_{\rm R} = \frac{N}{4}\;,~~ j_{\rm R} = \frac{N_a}{2}+\left(m+q\right) N_b\;,~~\bar{j}_{\rm R} = \frac{N_a}{2}\;
\end{equation}
respectively, and, as we will see in section \ref{sec:conservedcharges} these results will match precisely the momentum and the angular momenta of the dual supergravity solution.

\section{Supergravity setup}
\label{Sect:sugra}

Our final goal is to find the supergravity solutions dual to the CFT
states \eqref{eq:stateOkmnq} and \eqref{eq:RRsmQ}.
As an illustration in this section we focus on the $q=0$ case,
{\em i.e.}, we review the original superstrata \cite{Bena:2015bea,
  Bena:2016ypk, Bena:2017xbt}, which only involve bosonic
generators. In later sections we will generalise this approach to the
$q=1$ case, the new superstrata, which involve fermionic generators.

\subsection{The BPS ansatz}
\label{Sec:BPSansatz}

We work in Type IIB string theory on $\mathbb{R}^{1,4}\times
S^1 \times \mathcal{M}$, where
the internal space ${\mathcal{M}}$
is either $K3$ or $T^4$.
The D1-D5-P black hole is 1/8-BPS, which means
that it preserves 1/8 of the total 32 supercharges. All of its
microstates must also preserve the same amount of supercharges, meaning
that all of them are 1/8 BPS\@.  Such solutions of type IIB
supergravity that are in addition independent of ${\mathcal{M}}$ are described
by the following ansatz~\cite{Giusto:2013rxa, Giusto:2013bda}
\begin{subequations}\label{ansatzSummary}
 \begin{align}
d s^2_{10} & = \sqrt{\alpha}\,ds_6^2 + \sqrt{\frac{Z_1}{Z_2}}\,d\hat{s}^2_4,\label{10dmetric}\\
ds_6^2  &=-\frac{2}{\sqrt{\cP}}(d v+\beta)\Big[d u+\omega + \frac{\mathcal{F}}{2}(d v+\beta)\Big]+\sqrt{\cP}d s^2_4,\\
e^{2\Phi}&={Z_1^2\over \cP} ,\\
B_2&= -\frac{Z_4}{\cP}(d u+\omega) \wedge(d v+\beta)+ a_4 \wedge  (d v+\beta) + \delta_2, \label{Bform}\\ 
C_0&=\frac{Z_4}{Z_1} ,\\
C_2 &= -\frac{\alpha}{Z_1}(d u+\omega) \wedge(d v+\beta)+ a_1 \wedge  (d v+\beta) + \gamma_2,\\ 
C_4 &= \frac{Z_4}{Z_2} \widehat{\mathrm{vol}}_{4} - \frac{Z_4}{\cP}\gamma_2\wedge (d u+\omega) \wedge(d v+\beta)+x_3\wedge(d v + \beta) ,\\
C_6 &=\widehat{\mathrm{vol}}_{4} \wedge \left[ -{Z_1\over \cP}(d u+\omega) \wedge(d v+\beta)+ a_2 \wedge  (d v+\beta) + \gamma_1\right] \label{jtta10Apr18} 
\end{align}
\end{subequations}
with 
\begin{align}
\alpha \equiv \frac{Z_1 Z_2}{Z_1 Z_2 - Z_4^2},\qquad
\cP \equiv Z_1\,Z_2 - Z_4^2.
\end{align}
We have defined new asymptotically null coordinates $u$ and $v$ as
\begin{align}\label{sptvw}
u \equiv \frac{1}{\sqrt{2}}(t-y), \qquad \qquad v \equiv \frac{1}{\sqrt{2}}(t+y), 
\end{align}
where $t$ is the time coordinate and $y$ parametrises the $S^1$, which
has total length $2\pi R_y$. These new coordinates can be thought of
as world-volume coordinates of the dual CFT\@. In the above ansatz,
$ds_{10}^2$ denotes the string-frame metric of the ten-dimensional
spacetime, and $ds_6^2$ denotes the Einstein-frame metric of the
six-dimensional spacetime, which is a fibration over a 4-dimensional
base $\mathcal{B}$ with metric $ds_4^2$ which may depend on $v$.
The ansatz includes scalars $Z_1, Z_2, Z_4, \cF$; one-forms $\beta,
\omega, a_1, a_2, a_4$; two-forms $\gamma_1, \gamma_2, \delta_2$; and a
three-form $x_3$, all on $\mathcal{B}$.  These quantities can
depend on the coordinates of $\mathcal{B}$ and on $v$, but supersymmetry
requires them to be independent of $u$.  The RR potentials $C_p$ can have
an extra term proportional to a four-form $\cC$ on $\mathcal{B}$, but it
has been set to zero by using the gauge symmetries discussed in
\cite{Bena:2015bea}.

The advantage of the above ansatz is that the functions and forms used to express the  field content of the theory obey BPS differential equations which one can organise into layers and solve in successive steps, using the solutions to the previous layer as sources in the next one. 
The equations have three layers, which we call the zeroth, first, and second. The zeroth level gives the equations for the base space metric $ds_4^2$ and one-form $\beta$. This system of equations is non-linear and hard to solve. Hence we make an assumption that the base space $\mathcal{B}$ is $\mathbb{R}^4$ equipped with a flat metric $ds_4^2$  and that the one-form $\beta$ is $v$-independent. With these two assumptions, the zeroth-layer equations reduce only to the self-duality condition of $\beta$ on the base space,
\begin{align}
d\beta = *_4 d\beta,
\end{align}
where $*_4$ denotes the Hodge dual operator on
$\mathcal{B}=\mathbb{R}^4$.  The assumption of the simple base provides us
with a tractable class of solutions that can be explicitly written down,
although we must keep in mind that it limits how generic our solutions
are.

The first-layer equations are a set of linear equations
\begin{subequations}\label{layer1}
\begin{align}
*_4\cD\dot{Z}_1 = \cD \Theta_2, \qquad \cD *_4 \cD Z_1 = - \Theta_2 \wedge d\beta,\qquad \Theta_2 = *_4 \Theta_2,\\
*_4\cD\dot{Z}_2 = \cD \Theta_1, \qquad \cD *_4 \cD Z_2 = - \Theta_1 \wedge d\beta,\qquad \Theta_1 = *_4 \Theta_1,\\
*_4\cD\dot{Z}_4 = \cD \Theta_4, \qquad \cD *_4 \cD Z_4 = - \Theta_4 \wedge d\beta,\qquad \Theta_4 = *_4 \Theta_4,
\end{align}
\end{subequations}
where we have introduced gauge invariant\footnote{Invariant with respect to the gauge transformations given in (2.14) of \cite{Bena:2015bea}.} two-forms 
\begin{align}\label{theta2form}
\Theta_1 \equiv \cD a_1 + \dot{\gamma}_2, \qquad \Theta_2 \equiv  \cD a_2 + \dot{\gamma}_1, \qquad \Theta_4 \equiv \cD a_4 + \dot{\delta}_2, 
\end{align}
and defined a gauge covariant differential operator
\begin{align}
\cD \equiv d_4 - \beta \wedge \frac{\partial}{\partial v}, 
\end{align}
where $d_4$ denotes the exterior derivative on the base space. The above equations describe three systems of equations for $(Z_1, \Theta_2)$, $(Z_2, \Theta_1)$, and $(Z_4, \Theta_4)$, which have the same structure. Thus if, for example, we find a solution for the pair $(Z_4, \Theta_4)$ then we have also found a solution for the other two pairs. It is important to note that once we have solutions to the zeroth-layer equations, the first-layer equations can be fully solved at least in principle. 

The second layer of equations is a set of linear differential equations that determine  $\cF$ and  $\omega$. These are of special interest as their long distance behaviour allows us to read off the momentum and angular momentum charges of the geometry. The two equations are
\begin{subequations}\label{layer2eq}
\begin{align}\label{layer2eq1}
&\cD \omega + *_4 \cD \omega +\cF d\beta = Z_1 \Theta_1 + Z_2 \Theta_2 - 2 Z_4 \Theta_4,\\
&*_4 \cD *_4\left(\dot{\omega}- \frac12 \cD \cF\right) = \partial_v^2(Z_1 Z_2 - Z_4^2)- (\dot{Z}_1\dot{Z_2}- (\dot{Z}_4)^2)- \frac12 *_4(\Theta_1\wedge \Theta_2 - \Theta_4 \wedge \Theta_4).\label{layer2eq2}
\end{align}
\end{subequations}
The solutions to the first-layer equations serve as sources on the right-hand side of the differential equations. It is also important that these sources appear quadratically in the right-hand side of second-layer equations. Hence any first-order perturbation in the first-layer quantities becomes a second-order perturbation in the second layer. This fact is useful in the solution-generating technique that is used to build new solutions.


\subsection{Seed solution}
The construction of three-charge black holes via the superstratum
technology begins with a two-charge seed solution. All two-charge
solutions are known \cite{Lunin:2001jy, Lunin:2002bj, Lunin:2002iz, 
Kanitscheider:2007wq}, and can be obtained by a
systematic procedure using an auxiliary curve in $\mathbb{R}^8$, which
determines the values of the ansatz quantities of
\eqref{ansatzSummary}. For a short overview of the construction of
two-charge solutions and especially the one used in the subsequent
sections, see, for instance, Appendix A of \cite{Bena:2015bea}.

The seed solution used for the construction of superstrata solutions is
given as a perturbed round supertube solution \cite{Kanitscheider:2007wq}.
We start off by parameterising the base space $\mathbb{R}^4$ with a new set of
coordinates related to the Cartesian ones by
\begin{align}\label{newcoords}
x_1 + i x_2 = \sqrt{r^2 + a^2}\sin\theta\, e^{i\phi}, \qquad x_3 + i x_4 = r \cos\theta\, e^{i\psi},
\end{align}
where $\theta\in[0,{\pi\over 2}]$ and $\phi,\psi\in[0,2\pi)$. 
The ansatz quantities in these coordinates are given by\footnote{Here we focus on the asymptotically AdS solutions, so we write $Z_1$ and $Z_2$ after taking the decoupling limit.} 
\begin{subequations}
\label{seed}
\begin{align}
& d s^2_4 = (r^2+a^2 \cos^2\theta)\left(\frac{d r^2}{r^2+a^2}+ d\theta^2\right)+(r^2+a^2)\sin^2\theta\,d\phi^2+r^2 \cos^2\theta\,d\psi^2\,,\label{stmetric}\\
&Z_1 = \frac{Q_1}{\Sigma}\,,\qquad Z_2 = \frac{Q_5}{\Sigma}\,,\qquad\\
& \beta =  \frac{R\,a^2}{\sqrt{2}\,\Sigma}\,(\sin^2\theta\, d\phi - \cos^2\theta\,d\psi)\equiv \beta_0\,,\qquad \omega=\frac{R\,a^2}{\sqrt{2}\,\Sigma}\,(\sin^2\theta\, d\phi + \cos^2\theta\,d\psi)\equiv \omega_0 \,,\label{seedbeta}\\
&\gamma_2 = - Q_5 \frac{(r^2 + a^2)\cos^2\theta}{\Sigma}d\phi \wedge d\psi,\\
&Z_4 = R_y b a^k \frac{\sin^k\theta \,e^{-ik \phi}}{(r^2 + a^2)^{k/2}\Sigma}\,,\\
&\delta_2 =- R_y a^{k}b\frac{\sin ^k\theta }{ \left(r^2+a^2\right)^{k/2}}  \left[\frac{\left(r^2+a^2\right)\cos^2\theta \,e^{-i k \phi }}{\Sigma} \, d\phi \wedge  d\psi +i \frac{\cos{\theta}}{\sin{\theta}}\, e^{-i k \phi }\,d\theta\wedge d\psi \right],\\
   & \mathcal{F}= a_{1,4} = x_3 = \gamma_1 = \Theta_{1,2,4} = 0,
\end{align}
\end{subequations}
where we have introduced $$\Sigma \equiv r^2 + a^2 \cos^2\theta\,,$$ and
the parameter $k$ is a positive integer. 
We take the parameter $b$ to be small and only keep~$\cO(b)$ terms.
Hence we can think of the solution \eqref{seed} as a combination of the
background supertube ($b = 0$) \cite{Balasubramanian:2000rt,Maldacena:2000dr,Lunin:2001fv}, with an added perturbation ($b \neq 0$),
which turns on form fields $B_2,$ $C_0$, and $C_4$.
Because $Z_4$ appears only quadratically in the
10-dimensional metric \eqref{10dmetric}, the perturbation does not
change the metric at linear order in $b$. In the same approximation the parameter $a$ is related to the D-brane charges $Q_1$ and $Q_5$ via
\begin{align}
a  = \frac{\sqrt{Q_1 Q_5}}{R_y}.
\end{align}
It is not difficult to check that the above ansatz satisfies the BPS
equations \eqref{layer1} and \eqref{layer2eq} to linear order.  One
could consider the finite $b$ version of the above seed solution
\cite{Bena:2015bea}, but that is not necessary for our purposes.

Performing a coordinate transformation
\begin{align}
\label{spectralflowcoord}
\phit = \phi - \frac{t}{R_y}, \qquad \psit = \psi - \frac{y}{R_y},
\end{align}
one finds that the six-dimensional metric becomes 
\begin{align}\label{stmetric1}
 ds_6^2 ~=\; \sqrt{Q_1 Q_5}
 \left(- \;\! \frac{r^2+a^2}{a^2 R_y^2}dt^2 +\frac{r^2}{a^2 R_y^2}dy^2 
 +\frac{dr^2}{r^2+a^2}
 +d\theta^2+\sin ^2\theta\, d\phit^2+\cos ^2\theta \,d\psit^2 
 \right)\,,
\end{align}
which is that of global AdS$_3\times S^3$
with radius $R_{\rm AdS}^2 = \sqrt{Q_1 Q_5}$. 
Therefore, our seed solution represents AdS$_3 \times S^3 \times
T^4$ with a linear perturbation on top of it.  On the CFT side, this
corresponds to the NS-NS vacuum with a small excitation on
top\footnote{For this reason we will refer to the case with $b = 0$ as
the vacuum or the background geometry, while reserving the term seed
solution for the perturbed vacuum with $b \neq 0$.}.  Since the 
NS-NS vacuum is invariant under the action of $SL(2, \mathbb{C})\cong SL(2, \mathbb{R})_L\times SL(2, \mathbb{R})_R$ and
$SU(2)_L\times SU(2)_R$ transformations, acting on the excited state
with the generators of these symmetries generates a new state, which is
again the NS-NS vacuum with a different small excitation added to
it. Performing the corresponding transformations on the seed solution on the
gravity side similarly generates pure AdS$_3 \times S^3 \times T^4$ with
a different linear perturbation on top of it. This new solution will
again satisfy the BPS equations and, in addition to it, we will precisely
know the CFT state dual to the new geometry.

It is important to note that the transformation
\eqref{spectralflowcoord} involves a change of the
coordinates~$t$~and~$y$, which parametrise the boundary region of
AdS$_3$. This means that this transformation also affects the dual CFT
theory. In fact this transformation is dual to the spectral flow
transformation that changes the RR sector into the NS-NS sector on the CFT side
\eqref{eq:spfl}. For that reason we will refer to the coordinates
$(r,t,y, \theta, \phi, \psi)$ as the RR coordinates and $(r,t,y, \theta,
\phit, \psit)$ as the NS-NS coordinates.

At this point we make the first identification between the supergravity
solution and a dual CFT state. Using the standard AdS/CFT dictionary
the seed \eqref{seed}
expressed in the RR coordinates is dual to a RR ground state.
After performing the spectral flow transformation
\eqref{spectralflowcoord} to the NS-NS coordinates, the geometry is dual
to an anti-chiral primary with the conformal dimensions $h =
\overline{h}= \frac{k}{2}$ and $j = \overline{j} = - \frac{k}{2}$. We
thus identify the supertube ansatz expressed in the NS-NS coordinates is
dual to the state \eqref{eq:stateOk}. As usual, the gravity and the free CFT descriptions are valid in different points of the moduli space, so the dictionary mentioned above is meaningful when applied to protected quantities such as the 3-point correlators \cite{Baggio:2012rr}.

\subsection{Original superstrata}

This work follows up on the work of \cite{Giusto:2013bda, Bena:2015bea,
Bena:2017xbt}, where new superstrata solutions were obtained by acting
on the seed solution \eqref{seed} with the gravity realisations of the
bosonic symmetry generators $J_0^{+}$ and $L_{-1}$, thus finding the geometries
dual to the CFT states \eqref{eq:stateOkmnq} with $q = 0$. Below we will
review the construction of this class of superstrata.

As mentioned above, the background is invariant under the action of $SL(2,
\mathbb{R})_L \times SL(2, \mathbb{R})_R$ and $SU(2)_L \times SU(2)_R$,
while on the other hand the perturbation is not. Acting on the perturbed
geometry with these generators will give us a new perturbed
geometry. Focusing on the left sector of the theory and the
corresponding symmetry groups $SL(2, \mathbb{R})_L \times SU(2)_L$, one
can show that the generators of these symmetries can be explicitly
realised in the in the NS-NS coordinates as \cite{Maldacena:1998bw,
Bena:2017xbt}
\begin{gather}
\begin{split}
 L_0&={i R_y \over 2}(\partial_t+\partial_y),\\
 L_{\pm 1}
 &=ie^{\pm {i\over R_y}(t+y)}
 \biggl[
 -{R_y\over 2}\biggl({r\over \sqrt{r^2+a^2}}\partial_t+{\sqrt{r^2+a^2}\over r}\partial_y\biggr)
 \pm {i\over 2}\sqrt{r^2+a^2}\,\partial_r
 \biggr],
\end{split} 
\label{sl(2,R)_gen_NS}
\\
 J_0^3=-{i\over 2}(\partial_{\tilde\phi}+\partial_{\tilde\psi}),\quad
  J_0^\pm ={i\over 2}e^{\pm i(\tilde\phi+\tilde\psi)}
 (\mp i\partial_\theta+\cot\theta\, \partial_{\tilde\phi}-\tan\theta\, \partial_{\tilde\psi})\,,
\label{su(2)_gen_NS}
\end{gather}
which can be shown to satisfy the algebra \eqref{n4algc0}.

Acting on the seed solution \eqref{seed} with \eqref{su(2)_gen_NS} $m$ times and with \eqref{sl(2,R)_gen_NS} $n$ times leaves all ansatz quantities unchanged at linear order\footnote{{ See after \eqref{eq:LIstate} for a comment on the precise relation between $b_4$ and $b$}.} in $b_4 \sim b$ except for the function $Z_4$ and the two-form $\Theta_4$, which are now given by
\begin{align}
\label{Z4Th4_solngen_O(b)}
Z_4 &= b_4 \,z_{k,m,n}\,, \qquad \Theta_4 = b_4\, \vartheta_{k,m,n}\, 
\end{align} 
where we have introduced the notation
\begin{subequations}
\begin{align}
 z_{k,m,n} &=\,R_y\,\frac{\Delta_{k,m,n}}{\Sigma}\, \cos{\hat{v}_{k,m,n}},
\\
 \vartheta_{k,m,n}&=
 -\sqrt{2}\,
 \Delta_{k,m,n}
\biggl[\left((m+n)\,r\sin\theta +n\left({m\over k}-1\right){\Sigma\over r \sin\theta}  \right)\Omega^{(1)}\sin{\hat{v}_{k,m,n}}\notag\\
&\hspace{20ex}
 +\left(m\left({n\over k}+1\right)\Omega^{(2)} +\left({m\over k}-1\right)n\, \Omega^{(3)}\right) \cos{\hat{v}_{k,m,n}} \biggr]
\,,\label{oldTheta}
\end{align} 
\end{subequations}
with
\begin{align}
\begin{aligned}
  \Delta_{k,m,n} &\equiv
 \left(\frac{a}{\sqrt{r^2+a^2}}\right)^k
 \left(\frac{r}{\sqrt{r^2+a^2}}\right)^n 
 \cos^{m}\theta \, \sin^{k-m}\theta \,, 
 \\
 \hat{v}_{k,m,n} &\equiv (m+n) \frac{\sqrt{2}\,v}{R_y} + (k-m)\phi - m\psi \,,
\end{aligned} 
\label{Delta_v_kmn_def}
\end{align}
and we have expanded $\vartheta_{k,m,n}$ on a basis of self-dual two-forms  $\Omega^{(i)}$ ($i=1,2,3$) on
$\mathbb{R}^4$, given by:
\begin{equation}\label{selfdualbasis}
\begin{aligned}
\Omega^{(1)} &\equiv \frac{dr\wedge d\theta}{(r^2+a^2)\cos\theta} + \frac{r\sin\theta}{\Sigma} d\phi\wedge d\psi\,,\\
\Omega^{(2)} &\equiv  \frac{r}{r^2+a^2} dr\wedge d\psi + \tan\theta\, d\theta\wedge d\phi\,,\\
 \Omega^{(3)} &\equiv \frac{dr\wedge d\phi}{r} - \cot\theta\, d\theta\wedge d\psi\,.
\end{aligned}
\end{equation}
One of the important aspects of these solutions is that they now depend
on the variable $v$, unlike the seed solution \eqref{seed}.

The above solution-generating technique allows us to generate a family
of solutions, parametrised by the quantum numbers $(k,m,n)$. Although
the above solutions \eqref{Z4Th4_solngen_O(b)} only involve one mode at
a time, using the linearity of the first-layer BPS equations, we can
write down the general solution given by an arbitrary superposition of
modes with different quantum numbers. This superposition will
not only solve the first-layer BPS equations for the $Z_4$ and
$\Theta_4$, but also for the other two pairs $(Z_1, \Theta_2)$ and
$(Z_2, \Theta_1)$, because the structure of their equations is
identical. Hence the general class of solutions with $q = 0$ for the
first-layer equations are given by
\begin{align}
 \label{eq:ZIThI_gen_infinitesimal}
 \begin{aligned}
 Z_1 &= \frac{Q_1}{\Sigma} + \sum\limits_{k,m,n} b_1^{k,m,n}\, z_{k,m,n} \,,  &
 Z_2 &= \frac{Q_5}{\Sigma} + \sum\limits_{k,m,n} b_2^{k,m,n}\, z_{k,m,n}\,, &
 Z_4 & = \sum\limits_{k,m,n} b_4^{k,m,n}\, z_{k,m,n}\,,
\\
 \Theta_1 &= \sum_{k,m,n} b_{2}^{k,m,n}\, \vartheta_{k,m,n}\,, &
 \Theta_2 &= \sum_{k,m,n} b_{1}^{k,m,n}\, \vartheta_{k,m,n}\,, &
 \Theta_4 &=  \sum_{k,m,n} b_{4}^{k,m,n}\, \vartheta_{k,m,n}\,,
 \end{aligned}
\end{align}
where $Z_1$ and $Z_2$ also include the zero modes, which appear in the
seed solution. In these superpositions the coefficients $b_I^{k,m,n}$
are still taken to be infinitesimal.  In this case they generate
second-order source terms on the right-hand side of the second-layer BPS
equations, which govern the change of $\cF$ and $\omega$. Thus to linear
order in the perturbation parameter, these two ansatz quantities remain
the same. This infinitesimal solution corresponds to an infinitesimal
deformation of the empty global AdS$_3\times S^3 \times T^4$ spacetime.

However, we can again use the linearity of the first-layer BPS equation
to make all the coefficients $b_I^{k,m,n}$ finite, and the solution
\eqref{eq:ZIThI_gen_infinitesimal} will still remain a solution to the
first-layer equations. With the coefficients being finite, we have
non-vanishing source terms on the right-hand side of the second-layer
BPS equations. Solving these represents a challenge. In
\cite{Bena:2017xbt}, a general solution to the second-layer
equations for single-mode superstrata was found. These solutions
correspond to configurations with a single non-trivial coefficients
$b_4^{k,m,n} $ in \eqref{eq:ZIThI_gen_infinitesimal}.
The lesson from the non-linear solution-generating
technique employed in \cite{Giusto:2013rxa,Giusto:2013bda}
is that the descendant states have $b_2^{k,m,n} = 0$ for all values
of $k,m,n$ and this is also the case for the single-mode superstrata, where
one can consistently set $b_2 = 0$ \cite{Bena:2017xbt}.
Furthermore, $b_1$ was determined by ``coiffuring'',
which in the single-mode case requires that the $v$-dependent
source terms on the right-hand side of second-layer equations
vanish. In the case of $b_2 = 0$,
this corresponds to setting $b_1 = b_4^2$, thus having the solutions to
the first layer given by
\begin{subequations}
\begin{align}
&Z_1 ~=~  \frac{Q_1}{\Sigma} + \frac{b_4^2\, R_y^2 }{2 Q_5} \,
\frac{\Delta_{2k,2m,2n}}{\Sigma} \cos \vh_{2k,2m,2n} \,,   \qquad Z_2 ~=~ \frac{Q_5}{\Sigma} \,,\nonumber  \\ 
& \hspace{5ex} Z_4~=~    R_y  \, b_4\, \frac{ \Delta_{k,m,n} }{\Sigma} \cos \vh_{k,m,n},  \, \label{eq:ZIAdSsinglemode} \\
& \Theta_1 =0\,,\qquad
 \Theta_2 = \frac{b_4^2 R_y}{2 Q_5}\, \vartheta_{2k,2m,2n}\,,\qquad
 \Theta_4 = b_4\, \vartheta_{k,m,n}\,.\label{eq:ThetaIAdSsinglemode}
\end{align}
\end{subequations}
By using these terms in the source terms of the second-layer equations, one finds that the second-layer quantities are given by 
\begin{align}
\omega^{\rm orig}_{k,m,n} = \omega_0 + \omega^{\rm orig,RMS}_{k,m,n}\,, \qquad \cF = \cF^{\rm orig}_{k,m,n} \,, 
\end{align}
where we further decompose 
\begin{align}
\omega^{\rm orig,RMS}_{k,m,n}  = \mu_{k,m,n}^{\rm orig} (d\psit + d\phit) + \nu_{k,m,n}^{\rm orig}(d\psit - d\phit)\,.
\end{align}
One can show that the solutions for the second-layer equations are given by
\begin{align}
\label{cF}
\cF_{k,m,n}^{\rm orig} &= 4b_4^2\biggl[\frac{m^2 (k+n)^2}{k^2}\,F_{2k,2m,2n}+\frac{n^2 (k-m)^2}{k^2}\,F_{2k,2m+2,2n-2}\biggr],
 \\
\mu_{k,m,n}^{\rm orig}&= \frac{R_y\,b_4^2}{\sqrt{2}}\,\biggl[ 
\frac{(k-m)^2(k+n)^2}{k^2} F_{2k,2m+2,2n}
+\frac{m^2 n^2}{k^2} F_{2k,2m,2n-2}
\notag\\
&\hspace{15ex}
 -\frac{r^2+a^2\,\sin^2\theta}{4\,\Sigma}\,b_4^{-2}\mathcal{F}_{k,m,n}
-\frac{\Delta_{2k,2m,2n}}{4\,\Sigma}
+\frac{x_{k,m,n}}{4\,\Sigma}
\biggr]\,,
\label{mu}
\end{align} 
where the function $F_{2k, 2m,2n}$ is defined in \eqref{Ffun}, and the
functions $\nu_{k,m,n}^{\rm orig}$ are given by differential equations
\cite[(4.13)]{Bena:2017xbt}, and can be solved for each case
individually. We have put the superscript ``orig'' to distinguish these
original superstrata solutions from the new superstrata we are presenting below.
The solutions obtained this way are asymptotically
AdS$_3\times S^3 \times T^4$. To obtain asymptotically flat solutions,
one needs to add ``1'' to the warp factors $Z_1$ and $Z_2$. This
alters the right-hand side of second-layer equations and induces
new $v$-dependent terms into $\cF$ and $\omega$~\cite{Bena:2017xbt}.
As it will turn out that our new superstrata do not generate these additional
$v$-dependent terms in the case of asymptotically flat case, the
asymptotically flat extension of the new solutions will be simpler
than those of the original superstrata.

\section{Killing spinors of the AdS$_3\times S^3$ background}
\label{Sect:KillingSpinors}

In the previous section, we reviewed the construction of the original
superstrata, which are dual to the states \eqref{eq:stateOkmnq} with $q
=0$ and are generated by the supergravity realisation~\eqref{sl(2,R)_gen_NS} and~\eqref{su(2)_gen_NS} of the bosonic generators
$J_0^+$ and $L_{-1}$. We now proceed to the construction of new
superstrata, which have $q=1$ and involve fermionic generators
$G_{-\frac12}^{+A}$.  In supergravity, these fermionic generators
correspond to the Killing spinors of the AdS$_3\times S^3 \times T^4$
background.
To begin with, in this section, we work out the explicit form of these 
Killing spinors, and give a precise map between their components and the
fermionic generators $G_{\pm\half}^{\alpha A}$ in the CFT\@.

\subsection{Supersymmetry variations}
\label{Sect:nscoord}

The supersymmetry variations for the dilatino $\lambda$ and the
gravitini fields $\psi_M$ in type IIB supergravity are given
by\footnote{Here we follow~\cite{Bergshoeff:2001pv}: a brief summary is included in Appendix \ref{app:susyv} where more details on the spinor conventions can be found.}
\begin{subequations}\label{fervar1}
\begin{align}
\delta \lambda& = \left(\slashed{d}\phi -\frac12 \slashed H\,\sigma^3  - e^\phi \slashed F_1 \,(i\sigma^2) -{1\over 2} e^\phi\, \slashed F_3\, \sigma^1 \right) \epsilon \,, \\
\delta \psi_M&=  \left[ \nabla_M- \frac18 H_{MNP}\Gamma^{NP} \,\sigma^3  + \frac18 e^\phi \left(\slashed F_1\, (i\sigma^2) + \slashed F_3 \, \sigma^1 + \frac12 \slashed F_5\, (i\sigma^2)\right)\Gamma_M\right]  \epsilon  \, ,
\end{align}
\end{subequations}
see~\eqref{eq:conf1} and \eqref{eq:conf2} for our conventions on the form fields. In type IIB supergravity, each fermionic field appears in two copies, which we combine into a doublet $\epsilon \equiv (\epsilon^1, \epsilon^2)$.
The Pauli matrices $\sigma^i$ in the variations above act on the doublet indices, which will be made explicit in the following calculations when relevant.
The variations \eqref{fervar1} hold in a generic coordinate system in Type IIB supergravity. In our previous discussion, we have introduced two parametrisations of the unperturbed background: the NS-NS and the RR coordinates. In what follows we will focus on the NS-NS description and derive an explicit expression for the Killing spinors \eqref{summary2}, while in Appendix \ref{app:RRspinors} we derive the Killing spinors~\eqref{epsilonpm_RR} in the RR coordinates and present the map between the two sets of solutions.

As mentioned before, in the seed \eqref{seed}, the metric, the dilaton,
and $C_2$ do not change at $\cO(b)$, while $B_2,C_0,C_4$
get excited at $\cO(b)$.  In the NS-NS coordinates, the $\cO(b^0)$
fields are the metric and
\begin{subequations}
 \begin{align}
 e^{2\Phi}&={Q_1\over Q_5},\\
 C_2&=-\frac{r^2+a^2}{Q_1}du\wedge dv
 -Q_5 \cos ^2\theta\, d\phit \wedge d\psit
 -\frac{Q_5}{\sqrt{2} R_y} (du-dv)\wedge d\phit,
 \intertext{while the $\cO(b)$ fields are}
 B_2&=-b \Delta_{k,0,0}\,e^{-i\vh_{k,0,0}}\biggl[\frac{r^2+a^2}{R_y a^2}du\wedge dv+{1\over\sqrt{2}}(du-dv)\wedge \left(\frac{i\cos \theta\,  d\theta }{\sin\theta}+d\phit\right)\notag\\
 &\hspace{40ex}
 +R_y \cos\theta \left(\frac{i\, d\theta}{\sin\theta }+\cos\theta  d\phit\right)\wedge   d\psit \biggr],\\
 C_0&= \frac{b Q_5}{a^2 R_y} \Delta_{k,0,0}\,
 e^{-i\vh_{k,0,0}},\\
 C_4&=
 {b\over a^2 R_y} \Delta_{k,0,0}\,e^{-i\vh_{k,0,0}}
  \left[Q_1\widehat{\mathrm{vol}}_{4}
 +Q_5(r^2+a^2) \cos^2\theta\,du\wedge dv\wedge d\phit\wedge d\psit\right]
 \end{align}
\end{subequations}
where
\begin{align}
 \vh_{k,0,0}=k \biggl(\frac{u+v}{\sqrt{2} R_y}+\phit\biggr).
\end{align}
On the other hand, the field strengths are $F_3=\cO(b^0)$;
$H_3,F_1,F_5=\cO(b)$; and $d\phi =\cO(b^2)$.  Therefore, the
supersymmetry variations \eqref{fervar1} for the seed solution
\eqref{seed} split into the $\cO(b^0)$ part
\begin{subequations}
\label{fervar2}
\begin{align}
\delta \lambda_0 & = 
 -{1\over 2} e^\phi\, \slashed F_3\, \sigma^1 \, \epsilon\label{fervar20d}\,,\\
\delta \psi_{M,0} &  = \nabla_{\!M}^{}\epsilon + \frac18 e^\phi \slashed F_3 \Gamma_M\sigma^1\epsilon\label{fervar20g}\,,\\
\intertext{and the $\cO(b)$ part}
\delta \lambda_b & = -\frac12 \slashed H\,\sigma^3 \epsilon - e^\phi \slashed F_1 \,(i\sigma^2) \epsilon\label{fervar2bd}\,,\\
\delta\psi_{M,b} & = - \frac18 H_{MNP}\Gamma^{NP} \,\sigma^3 \epsilon + \frac18 e^\phi \left(\slashed F_1 + \frac12 \slashed F_5\right) \Gamma_M \, (i\sigma^2)\epsilon\label{fervar2bg}\,.
\end{align}
\end{subequations}

In this section, we are interested in the Killing spinors in the
unperturbed ($b=0$) background, which is nothing but AdS$_3\times
S^3\times T^4$.  In the NS-NS coordinates its metric takes the form
\eqref{stmetric1}. We will work in the $u,v$ coordinates \eqref{sptvw},
in which the metric becomes
\begin{align}
 ds_6^2 &= \sqrt{Q_1 Q_5}\left[ - \frac{a^2 du^2 + 2(a^2 +2r^2) du\,dv + a^2 dv^2}{2a^2 R_y^2}+\frac{dr^2}{r^2+a^2}
 +d\theta^2+\sin ^2\theta d\tilde\phi^2+\cos ^2\theta d\psit^2 
 \right],
\end{align}
which suggests the following choice of 10-dimensional vielbeine
\begin{subequations}\label{vielbeine}
\begin{align}
E^{\uu} &= \frac{1}{2\sqrt{a R_y}} \left[  \left(  \sqrt{r^2 + a^2}+r \right) du + \left( \sqrt{r^2 + a^2} - r \right) dv \right],\\
E^{\vu} &= \frac{1}{2\sqrt{a R_y}}  \left[ \left(  \sqrt{r^2 + a^2} - r \right) du + \left(\sqrt{r^2 + a^2} + r \right) dv\right], \\
E^\ru & = \sqrt{a R_y}\, \frac{dr}{\sqrt{r^2 + a^2}}, \qquad  E^\thet = \sqrt{a R_y}\, d\theta, \\
 E^\phitu & = \sqrt{a R_y}\, \sin\theta \,d\phit, \quad \qquad E^\psitu = \sqrt{a R_y}\, \cos\theta \,d \psit,\\
E^{\underline{k}} &= \left( \frac{Q_1}{Q_5}\right)^{\frac{1}{4}} dx^k,
\end{align}
\end{subequations}
where $x^k$, $k = 6,7,8,9$ are the coordinates of the internal $T^4$.
With this choice the metric can be written as 
\begin{align}
ds_{10}^2 = - 2 E^\uu E^\vu + \delta_{\underline{a} \underline{b}}E^{\underline{a}} E^{\underline{b}},
\end{align}
with $a = r,\theta, \phit, \psit, 6,7,8,9$. 
The variations \eqref{fervar20d} and \eqref{fervar20g} can be written out explicitly as
\begin{subequations}\label{KSeq1}
\begin{align}
 \delta \lambda^1_0 &= \frac{1}{\sqrt{a R_y}} \Gamma^{\underline{r}}\Gamma^{\underline{uv}}\left(1 -  \Gamma^{\underline{6789}}\right)\epsilon^2,\\
\delta \psi_{u,0}^1 &=  \partial_u \epsilon^1 + \frac{1}{8a R_y}\left(\sqrt{r^2 + a^2}- r\right) \Gamma^{\uu \ru} \left[ 2 \epsilon^1 + \Gamma^{\uu \vu}\left(1 + \Gamma^{\underline{6789}}\right)\epsilon^2\right] \\* & \hspace{7.5ex} - \frac{1}{8a R_y}\left( \sqrt{r^2 + a^2}+r \right) \Gamma^{\vu \ru} \left[  2\epsilon^1 - \Gamma^{\underline{uv}}\left(1 + \Gamma^{\underline{6789}}\right)\epsilon^2\right],\\ 
\delta \psi_{v,0}^1 &= \partial_v\epsilon^1 +  \frac{1}{8a R_y}\left(\sqrt{r^2 + a^2}- r\right) \Gamma^{\vu \ru} \left[ 2 \epsilon^1 + \Gamma^{\uu \vu}\left(1 + \Gamma^{\underline{6789}}\right)\epsilon^2\right] \\* & \hspace{7.5ex} - \frac{1}{8a R_y}\left(  \sqrt{r^2 + a^2} + r\right) \Gamma^{\uu \ru} \left[  2\epsilon^1 - \Gamma^{\underline{uv}}\left(1 + \Gamma^{\underline{6789}}\right)\epsilon^2\right],\\
 \delta \psi_{r,0}^1 &= \partial_r \epsilon^1 - \frac{1}{4 \sqrt{r^2 + a^2}}\Gamma^{\underline{uv}}\left(1 + \Gamma^{\underline{6789}}\right)\epsilon^2,\\
\delta \psi_{\theta,0}^1 &= \partial_\theta \epsilon^1 - \frac{1}{4}\Gamma^{\underline{r\theta}}\,\Gamma^{\underline{uv}}\left(1 + \Gamma^{\underline{6789}}\right)\epsilon^2,\\
\delta \psi_{\phit,0}^1 &= \partial_\phit\, \epsilon^1 + \frac{\cos \theta}{2} \Gamma^{\underline{\phit \theta}}\, \epsilon^1 - \frac{\sin\theta}{4}\Gamma^{\underline{r\phit}}\Gamma^{\underline{uv}}\left(1 + \Gamma^{\underline{6789}}\right)\epsilon^2,\\
\delta \psi_{\psit,0}^1 &= \partial_{\psit} \, \epsilon^1 - \frac{\sin\theta}{2}\Gamma^{\underline{\psit\theta}}\,\epsilon^1 - \frac{\cos\theta}{4}\Gamma^{\underline{r\psit}}\Gamma^{\underline{uv}}\left(1 + \Gamma^{\underline{6789}}\right)\epsilon^2,\\
\delta \psi_{k,0}^1 &= \partial_k \epsilon^1 - \frac{Q_1^{\frac14} Q_5^{-\frac14}}{4\sqrt{a R_y}}\,\gu{uvrk} \left( 1 - \gu{6789}\right)\epsilon^2.
\end{align}
\end{subequations}
In these equations we have made the doublet index of the fermionic fields explicit. We have only given the variations for $\delta \lambda^1_0$ and $\delta \psi_{M, 0}^1$, as the variations for $\delta \lambda^2_0$ and $\delta \psi_{M, 0}^2$ can be obtained simply by interchanging the doublet indices $1\leftrightarrow 2$ on all fermions in the above variations.  

\subsection{The Killing spinors}

Killing spinors of the AdS$_3\times S^3\times T^4$ background are
non-trivial spinors that satisfy the equations
\begin{align}\label{eq:KSeq2}
\delta \lambda^{1,2} = \delta \psi^{1,2}_M = 0.
\end{align}
We find that the spinors that solve the equations \eqref{eq:KSeq2} are given by 
\begin{subequations}\label{summary1}
\begin{align}
\epsilon^1 &= \frac{1}{2}R_-\, Y_-\, (\etat + \eta ) + \frac12 R_+ \,Y_+ \,(\xit + \xi)\,,\\
\epsilon^2 &= \frac{1}{2}R_-\, Y_-\, (\etat - \eta ) + \frac12 R_+ \,Y_+ \,(\xit - \xi)\,,
\end{align}
\end{subequations}
where in the above spinors, the following definitions are used\footnote{In \eqref{summary1} we suppress the functional dependencies to avoid cluttering. Note that the angular parts can be solved with the help techniques developed in \cite{Lu:1996rhb, Lu:1998nu}, where similar differential equations have been considered, however in a different coordinate system.}
\begin{subequations}\label{summary2}
\begin{align}
Y_\pm(\theta, \phit,\psit) &\equiv \exp\left(\pm\frac{1}{2}\theta \Gamma^{\underline{r\theta}}\right) \exp\left(\frac12 \phit \gu{\theta \phit}\right) \exp\left(\pm\frac12 \psit \gu{r \psit}\right)\,,\label{eq:Rpmdef}\\
R_\pm(r) &\equiv \biggl(\frac{\sqrt{r^2 + a^2} \pm r}{a}\biggr)^\frac12\,,\\
\xit(u) &= \zetat_+ e^{-\frac{iu}{\sqrt{2}R_y}} + \zetat_- e^{\frac{iu}{\sqrt2R_y}} ,\qquad 
\etat(u) = i\guh{ru}\Bigl(\,\zetat_- e^{-\frac{iu}{\sqrt{2}R_y}} - \,\zetat_+ e^{\frac{iu}{\sqrt2R_y}}\Bigr),\\
\xi (v) &= \zeta_- e^{-\frac{iv}{\sqrt2R_y}}+ \zeta_+  e^{\frac{iv}{\sqrt2R_y}},\qquad 
\eta(v) = i \guh{rv} \Bigl(\zeta_-e^{-\frac{iv}{\sqrt{2}R_y}} - \, \zeta_+ e^{\frac{iv}{\sqrt2R_y}}\Bigr),\label{eq:chidef}
\end{align}
\end{subequations}
and we have defined
\begin{align}
\guh{rv} \equiv \frac{1}{\sqrt2} \gu{rv}, \qquad \guh{ru} \equiv \frac{1}{\sqrt{2}}\gu{ru}\,.
\label{def_ghat}
\end{align}
The spinors $\zeta_\pm,\zetat_\pm$ are constant spinors that do not depend on any coordinates.
As standard in type IIB supergravity, the $\epsilon^{1,2}$ in
\eqref{summary1} are Majorana-Weyl spinors. The Weyl condition is
\begin{align}
\label{WeylCond}
\Gamma_{(10)}\,\epsilon^{1,2} = \epsilon^{1,2}\,,
\end{align}
with $\Gamma_{(10)} \equiv \Gamma^{\underline{uvr\theta\phit\psit
6789}}$.  In our convention in which the charge conjugation matrix 
is $\cC=\gu{t}$, the Majorana condition reads
\begin{align}
\label{MajoranaCond}
\epsilon^* = \epsilon\,.
\end{align}
We can now spell out the constraints following from~\eqref{MajoranaCond} on the spinors in \eqref{summary1} and \eqref{summary2}.
The factors $Y_{\pm}(\theta, \phit,\psit)$ and $R_{\pm}(r)$ are real
functions containing the explicit dependence of the spinors on the
angular and radial parts respectively. Then in order for
$\epsilon^{1,2}$ to be real, spinors $\zeta_{\pm}$, $\zetat_{\pm}$
must satisfy
\begin{align}\label{Majoranacond2}
\zeta_{\pm}^* = \zeta_{\mp}\,,\qquad \zetat_{\pm}^* = \zetat_{\mp}\,.
\end{align}
This means that $\zeta_{\pm}$ and $\zetat_{\pm}$ are complex.  On the
other hand, $\xi(v)$, $\eta(v)$, $\xit(u)$ and $\etat(u)$ are real spinors.
Therefore, $\xi(0)$, $\eta(0)$, $\xit(0)$ and $\etat(0)$ are constant,
real spinors, each containing 4 independent real degrees of freedom
and can be used to parametrise the variations.

Furthermore the dilatino variation and the variation in the $T^4$ subspace impose 
\begin{align} \label{cond2}
\Gamma^{\underline{6789}}\,\epsilon^{1,2} = \epsilon^{1,2}\,.
\end{align}
All spinors in \eqref{summary1} and \eqref{summary2},
$\xi,\xit,\eta,\etat,\zeta_\pm,\zetat_\pm$, must satisfy the conditions
\eqref{WeylCond} and \eqref{cond2}.  In addition, they have the
following chirality for the matrix $\gu{uv}$:
\begin{align}\label{zetauveigen}
\begin{aligned}
 \gu{uv}\,\xi(v) &= - \xi(v)\, ,
 \quad&
 \gu{uv}\,\eta(v) &= + \eta(v)\, ,
 \quad&
 \gu{uv}\,\zeta_{\pm} &= - \zeta_{\pm}\, , 
 \\
 \gu{uv}\,\xit(u) &= +\xit(u)\, ,
 &
 \gu{uv}\,\etat(u) &=- \etat(u)\, ,
 &
 \gu{uv}\,\zetat_{\pm} &= +\zetat_{\pm}\, .
\end{aligned}
\end{align}

As mentioned above each one of $\eta,\etat,\xi,\xit$ contains 4 real
degrees of freedom.  On the other hand, each one of $\zeta_\pm,\zetat_\pm$
contains 4 complex degrees of freedom, but only half of them are
independent.  Therefore, either way, each spinor $\epsilon^{1,2}$
contains 8 real degrees of freedom, combining to 16 in total. This is
what we expected, as global AdS$_3\times S^3$ should preserve half of
the total 32 supercharges.

In order to generate new solutions, we need to identify the spinor
components that correspond to different modes of the fermionic
generators $G_{\pm \frac12}^{\alpha A}$ and $\widetilde G_{\pm
\frac12}^{\dot \alpha A}$. We will only be interested in generating left-moving excitations, which are generated by $G$ and not $\widetilde G$,
hence we will limit ourselves only to the discussion around the left-moving sector. However, the discussion on the right-moving sector is
completely analogous.
The relation between the supergravity Killing spinors $\zeta_{\pm}^{\alpha A}$ and the CFT supercurrent $G_{\pm \frac12}^{\alpha A}$ can be encoded in terms of the projectors
\begin{align}
  \label{STprojtxt}
  \cP_S^{\pm} \equiv \frac12\pm  J_{\psit}\;,\quad \qquad
\cP_T^{A} \equiv  \frac{1 + (-1)^A i \gu{67}}{2}, \quad A = 1,2,
\end{align}
where $J_{\psit}$ is defined in~\eqref{eq:SU(2)Jgen1}. We leave the discussion of this point to Appendix~\ref{sec:aligning} and here quote just the final result
\begin{align}
\label{eq:identificationtxt}
\zeta_\pm^{\alpha A} \equiv \cP_T^A\, \cP_S^\alpha\, \zeta_\pm\,
 ~\longleftrightarrow ~
 G_{\pm \frac12}^{\alpha A},
 \qquad \qquad 
 \alpha =\pm,\quad  A  = 1,2.
\end{align}
With this identification we can proceed to generate the supergravity
solution corresponding to the state \eqref{eq:stateOkmnq} in CFT\@.

\section{Fermionically generated superstrata: linear analysis}
\label{Sect:QQsol}

In this section we derive a linearised classical solution to the
supergravity equations using the Killing spinors for AdS$_3 \times S^3
\times T^4$ obtained in the previous section. We do so by performing two
supersymmetry variations, generated by the spinors corresponding to
$G_{- \frac12}^{\alpha A}$, on the $b \neq 0$ seed solution \eqref{seed}. Single
variations of bosonic fields vanish, as these are proportional to the
fermionic fields, which are set to zero for classical supergravity
solutions. Double variations of the bosonic fields, however, are
non-vanishing, as these include terms which are proportional to the
variations of the fermionic fields. Note that, by definition, Killing
spinors are non-trivial spinors for which the variations of fermionic
fields vanish. However, here we will use the Killing spinors for the
unperturbed case ($b = 0$) on the perturbed background ($b \neq 0$),
which will generate fermionic variations that are non-vanishing. We will
limit ourselves to terms at linear order in parameter $b$. Using these
non-vanishing fermionic variations we can generate new solutions at
linear order in the parameter $b$.

This section will give a step-by-step procedure of how to obtain the
new solutions. We begin by presenting the variations of the fermionic
fields of the seed solution, generated by the Killing spinors
\eqref{summary1}. We then present the double variations of the bosonic
fields generated, thus obtaining the solution dual to the state
$G_{-\frac12}^{+1}G_{-\frac12}^{+2}|O^{--}\rangle_k\,$. In this
derivation we treat the complex spinors $\zeta^{\alpha A}_+$ and
$\zeta^{\alpha A}_-$ as independent, although they really are not, due
to the relation \eqref{Majoranacond2}. Because of this, the spinors
$\epsilon^{1,2}$ become complex and we will obtain a complex-valued
perturbation: as usual, at the linear level, we can derive a standard
supergravity solution by taking the real part of the final result. While
this is indeed a new solution to the first-layer BPS equation it is not 
linearly independent from the ones already known, as discussed
in section \ref{Sect:CFT}. We therefore present the new, linearly
independent solution, dual to the state
$\bigl(G_{-\frac12}^{+1}G_{-\frac12}^{+2} + \frac{1}{k} J^+_0
L_{-1}\bigr) |O^{--}\rangle_k\,$, and further give the solution dual to
the state \eqref{eq:stateOkmnq} with $q =1$.


\subsection{Variations of fermionic fields}

As a preliminary step, we use the Killing spinors found in the previous
section to calculate the supersymmetry variations of the fermionic fields.
The result can be used later when we consider double variations of
bosonic fields.  Furthermore, it also serves as a consistency check of
the identification \eqref{eq:identificationtxt} between the fermionic
generators in supergravity and CFT\@.

As discussed in previous sections, the seed solution \eqref{seed}
expressed in the NS-NS coordinates is dual to an anti-chiral primary
state, which should be annihilated by $G_{-\half}^{-A}$ but not by
$G_{-\half}^{+A}$.  According to the identification
\eqref{eq:identificationtxt}, this implies that the supersymmetry variation
of the seed by $\zeta_{-}^{-A}$ will vanish while the supersymmetry
variation by $\zeta_{-}^{+A}$ will not. In performing this check,
we set $\zeta^{\alpha A}_+ = 0$ and look at the variations of the
fermionic fields generated by the components in the spinor
$\zeta^{\alpha A}_-$. With this choice, the Killing spinor
\eqref{summary1} simplifies to
\begin{align}\label{spinorzeta-}
\epsilon^1 = - \epsilon^2 =\frac12 \left(R_+ - i R_- \guh{vr}\right) e^{-\frac{iv}{\sqrt2 R_y}}\, Y_+ \zeta^{\alpha A}_-\,.
\end{align}
The dilatino variation generated by this spinor is given by 
\begin{align}\label{dilvar-}
\delta \lambda_b^1 = \delta \lambda_b^2 &=
\frac{b a^k \,k \sin^{k-1}\theta \,}{\sqrt{a R_y}\left(a^2+ r^2\right)^{\frac{k+1}2}} \notag\\
 &\quad\times\left( R_- - i R_+ \guh{vr}\right)e^{-ik\bigl( \frac{u+v}{\sqrt2 R_y} + \phit\bigr)-  \frac{iv}{\sqrt2 R_y}}  \left( \cos\theta\, \gu{\theta} + \sin\theta\, \gu{r} - i \gu{\phit}\right) Y_+ \zeta^{\alpha A}_-\,.
\end{align}
This expression is correct for a generic component $\zeta^{\alpha A}_-\,$. However, as discussed above, this variation should distinguish between the $\zeta^{+  A}_-\,$ and the $\zeta^{- A}_-\,$ components of the spinor. Indeed, this is the case here, as one can see from the factors that contain the gamma matrices with components along $S^3$. Using the definition of $Y_+$ in \eqref{summary2} one can show that
\begin{align}\label{projection}
 \left( \cos\theta \,\gu{\theta} + \sin\theta\, \gu{r} - i \gu{\phit}\right) Y_+ \zeta^{\alpha A}_-  = 2 \gu{\theta}\, e^{-\theta \gu{r\theta} }\,  Y_+\zeta^{+ A}_-\,
\end{align}
because, when we commute the factor in the brackets through $Y_+$, we
generate a projector $\cP_S^+\,$, which projects out the $\zeta^{-A}_-$
component. Therefore, as expected, the supersymmetry variation of the
seed by $\zeta_-^{-A}$ vanishes, while the supersymmetry
variation by $\zeta_-^{+A}$ does not.
The variations of the gravitino components $\psi_M$ are calculated in
the same manner and discussed in Appendix~\ref{app:ExplRes}.

\subsection{Solution-generating technique}

The first step to finding the geometry dual to the state
\eqref{eq:stateOkmnq} with $q =1$ is to find the geometry dual to
$G_{-\frac12}^{+1}G_{-\frac12}^{+2}|O^{--}\rangle_k\,$. In order to do
so, we do a double variation of the bosonic fields. These variations
generically have two kinds of terms: either they are proportional to
fermionic fields or the variation of fermionic fields. In a classical
solution, fermionic fields vanish and hence we are left only with the
second type of terms. Using the variations summarised in
Appendix~\ref{app:susyv}, we get that, for example, the variation of the
axion field $C_0$ is given by
\begin{align}\label{doubleaxionvar1}
\delta'\delta C_0 &= \frac12 e^{-\phi}\, \epsilon^T\Gammau{t} (i\sigma^2) \delta' \lambda
\end{align}
where $\delta$ denotes the variation generated by spinors $\epsilon^{1,2}$ and
$\delta'$ denotes the variation with different spinors
$\epsilon'{}^{1,2}$.  In our case the constant components of these spinors will be  $\zeta_-^{+ 1}$ for the
variation $\delta$ and $\zeta_-^{\prime + 2}$ for the variation
$\delta'$. Following the procedure of \cite{Bena:2015bea, Bena:2017xbt} we are interested in calculating the infinitesimal
deformation from the seed solution in the ansatz function $Z_4$ and the
two-form $\Theta_4$.  The physical fields in which these two quantities
are appearing are the axion $C_0$ and the NS-NS gauge $B$-field, so
we are interested in their variations. Since these two calculations are
analogous, we will only present the detailed calculation for the axion
field, while the results of the $B$-field can be found in Appendix
\ref{app:ExplRes}.

As mentioned at the beginning of this section, because we treat complex
spinors $\zeta_+^{\alpha A}$ and $\zeta_-^{\alpha A}$ as independent,
the spinor $\epsilon$ becomes complex.  This is justified with the
understanding that we will take the real part in the final result.  In
the intermediate calculations, although $\epsilon^{1,2}$ are really
complex, we still treat them as real spinors.  In writing down
\eqref{doubleaxionvar1}, we used the relation \eqref{conjugation}, which
is valid only for real (Majorana) spinors, in order to rewrite
$\bar\epsilon$ appearing in the formula \eqref{hszi10Apr18} in terms of
$\epsilon^T$.  Another way to justify this manipulation is that, because
the first variation parameter $\epsilon$ and the second one $\epsilon'$
are on an equal footing, it is not possible for $\epsilon$ to enter in
the variation $\delta'\delta C_0$ being complex conjugated and
$\epsilon'$ without being complex conjugated.  They must both enter
without being complex conjugated,  as in 
\eqref{doubleaxionvar1}.

The variations we consider are generated by the Killing spinors
\eqref{spinorzeta-}. As mentioned above, the first variation is
generated by the component $\zeta_-^{+ 1}$ and the second by $\zeta_-^{\prime + 2}$. In this case, as we have seen in \eqref{spinorzeta-}, 
we have $\epsilon^1 = - \epsilon^2$.   Furthermore, the variations are
such that $\delta\lambda_0 = 0$ and $\delta\lambda_b^1 =
\delta\lambda_b^2$. With these, the axion  variation \eqref{doubleaxionvar1} simplifies to
\begin{align}\label{doubleaxionvar2}
\delta'\delta C_0 &=  e^{-\phi} \, (\epsilon^1)^T \Gammau{t}\,\delta'\lambda_b^1\,.
\end{align}
We insert the spinor \eqref{spinorzeta-} and the variation
\eqref{dilvar-} (since we use only $\zeta_-^{+ A}$ we use the projection~\eqref{projection} in the variation) into this expression to obtain
\begin{align}
\delta'\delta C_0 &= b\frac{\sqrt{a R_y}}{Q_1}\frac{ k a^k \sin^{k-1}\theta}{ \left(a^2 + r^2\right)^{\frac{k+1}{2}}}e^{-i\bigl[ k\bigl({u+v \over \sqrt2} + \phit\bigr) + \sqrt2 v - (\phit + \psit)\bigr]} \nonumber \\&\qquad\times {\left(\zeta^{+1}_-\right)}^T e^{- {\theta \over 2}\gu{r\theta} } \left(R_+ + i R_- \guh{vr} \right) \left( \guh{v} + \guh{u} \right) \left( R_- - i R_+ \guh{vr}\right) \gu{\theta}\, e^{- {\theta \over 2 }\gu{r\theta} } \zeta_-^{\prime + 2} \nonumber \\
&= 2kb a^{k-1} \frac{\sqrt{a R_y}}{Q_1} \frac{ r \sin^{k-1}\theta\,\cos\theta \, }{ \left(a^2 + r^2\right)^{\frac{k+1}{2}}}e^{-i\bigl[ k\bigl({u+v \over \sqrt2 R_y } + \phit\bigr) + \sqrt2 \frac{v}{R_y} - (\phit + \psit)\bigr]}  \left[{\left(\zeta^{+1}_-\right)}^T i\gu{r\theta} \, \zeta_-^{\prime + 2}\right].
\label{eq:C0dvar} 
\end{align}
In the second equality we have used the properties of the Clifford algebra and the $R_\pm$ functions together with the fact that $\gu{uv}\, \zeta^{\alpha A}_- = - \zeta^{\alpha A}_-$ and hence  $\gu{u}\, \zeta^{\alpha A}_- = (\zeta^{\alpha A}_-)^T \gu{v} = - (\gu{u}\,\zeta_-)^T = 0$. Furthermore, one uses the fact that $(\gu{r\theta}){}^2 = -1$ to expand
\begin{align}
\gu{r\theta}\, e^{-\theta \gu{r\theta}} = \cos\theta\, \gu{r\theta} + \sin\theta.
\end{align}
Finally noting that due to the projector property  $(\cP_S^{+})^T= \cP_S^{-}$, we have $(\zeta_-^{+1})^T\zeta_-^{\prime + 2} =0$, which leaves us with the result~\eqref{eq:C0dvar}. The calculation for the $B$-field follows along the same lines; see~\eqref{BdvarNS} for the final result. 

After the inverse spectral flow coordinate transformation (the inverse transformation of~\eqref{spectralflowcoord}), one finds that the double variation leaves all ansatz quantities unchanged at linear order in $b$, except for  
$Z_4$ and $\Theta_4$, which are now given by\footnote{The superscript $f$ is to denote that these solutions are obtained by acting with fermionic generators only and hence the state that this solution is dual to is  $G_{-\frac12}^{+1}G_{-\frac12}^{+2}|O^{--}\rangle_k\,$.}
\begin{subequations}
\begin{align}
\label{fermstate}
Z_4^f &= b k R_y\frac{\Delta_{k,1,1} }{\Sigma}\,\cos{\hat v_{k,1,1}}\,,\\
\Theta_4^f &= -\sqrt2 b k \Delta_{k,1,1}\left[\left((k-1)\frac{\Sigma}{r\sin\theta}+ 2r\sin\theta\right)\Omega^{(1)}\, \sin\hat{v}_{k,1,1} \right.\nonumber\\*
& \hspace{24ex} \left. +\left((k+1)\Omega^{(2)} + (k-1)\Omega^{(3)}\right) \cos \hat{v}_{k,1,1} \right]\,,
\end{align}
\end{subequations}
where we have again expanded $\Theta_4$ in the self-dual basis \eqref{selfdualbasis}. As in mentioned above, in the final result, we selected the real part of the perturbation. We also normalised the spinors as follows: $[(\zeta_-^{+1})^T i\gu{r\theta}\,  \zeta_-^{\prime +2}] \to \frac12 \sqrt{a R_y}\,$, which is natural since in our conventions, spinors have a mass dimension of $-1/2$ and so a spinor bilinear should have mass dimension $-1$ or should have units of length. One can check explicitly that the result~\eqref{fermstate} satisfies the first-layer BPS equations \eqref{layer1}.

To get a feeling for this solution we can compare it with the old superstrata solution \eqref{Z4Th4_solngen_O(b)}, for $(k,m,n) = (k,1,1)$, which is obtained by using the bosonic symmetry generators \eqref{su(2)_gen_NS} and~\eqref{sl(2,R)_gen_NS} only. The solutions are similar and this may not be unexpected, as they both introduce the same amount of momentum and angular momentum into the geometry. We notice that the form of the function $Z_4$ is the same, while $\Theta_4$ slightly differs in the relative factors multiplying the basis elements $\Omega^{(i)}$. 

As shown on the CFT side, the state $G_{-\frac12}^{+1}G_{-\frac12}^{+2}|O^{--}\rangle_k\,$ is not linearly independent from the state $J_0^+ L_{-1} |O^{--}\rangle_k\,$. The proper combination which contains the information about the new, linearly independent CFT state is given by the combination
\begin{align}\label{eq:LIstate}
\left( G_{-\frac12}^{+1} G_{-\frac12}^{+2} + \frac1{k} J_0^+ L_{-1}\right)  |O^{--}\rangle_k\,.
\end{align}
In order to write the supergravity solution dual to~\eqref{eq:LIstate}, let us briefly discuss the relation between the parameter $b_4$ in~\eqref{Z4Th4_solngen_O(b)} and the original parameter $b$ appearing in the seed solution. By keeping track of the overall normalisation when acting with bosonic generators~\eqref{su(2)_gen_NS} and~\eqref{sl(2,R)_gen_NS}, we have $b_4 = (-1)^n k  \frac{\left(k + (n-1)\right)!}{(k-m)!} b$.
Because for $m = n =1$ this gives $b_4=- k^2b$, one finds that the
solution dual to the state \eqref{eq:LIstate} is given by the following
new linear perturbation
\begin{subequations}\label{mainbit}
\begin{align}
Z_4 &= 0,\\
\Theta_4 &= -\sqrt2 \,\bhat\, \Delta_{k,1,1}\left[\frac{\Sigma}{r\sin\theta}\Omega^{(1)} \sin{\hat v_{k,1,1}}+ \left(\Omega^{(2)} + \Omega^{(3)}\right)\cos{\hat v_{k,1,1}}\right]\,,
\end{align}
\end{subequations}
where we introduced $\bhat = (k^2-1)b$.
The new solution has a vanishing $Z_4$
function, which means that both the axion field $C_0$ and also the
component $B_{uv}$ of the $B$-field are vanishing. However, because
$\Theta_4$ is non-vanishing, the components of $B_{\mu \nu}$ with one
leg in AdS$_3$ and one in $S^3$ are non-vanishing, which agrees with the
spectrum calculated by \cite{Deger:1998nm}. For a brief summary of their
results and the connection to the present work, see Appendix
\ref{app:DegerEtal}\@.
One can show that $Z_4$ and $\Theta_4$ given in \eqref{mainbit} satisfy
the first-layer of BPS equations \eqref{layer1}. The solution is dual to
the CFT state with quantum numbers $(k,m,n,q) = (k,0,0,1)$.
Notice that for $k=1$ both the above supergravity perturbation and the
corresponding CFT state are trivial, which provides a further check on
the identification proposed.

We can generalise this approach and use the geometry~\eqref{mainbit}
as a new seed solution and act on
it with the bosonic symmetry generators \eqref{sl(2,R)_gen_NS} and
\eqref{su(2)_gen_NS} to obtain the geometry dual to the state $(k,m,n,q=1)$.
One finds that the geometry dual to the state
\eqref{eq:stateOkmnq} with $q =1$ is again unchanged at linear order
from the original seed solution \eqref{seed} in all the ansatz
quantities except for\footnote{We reabsorb all overall normalisations in the parameter $\bhat_4^{k,m,n}$.}
\begin{equation}\label{GGdesc}
  Z_4 = 0\,,\qquad \Theta_4 =  \bhat_4^{k,m,n}\, \widehat{\vartheta}_{k,m,n}\,,
\end{equation}
with
\begin{equation}
   \widehat{\vartheta}_{k,m,n} =
             \Delta_{k, 1+m, 1+n}\left[\frac{\Sigma}{r\sin\theta}\Omega^{(1)} \sin{\hat v_{k,1+m,1+n}}+ \left(\Omega^{(2)} + \Omega^{(3)}\right)\cos{\hat v_{k,1+m,1+n}}\right]\,.
\end{equation}
Again it is not difficult to show explicitly that these satisfy the first-layer BPS equations.
Also in this more general case $Z_4$ remains trivial and the structure of $\Theta_4$ is the same as in \eqref{mainbit}, apart from a change in the argument $\hat v_{k, 1+m, 1+n}$ of the trigonometric functions and in the quantum numbers in the $\Delta_{k, 1+m, 1+n}$ function.

\section{Non-linear completion}
\label{sect:NonLin}

In the previous sections we generated new solutions to the BPS equations at linear order in the perturbation parameter $b$.
Since the first-layer BPS equations are linear,
any linear combination of solutions~\eqref{Z4Th4_solngen_O(b)} and~\eqref{GGdesc}, with an allowed combination of the quantum numbers $(k,m,n)$, is also a solution of these equations. This general bulk configuration correspond to a CFT state containing various excitations \eqref{eq:stateOkmnq} with different values of $(k,m,n,q)$.
Up to this point we have taken the coefficients $b_I^{k,m,n}$ and $\bhat_I^{k,m,n}$ to be infinitesimally small, making any solution only an infinitesimal deformation of the empty AdS$_3\times S^3 \times T^4$ space. In principle we could promote all these coefficients to be finite\footnote{See \cite{Bena:2017xbt} for discussion on
the technical difficulty in superposing modes with completely general
$(k_1,m_1,n_1)$ and $(k_2,m_2,n_2)$.}, which corresponds, on the CFT side, to taking many copies of the same excitation.
With the parameters $\bhat_I^{k,m,n}$ being finite, the scalars $Z_I$ and the two-forms~$\Theta_I$ become sources on the right-hand sides of the second-layer equations. Thus, once we have a finite solution to the first-layer equations, we can determine the deformation of $\cF$ and $\omega$ from their seed values by solving the second-layer equations.

Here we will not tackle this general problem and will focus on a single-mode superstratum, {\em i.e.}, we make a single, arbitrary mode coefficient to be finite and the solve the second-layer equations with the corresponding source terms. We find that a special feature of this new class of non-linear solutions is that their extension to asymptotically flat solutions is trivial. We close this section by calculating the conserved charges of the newly obtained solutions and comparing them to the CFT states \eqref{eq:stateOkmnq} to find perfect agreement between the two results.

\subsection{Solving the second-layer equations}

The second-layer equations \eqref{layer2eq} are in general coupled
second-order partial differential equations for the components of the
anti-self-dual one-form $\omega$ and the scalar $\cF$. These
quantities encode the information about the
conserved charges of the geometry, {\em i.e.}, the momentum charge
$Q_p$ and the angular momenta $J$ and $\widetilde J$.
As mentioned above, our goal here is to calculate the backreaction on
$\omega$ and $\cF$ that the new finite deformations cause on the
geometry. We will limit ourselves to the case of single-mode
superstrata involving the new $q=1$ excitations, meaning that
our initial ansatz for the finite first-layer solution is
\begin{align}\label{eq:lincombsol2}
\begin{aligned}
 Z_1 & = \frac{Q_1}{\Sigma}\,, \quad& \Theta_1 &=  \bhat^{k,m,n}_2\, \widehat{\vartheta}_{k,m,n}\,,\\
 Z_2 & = \frac{Q_5}{\Sigma}\,, \quad& \Theta_2 &=  \bhat^{k,m,n}_1\, \widehat{\vartheta}_{k,m,n}\,,\\
 Z_4 &= 0\,,\quad& \Theta_4 &=  \bhat_4^{k,m,n}\, \widehat{\vartheta}_{k,m,n}\,,
\end{aligned}
\end{align} 
where $\bhat_{1,2}^{k,m,n}$ should be determined as a function of $\bhat_4^{k,m,n}$, which, in turn, is related to number of excitations \eqref{eq:stateOkmnq} in the corresponding CFT state. We further simplify our ansatz by
recalling that the bosonic descendants of an anti-chiral primary state
have  a trivial $\Theta_1$ \cite{Giusto:2013bda}. The same feature
is shared by the bosonic superstrata obtained by using the
excitations \eqref{eq:stateOkmnq} with $q=0$ \cite{Bena:2017xbt}. 
We assume that a similar property holds also for $q=1$ excitations and
set to zero the coefficients~$\bhat_2^{k,m,n}$.
Furthermore, in all previous work \cite{Bena:2015bea,
Bena:2017xbt,Niehoff:2013kia}, the coefficients $\bhat_1^{k,m,n}$ were
tuned in such a way that the singularity-causing $v$-dependent terms
vanished.  This procedure was called ``coiffuring''. However, we find that, with
the solutions generated with the new solution-generating technique, the
right-hand sides of the second-layer equations are automatically 
$v$-independent. For this reason, we assume that all $\bhat_1^{k,m,n}$ are also vanishing.
In this case the only non-trivial source in the second-layer equations
is $\Theta_4 \wedge \Theta_4$ which takes the simple form written
on the right-hand side of \eqref{eq:2ndlaySimp}. Since there are
no potentially dangerous terms to be taken care of by coiffuring,
the ansatz we use for the first-layer solution of a single-mode
superstrata is simply
\begin{subequations}
\label{eq:singlemodesuperstrata}
\begin{align}
&Z_1 = \frac{Q_1}{\Sigma}\,, \quad Z_2 = \frac{Q_5}{\Sigma}\,, \quad Z_4 = 0\,,\\
&\Theta_1 = \Theta_2 = 0\,, \qquad \Theta_4 = \bhat\, \widehat{\vartheta}_{k,m,n}.
\end{align}
\end{subequations}

We now want to solve the second-layer equations where the source
on the right-hand side is determined by
\eqref{eq:singlemodesuperstrata}. By following~\cite{Bena:2017xbt}
we take the following ansatz for $\cF$ and $\omega$
\begin{align}\label{RMSdef}
\cF  =  \cF^{\rm RMS}(r,\theta),\qquad 
\omega = \omega_0 + \omega^{\rm RMS} (r, \theta),
\end{align}
where $\omega_0$ is given in \eqref{seed}, and the RMS superscript
denotes that the ansatz functions are independent of the coordinate~$v$
and are thus non-oscillating. We assume that $\cF^{\rm RMS}$ and
the components of $\omega^{\rm RMS}$ only depend on the coordinates $r$
and $\theta$ and that $\omega^{\rm RMS}_r = \omega^{\rm RMS}_\theta =
0$. Using~\eqref{eq:singlemodesuperstrata} on the right-hand side of the
second-layer equations, we see that the RMS parts of~\eqref{RMSdef} are
governed by the differential equations
\begin{subequations}\label{eq:2ndlaySimp}
\begin{align}
&d_4 \omega^{\rm RMS} + *_4 d_4 \omega^{\rm RMS} +\cF^{\rm RMS} d\beta = 0,\\
&\flap \cF^{\rm RMS} =  4\,\bhat^2\, \frac{\Delta_{2k, 2m + 2, 2n+2} + \Delta_{2k, 2m+4, 2n}}{\left(r^2+a^2\right)\cos^2\theta\,\, \Sigma}\,,\label{eq:2ndlaysimpeq2}
\end{align} 
\end{subequations}
where we have used the fact that $d_4\omega_0$ is anti-self-dual and have introduced $\flap \equiv - *_4 d_4 *_4 d_4$ as the Laplace operator in the four-dimensional Euclidean base space. Notice that the top differential equation has a vanishing right-hand side, which was not the case in previously known examples. 

One can show that when acting on a scalar function that depends only on $r$ and $\theta$, the Laplace operator can be written as 
\begin{equation}\label{scalarflap}
\flap F \equiv \frac{1}{r\Sigma}\, \partial_r \big( r (r^2 + a^2) \, \partial_r F  \big)  +    \frac{1}{\Sigma\sin \theta \cos \theta}\partial_\theta \big( \sin \theta \cos \theta\, \partial_\theta F  \big)\,.
\end{equation}
We then note that the differential equation of the form
\begin{align}\label{solvableeq}
 \widehat{\cL}F_{2k,2m,2n}={\Delta_{2k,2m,2n}\over (r^2+a^2)\cos^2\theta\,\, \Sigma} 
\end{align}
is solved by the function\footnote{For the proof see Appendix A of \cite{Bena:2017xbt}.}
\begin{align} \label{Ffun}
F_{2k,2m,2n}&=-\!\sum^{j_1+j_2+j_3\le k+n-1}_{j_1,j_2,j_3=0}\!\!{j_1+j_2+j_3 \choose j_1,j_2,j_3}\frac{{k+n-j_1-j_2-j_3-1 \choose k-m-j_1,m-j_2-1,n-j_3}^2}{{k+n-1 \choose k-m,m-1,n}^2}\notag\\
&\qquad\qquad\qquad\qquad\qquad\qquad
 \times
\frac{\Delta_{2(k-j_1-j_2-1),2(m-j_2-1),2(n-j_3)}}{4(k+n)^2(r^2+a^2)}\,,
\end{align} 
with 
\begin{equation} 
{j_1+j_2+j_3 \choose j_1,j_2,j_3}\equiv \frac{(j_1+j_2+j_3)!}{j_1!\, j_2!\, j_3!}\,.
\end{equation} 
Notice that \eqref{eq:2ndlaysimpeq2} is already in the form \eqref{solvableeq}, and can be thus solved by a linear combination of the $F$ functions.

To get the solution for $\omega$, we introduce the ansatz\footnote{Notice that in previous work $\nu_{k,m,n}$ was named $\zeta_{k,m,n}$. Here we change the notation to avoid confusion with the complex spinor components $\zeta_{\pm}^{\alpha A}$.} \cite{Bena:2017xbt}
\begin{align}\label{omegaansatz}
\omega^{\rm RMS} = \mu_{k,m,n} (d\psi + d \phi) + \nu_{k,m,n} (d\psi - d\phi)\,,
\end{align} 
and define a new function
\begin{align}
\label{muhat}
\muh_{k,m,n} \equiv \mu_{k,m,n} + \frac{R_y}{4\sqrt2}\,\frac{r^2 + a^2 \sin^2 \theta}{\Sigma}\, \cF_{k,m,n},
\end{align}
where $\cF_{k,m,n} \equiv \cF$ is the solution of \eqref{eq:2ndlaysimpeq2}. One can show that $\muh_{k,m,n}$ satisfies the differential equation
\begin{align}\label{eq:muhateq}
\flap \muh_{k,m,n} = \frac{\bhat^2  R_y}{\sqrt2}\frac{\Delta_{2k, 2m+2, 2n} + \Delta_{2k, 2m+4, 2n+2}}{\left(r^2+a^2\right)\cos^2\theta\,\, \Sigma}\,,
\end{align}
which can be again solved by a linear combination of the functions \eqref{Ffun}. One is thus able to determine $\cF_{k,m,n}$ and $\muh_{k,m,n}$ as the solution of their respective second-order differential equations. By following the same approach of~\cite{Bena:2017xbt}, one can show that the explicit forms of the ansatz quantities are given by 
\begin{subequations}
\label{eq:2ndlaysol}
\begin{align}
\mathcal{F}_{k,m,n} &= 4\bhat^2 \, \left( F_{2k, 2m+2, 2n+2} + F_{2k, 2m+4,2n}\right)\,,\label{eq:solcalF}\\
\mu_{k,m,n} &= \frac{\bhat^2 R_y}{\sqrt2}\, \biggl[ F_{2k, 2m+2, 2n} + F_{2k, 2m+4, 2n+2}\nonumber \\*
& \hspace{10ex}
 - \frac{r^2 + a^2 \sin^2\theta}{\Sigma}\left( F_{2k, 2m+2, 2n+2} + F_{2k, 2m+4,2n}\right) + \frac{\xh_{k,m,n}}{4\Sigma}\biggr] \label{eq:solmu},
\end{align}
\end{subequations}
where in \eqref{eq:solmu} we have added an additional harmonic piece that is left undetermined by the differential equations. Its form is chosen so that the solution is regular at the centre of the coordinate system and will be determined in the next subsection.
The remaining $\nu_{k,m,n}$ functions are obtained by solving first-order differential equations, which contain $\cF_{k,m,n}$ and $\mu_{k,m,n}$ as sources. These equations are 
\begin{subequations}
\label{eq:zetasol}
\begin{align}
\p_r\nu_{k,m,n} &= - \frac{\left(a^2 + r^2\right) \sin^2\theta - r^2 \cos^2\theta}{r^2 + a^2\sin^2\theta}\,\p_r\mu_{k,m,n}  - \frac{2r\sin\theta \cos\theta}{r^2 + a^2 \sin^2\theta}\,\p_\theta\mu_{k,m,n}\nonumber \\*
&\quad - \frac{\sqrt2 a^2 R_y \left(a^2 + 2r^2\right)r \sin^2\theta \cos^2\theta}{\left(r^2 + a^2 \sin^2\theta\right) \Sigma^2}\,\cF_{k,m,n} ,\\
\p_\theta \nu_{k,m,n} &=   \frac{2\left(a^2 + r^2\right)r \sin\theta \cos\theta}{r^2 + a^2\sin^2\theta}\,\p_r\mu_{k,m,n}+ \frac{r^2 \cos^2 \theta - \left(a^2 + r^2\right) \sin^2\theta}{r^2 + a^2 \sin^2 \theta}\,\p_\theta \mu_{k,m,n}\nonumber\\*
&\quad +\frac{\sqrt2 a^2 R_y \left(a^2 + r^2\right) r^2 \sin\theta \cos\theta \cos 2\theta}{\left(r^2  + a^2 \sin^2\theta\right) \Sigma^2}\,\cF_{k,m,n}\,.
\end{align}
\end{subequations}
These equations can be solved by integration on a case-by-case basis for each set of quantum numbers.

\subsection{Regularity}

We want our solutions to be free of any singularities. However, the
coordinates introduced in~\eqref{newcoords}, which are used to describe
the solutions, have points at which they degenerate. Hence using these
coordinates we have to take special care of the functions and components
of the forms in order for our solutions to be regular.  There are
regions of the base manifold at which the coordinates degenerate. The
first one is at $\theta = 0$, where the $\phi$ circle degenerates. The
second one is at $\theta = \frac{\pi}{2}$, where the $\psi$ circle
shrinks. Finally at the locus $(r = 0, \theta = 0)$ the entire sphere
$S^3_{\theta\phi\psi}$ shrinks. Imposing regularity at these regions
imposes constraints on our solutions. We will look at each of the
constraints separately. Since now the scalar functions $Z_1$ and $Z_2$
are the same as the seed solutions and $Z_4$ is vanishing, we only focus
on the regularity of the forms. The standard procedure to check the
regularity is to express the forms in a coordinate system without any
degenerate points. However, we will instead impose the equivalent
condition that form components along $d\phi$ and $d\psi$ vanish at
a degenerate locus.

We start by looking at the region where $(r =0, \theta = 0)$. Focus on
the regularity of the $1$-form $\omega$, especially on the $d\psi +
d\phi$ component, which explicit expression is given in
\eqref{eq:solmu}. In order to cancel out the singularity caused by the
form legs, $\mu_{k,m,n}$ needs to vanish at the point of interest. This
determines the constant
multiplying the harmonic term in $\mu_{k,m,n}$, which must be
\begin{align}\label{eq:xhatdef}
\xh_{k,m,n} = k \frac{(k-m-2)!\, m!\, n!}{(k+n+1)!}\,.
\end{align}
Again the $\xh$ notation is used to distinguish the normalisation from the $q=0$ case.
The same analysis needs to be repeated for the $d\psi - d\phi$ component of the one-form, which also needs to be vanishing at the centre of the base manifold. However, as we are currently lacking a closed expression for generic quantum numbers, the analysis needs to be done on a case-by-case basis.

Let us now look at the points of the location of the supertube, namely
$(\theta = \frac{\pi}{2})$. This is the same as looking at $\Sigma = 0$,
where the scalars $Z_1$ and $Z_2$ diverge. To check the regularity of
the solution, we look at the $(d\phi + d\psi)^2$ component of the metric
and demand that it is regular at the position of the supertube. Imposing
regularity at this locus of spacetime changes the relation between the
brane charges $Q_1$, $Q_5$ and the parameters defining the supertube
ansatz $a$ and $b$, which is now given by
\begin{align}\label{eq:constgrav}
\frac{Q_1 Q_5}{R_y^2}	= a^2 + \frac{\bhat^2}{2} \xh_{k,m,n}\,,
\end{align}
where $\xh_{k,m,n}$ is defined in \eqref{eq:xhatdef}.

\subsection{Asymptotically flat solution}
\label{sect:AsyFlat}

Up to this point, the solutions we presented are asymptotically
AdS$_3\times S^3 \times T^4$, which allows for the identification with
the dual CFT states discussed in section \ref{Sect:CFT}. It turns out
that these solutions can be extended to asymptotically flat
configurations (in our case that is asymptotically equal to
$\mathbb{R}^{4,1}\times S^1 \times T^4$) in a straightforward way and so
can be identified with microstates of asymptotically flat black holes.
As usual, this is done by adding back ``$1$'' in the functions $Z_1$ and
$Z_2$. By introducing this extra constant, one usually obtains
additional $v$-dependence on the right-hand side of second-layer
equations. The effect of these additional terms is that we get new
$v$-dependent terms in \eqref{RMSdef}, with the differential equations
determining these new terms usually being cumbersome to solve. Luckily,
the novel feature of the solution \eqref{eq:singlemodesuperstrata} is
that including the constant term in $Z_{1,2}$ does not add any
additional source terms into the second-layer equations.

Thus focusing on the single-mode superstrata, the asymptotically
flat extension of the solution \eqref{eq:singlemodesuperstrata},
\begin{subequations}
\label{eq:asy-flatsinglemodesuperstrata}
\begin{align}
Z_1 &= 1 + \frac{Q_1}{\Sigma}, \qquad  Z_2 = 1 + \frac{Q_5}{\Sigma}, \qquad Z_4 = 0\,,\\
\Theta_1 &= \Theta_2 = 0,\qquad \Theta_4 =  \bhat\, \widehat{\vartheta}_{k,m,n}\,,
\end{align}
\end{subequations}
gives the same second-layer equations for $\cF$ and $\omega$ as in the asymptotically AdS case.
So, both the asymptotically AdS and the full asymptotically flat
solution are $v$-independent.

\subsection{Conserved charges}
\label{sec:conservedcharges}

One can extract the conserved charges of our new solutions from their
large-distance behaviour. This gives a consistency check of the proposed
identification between the new superstrata and the CFT states \eqref{eq:RRsmQ}.

The angular momentum charges $J$ and $\widetilde J$ associated with the left-moving and right-moving sector of the CFT respectively are related to the $J_{\phi}$ and $J_{\psi}$ components of the supergravity angular momentum through 
\begin{align}
J = \frac{J_{\phi} + J_{\psi}}{2}\,, \qquad \widetilde{J} = \frac{J_{\phi} - J_{\psi}}{2}\,.
\end{align} 
These charges can be found by analysing the $g_{\phi \psi}$ component of the ten-dimensional metric. In our
ansatz, this component is obtained by looking at the $\phi+\psi$
components of the one-forms $\beta$ and $\omega$. It is straightforward
to adapt the general prescription of \cite{Myers:1986un} to this case
(see for instance~\cite{Giusto:2013bda}), and one obtains
\begin{align}\label{angularmom}
\beta_\phi + \beta_\psi + \omega_\phi + \omega_\phi &\sim \sqrt2\, \frac{J - \widetilde J \, \cos2\theta}{r^2}+ \cO(r^{-3})\,.
\end{align} 
Thus it is possible to read off the angular momenta of our newly obtained solutions from the knowledge of $\beta$ (which is unchanged from~\eqref{stmetric}) and $\mu$~\eqref{eq:solmu}. One finds that these are given by
 \begin{align}
J &= \frac{R_y}{2}\left( a^2 + \bhat^2\frac{m+1}{k}\xh_{k,m,n}\right)\,,\qquad \widetilde J = \frac{R_y a^2}{2}\,.
\end{align}

Similarly, the momentum charge $Q_p$ can be extracted from the
large-distance behaviour of the function $\cF$:
\begin{align}\label{momcharge}
\cF & \sim - \frac{2Q_p}{r^2} + \cO(r^{-3})\,.
\end{align}
For our new solutions, we find
\begin{align}
Q_p &= \bhat {}^2\frac{ m+n+2}{2k}\xh_{k,m,n}\,.
\end{align}

Finally, we note that the brane charges $Q_1$ and $Q_5$ that appear in the scalar functions $Z_1$ and $Z_2$ do not change, as the two functions remain unchanged from the seed solutions \eqref{seed}. 

The supergravity charges calculated above are proportional to the
quantised charges calculated on the CFT side. The brane charges $Q_1$
and $Q_5$ are related to the D-brane numbers $n_1$ and $n_5$ through
\begin{equation}
Q_1 = \frac{(2\pi)^4\,g_s\,\alpha'^3}{V_4}\,n_1
~~,~~~~ 
Q_5 = \,g_s\,\alpha'\,n_5\,,
\end{equation}
with $V_4$ being the coordinate volume of $T^4$, $g_s$ the string coupling, and $\alpha'$ the Regge slope.
The relation between  the $Q_p$ obtained in \eqref{momcharge} and the quantised momentum number $n_p$ is
\begin{equation}
Q_p = \frac{(2\pi)^4\,g_s^2\,\alpha'^4}{V_4 R_y^2}\,n_p\ = \frac{Q_1 Q_5}{R_y^2 N} n_p.
\end{equation}
The angular momenta $J$, $\tilde{J}$ obtained from the supergravity calculation in
\eqref{angularmom} are related to the quantised ones $j$, $\bar
j$ by
\begin{equation}
J= \frac{(2\pi)^4 g_s^2 \alpha'^4}{V_4\,R_y}\, j = \frac{Q_1 Q_5}{R_y N}\, j \,,\qquad \tilde J= \frac{(2\pi)^4 g_s^2 \alpha'^4}{V_4\,R_y}\, \bar j = \frac{Q_1 Q_5}{R_y N}\, \bar j\,.
\end{equation}

We are now able to compare the charges obtained from the supergravity solutions with the ones calculated on the CFT side. The latter are given by \eqref{eq:QQchcft}. The crucial point at this step is to identify the supergravity constraint obtained from the regularity condition at the position of the supertube \eqref{eq:constgrav} and the CFT constraint for the total number of strands \eqref{eq:RRsmQ}. The parameter $a$ is connected with the number of untwisted strands $N_a$ and the number of twisted strands $N_b$ with the parameter $b$. We find that these quantities are connected by 
\begin{align}\label{eq:abNaNb}
a^2 = \frac{Q_1 Q_5}{R_y^2}\, \frac{N_a}{N}\,, \qquad b^2 = \frac{2 Q_1 Q_5}{R_y^2} \, \frac{k N_b}{N} \xh_{k,m,n}^{\,-1}\,.
\end{align}
Using this identification together with the relations between the supergravity and quantised momenta, we get that the CFT charges corresponding to the supergravity solutions obtained above are given by\footnote{We reinstated the subscript R for $j$ and $h$ to stress that we are listing the results in the in Ramond sector.}
\begin{subequations}\label{eq:CFTchargesj}
\begin{align}
\bar j_{\rm R} &= \frac{R_y N}{Q_1 Q_5} \widetilde J = \frac{N_a}2\,,\\
j_{\rm R} &= \frac{R_y N}{Q_1 Q_5}  J = \frac{ N_a}{2} + (m+1) N_b\,,
 \end{align}
\end{subequations}
for the angular momenta, and using that the momentum charge $n_p = h_{\rm R} - \bar h_{\rm R}$, we get 
\begin{align}
n_p = \frac{R_y^2 N}{Q_1 Q_5} Q_p = (m + n + 2)N_b\,.
\end{align}
We see that in all three cases these charges agree with the ones given in \eqref{eq:QQchcft}, if we set $q = 1$, as we should for our new supergravity solutions. This provides a check that the newly obtained solutions really are dual to the CFT states \eqref{eq:stateOkmnq}, with $q = 1$.

\section{Compendium of formulas and explicit solutions}
\label{Sect:Summary}

Here we collect the expressions for the new superstrata that
we constructed in this paper.

The new superstrata represent supersymmetric solutions of supergravity,
whose 10-dimensional fields are given by the ansatz
\eqref{ansatzSummary}.  The quantities that enter the ansatz satisfy two
layers of BPS equations, which are differential equations on a
four-dimensional base, which we took to be flat $\bbR^4$.
The ansatz quantities that solve the first-layer
equations~\eqref{layer1} are
\begin{subequations}
\label{eq:SumSollay1}
\begin{align}
&Z_1 = \frac{Q_1}{\Sigma}\,, \quad Z_2 = \frac{Q_5}{\Sigma}\,, \quad Z_4 = 0\,, \quad \Theta_1 = \Theta_2 = 0\,,\\*
& \Theta_4 = \bhat_4^{k,m,n}\,  \Delta_{k, 1+m, 1+n}\left[\frac{\Sigma}{r\sin\theta}\Omega^{(1)} \sin{\hat v_{k,1+m,1+n}}+ \left(\Omega^{(2)} + \Omega^{(3)}\right)\cos{\hat v_{k,1+m,1+n}}\right]\,,
\end{align}
\end{subequations}
where the definitions of $\Delta_{k,m,n}$ and $\hat v_{k,m,n}$ can be
found in \eqref{Delta_v_kmn_def} and the self-dual two-forms
$\Omega^{(i)}$ are given in \eqref{selfdualbasis}.  
The range of the integers $(k,m,n)$ is $k\ge 1$, $0\le m\le k-2$, $n\ge 0$.

Due to the linearity the BPS equations, an arbitrary superposition of
solutions~\eqref{eq:SumSollay1} is still a solution of the first-layer
equations (see \eqref{eq:lincombsol2}). However, in this paper, we
limited ourselves to single-mode superstrata, for which only one of the
coefficients $\widehat{b}_4^{k,m,n}\equiv \bhat$ is non-vanishing.  In
this case, the ansatz quantites that solve the second-layer BPS
equations \eqref{layer2eq} are given by
\begin{align}
\cF  =  \cF^{\rm RMS}(r,\theta),\qquad 
\omega = \omega_0 + \omega^{\rm RMS} (r, \theta)\,,
\end{align}
where $\omega_0$ is defined in \eqref{seedbeta},
\begin{equation}
\omega^{\rm RMS} = \mu_{k,m,n} (d\psi + d \phi) + \nu_{k,m,n} (d\psi - d\phi)\,,
\end{equation} 
and
\begin{subequations}
\begin{align}
\mathcal{F}^{\mathrm{RMS}} &= \mathcal{F}_{k,m,n} = 4\bhat^2 \, \left( F_{2k, 2m+2, 2n+2} + F_{2k, 2m+4,2n}\right)\,,\\
\mu_{k,m,n} &= \frac{\bhat^2 R_y}{\sqrt2}\, \biggl[ F_{2k, 2m+2, 2n} + F_{2k, 2m+4, 2n+2}\nonumber \\*
& \hspace{10ex}
 - \frac{r^2 + a^2 \sin^2\theta}{\Sigma}\left( F_{2k, 2m+2, 2n+2} + F_{2k, 2m+4,2n}\right) + \frac{\xh_{k,m,n}}{4\Sigma}\biggr]\,,
\end{align}
\end{subequations}
while $\nu_{k,m,n}$ has to be calculated on a case-by-case basis using
the differential equations~\eqref{eq:zetasol}.  The
functions $F_{2k,2m,2n}$ are given by \eqref{Ffun} and the coefficient
$\xh_{k,m,n}$ is  fixed by
regularity of the solution to be
\eqref{eq:xhatdef}.

In the dual CFT, the single-mode superstrata correspond to
states of the form
\begin{equation}
  \left|++\right\rangle^{N_a}\, \left(|k,m,n,q\rangle^{\rm R}\right)^{N_b}\;,~~~\mbox{with}~~~  N_a+k N_b = N \,\label{hkrj1Mar19}
\end{equation}
where $|k,m,n,q\rangle^{\rm R}$ is the spectral flow to the R sector of the NS state
\begin{equation}
   |k,m,n,q\rangle^{\rm NS} = (J^+_0)^m (L_{-1})^n  \left(G_{-\frac12}^{+1}G_{-\frac12}^{+2} + \frac{1}{k} J^+_0 L_{-1}\right)^q |O^{--}\rangle_k \;.
\end{equation}
with $q = 0,1$. States with $q=1$ are dual to the new superstrata while the states with $q=0$ are dual to the original superstrata constructed in~\cite{Bena:2017xbt}. The relation between the supergravity parameters $a,b$ and
the CFT parameters $N_a,N_b$ is given by \eqref{hkrj1Mar19}.

The above solutions are asymptotically AdS$_3$ but by simply setting
\begin{align}
Z_1 \rightarrow Z_1 +1\,,\qquad Z_2 \rightarrow Z_2 + 1\,,
\end{align}
the solutions become asymptotically flat. Such a transformation does not spoil the second-layer equations, meaning that the $\cF_{k,m,n}$ and $\omega_{k,m,n}$ are still valid solutions, even in the asymptotically flat case. 
In contrast, extending the original superstrata to asymptotically flat
ones required a non-trivial step of solving differential equations
\cite{Bena:2017xbt}.

\subsection{Explicit examples}

The solutions above are valid for any allowed set of quantum numbers
$(k,m,n, q=1)$. However, $\cF_{k,m,n}$ and $\mu_{k,m,n}$ contain linear
combinations of $F_{2k,2m,2n}$, which include non-trivial sums and are
generically hard to evaluate.
However, in certain cases, these sums can be evaluated explicitly, which
then makes it possible to find solutions to the second-layer BPS
equations in a closed form.  Here we present the explicit expression for
$\cF_{k,m,n},\omega_{k,m,n}$ for two classes of solutions.

\subsubsection{$(k,m,n,q) = (k, 0,0,1)$ class of solutions}

The simplest class of solutions is parametrised by the quantum numbers
$(k,m,n,q) = (k,0,0,1)$, with $k$ being an arbitrary positive
integer. These geometries already carry three charges, as momentum is
added through the action of the fermionic generators. They are given by
\begin{subequations}
\begin{align}
\label{eq:calFk00}
\cF_{k,0,0} &= -\frac{\bhat^2}{k (k^2 -1)^2 (X-1)^3 (a^2 +r^2)}\,\left[ P_{\cF}^{(0)}(X; k) +  P_{\cF}^{(1)}(X; k)\ Z\right]\,,\\
\label{eq:omegak00}
  \omega_{k,0,0}
 &=
 {\bhat^2\, R_y \over {\sqrt{2}\, k^2 (k^2-1)^2  (X-1)^4 \Sigma }} \, \left[ \left( P_{\phi}^{(0)}(X;k)  + P_{\phi}^{(1)}(X;k)\,Z + P_{\phi}^{(2)}(X;k)\,Z^2\right) \sin^2\theta \,d\phi\right. \nonumber\\
 & \hspace{31ex} +\left. \left( P_{\psi}^{(1)}(X;k)\, Z + P_{\psi}^{(2)}(X;k)\, Z^2\right) \cos^2 \theta\, d\psi   \right]\,,
 \end{align}
\end{subequations}
where we introduced the notation
\begin{align}\label{eq:defXZ}
X = \frac{a^2 \sin^2\theta}{r^2 + a^2}\,,
\qquad Z = \frac{r^2}{r^2 + a^2}\,,
\end{align}
and $P_{\cF}^{(l)}$, $P_{\psi}^{(l)}$, and $P_{\phi}^{(l)}$ denote polynomial functions in the variable $X$ with the parameter $k$. They are given by
\begin{subequations}\label{eq:Pfun}
\begin{align}
P_{\cF}^{(0)}(X;k) &  = 2(X-1)(X+1)(X^k -1) - k (X-1)^2 (1+3X) X^{k-1}\nonumber\\*
& \qquad  - 2k^2 (X-1)^2 (X^{k-1}-1) + k^3 (X-1)^3 X^{k-1}\,,\\
P_{\cF}^{(1)}(X;k) &  = 4(X+1)(X^{k} -1) - k(X-1) (1+ 6X + X^2) X^{k-1} \nonumber\\ & \qquad+ 2k^2 (X-1)^2 (X+1) X^{k-1}   - k^3 (X-1)^3 X^{k-1}\,,\\
P_{\phi}^{(0)}(X;k) &  = -2 (X-1)^2 (X^k -1) - k (X -1)^2(X+1 ) (X^{k-1} -1) \nonumber\\
& \qquad+ 2k^2 (X-1)^3 X^{k-1} + k^3 (X-1)^3 (X^{k-1}-1)\,,\\
P_{\phi}^{(1)}(X;k) &  =  - 4 (X-1)(X+2) (X^k-1) + 2k(X-1) (2X - X^{k-1} - 4 X^k + 3 X^{k+1}) \nonumber\\
& \qquad + 4k^2 (X-1)^2 X^{k-1} - 2k^3 (X-1)^3 X^{k-1}\,,\\
 P_{\phi}^{(2)}(X;k) &  =  -6(X+1)(X^k-1) + k(X-1)(2+ X^{k-1} + 8X^k + X^{k+1}) \nonumber\\
& \qquad  - 2k^2 (X-1)^2 (X+1) X^{k-1} + k^3 (X-1)^3 X^{k-1}\,,\\
P_{\psi}^{(1)}(X;k) &  =  -2 (X-1)(1+2X)(X^k -1) + k(X-1) (-1+2X +X^2 - 4X^k + 2X^{k+1})\nonumber\\
&\qquad  + 2k^2 (X-1)^2 X^k - k^3 (X-1)^3\,,\\
P_{\psi}^{(2)}(X;k) &  = -2 (1+ 4X + X^2)(X^k-1) + 2k(X-1)(X+1)(1+ 2X^k) - 2k^2 (X-1)^2 X^k\,.
\end{align}
\end{subequations}
The above gives regular solutions for any $k>1$.

It is interesting to see what happens for $k=1$.
If we treat $k$ as a continuous parameter, one can notice that all
functions $P$ in~\eqref{eq:Pfun} vanish linearly as $k\to 1$, and so
$\cF_{k,0,0}$ and
$\omega_{k, 0,0}$ have a simple pole at this value of $k$.  This means
that the solution is ill-defined for $k=1$,
which is consistent with the CFT result in section \ref{Sect:CFT} that
the states $\ket{1,0,n,1}^{\rm NS}$ are unphysical.
One may think that one could multiply $\bhat^2$ by $(k-1)$ to start with
in order to cancel this pole and get a physical solution.  However, one
can show that the solution with such modified normalisation would
contain a logarithmic divergence at $\theta \rightarrow 0$ and does not
represent a phyically allowed geometry.

\subsubsection{$(k,m,n,q) = (2,0,n,1)$ class of solutions}
The next simplest, physically meaningful class of three-charge examples is given by the quantum numbers $(k,m,n,q) = (2,0,n,1)$, where $n = 0,1,2,\ldots$. In the case where $n = 0$ the momentum charge is coming purely from fermionic generators, whereas for $n \neq 0$ the momentum charge is also added through the action of bosonic generators.
One finds
that for generic values of $n$ the solutions for the second-layer equations are given by
\begin{subequations}
\begin{multline}
\cF_{2,0,n} = \frac{\bhat^2}{a^2\,(n+1)^2 (n+2)(n+3)^2}\Biggl\{ \biggl[- 4\,  \frac{1-Z^{n+1}}{1-Z} - 2 (n+1)(n+3) \\*  + (n+1)\left((n+3)(n+4) +2\right) Z^{n+1} - (n+1)^2 (n+4) Z^{n+2}\biggr]- \biggl[ 4 - 8 \ \frac{1-Z^{n+1}}{1-Z} \\*
+ (n^3 + 8n^2 + 21n + 10) Z^{n+1} - 2 (n+1)^2 (n+4) Z^{n+2} + (n+1)^2 (n+2) Z^{n+3}\biggr] \sin^2 \theta\Biggr\}\,,
\end{multline}
\begin{multline}
\omega_{2,0,n} =   \frac{\bhat^2\, R_y }{\sqrt2 a^2 }\,\frac{1}{Z \sin^2 \theta + \cos^2 \theta}\, \frac{1}{(n+1)^2(n+2)(n+3)^2}\\*
\qquad\times \Biggr\{ \biggl[ 4Z \, \frac{1-Z^{n+1}}{1-Z} - (n+1)(n+5)\, Z^{n+2}
 + (n+1)^2 Z^{n+3} \biggr]\,(\cos^2\theta\, d\psi-\sin^2\theta\, d\phi)\\*
  + 2\, (n+1)(n+3)( 1- Z) \sin^2 \theta d\phi\Biggr\}\,,
\end{multline}
\end{subequations}
where we have used the same variable $Z$ as defined in~\eqref{eq:defXZ}.

\section*{Acknowledgments}

\vspace{-2mm}

We thank Stefano Giusto 
for useful discussions.
This work is supported in part by the Science and
Technology Facilities Council (STFC) Consolidated Grant ST/L000415/1
{\it String theory, gauge theory \& duality}.
The work of MS was supported in part by JSPS KAKENHI Grant Numbers
16H03979, and MEXT KAKENHI Grant Numbers 17H06357 and 17H06359.


\appendix

\section{Conventions}
\label{app:susyv}

Most of the conventions used in this paper follow directly the conventions used in \cite{Giusto:2013rxa}\footnote{See Appendix A of that paper.}. Here we present only a brief summary of the most important conventions used throughout the main text.
\subsection{Supersymmetry variations}

From \cite[(2.17)]{Bergshoeff:2001pv}, the supersymmetry transformations for bosonic fields in  type IIB supergravity are given by
\begin{subequations}
 \label{mgjw23Aug18}
 \begin{align}
 \delta e^{\underline{a}}_\mu &= \bar{\epsilon}\gu{a} \psi_\mu,\\
 \delta B_{\mu\nu} &= 2\bar{\epsilon}\Gamma_{[\mu}\sigma_3\psi_{\nu]},\\
 \delta \phi &={1\over 2}\bar{\epsilon}\lambda,\\
 \delta C^{(2n-1) }_{\mu_1\dots \mu_{2n-1}}
 &=-e^{-\phi} \bar{\epsilon}\Gamma_{[\mu_1\dots \mu_{2n-2}}\cP_{n}
 \left((2n-1)\psi_{\mu_{2n-1}]}-{1\over 2}\Gamma_{\mu_{2n-1}]}\lambda\right)
  \cr
  &\qquad
 +(n-1)(2n-1)C^{(2n-3)}_{[\mu_1\dots \mu_{2n-3}}\delta B_{\mu_{2n-2}\mu_{2n-1}]}\,,
  \label{hszi10Apr18}
 \end{align}
\end{subequations}
where $\mu,\nu,\dots$ are curved spacetime indices while
$\underline{a}, \underline{b}, \dots$ are local Lorentz indices.
The fermionic field variations read
\begin{align}
\label{eq:ferm}
 \delta \psi_\mu 
& = \left(\p_\mu + {1\over 4}\slashed{\omega}_\mu +{1\over 8}\cP H_{\mu\nu\rho}\Gamma^{\nu\rho} \right)\epsilon
 +{1\over 16}e^\phi \sum_{n} {1\over (2n)!} \slashed{F}_{2n}\Gamma_\mu \cP_n \epsilon,
 \\
 \delta \lambda &= \left(\slashed{\p}\phi + {1\over 12}\slashed{H}\cP\right)\epsilon 
 +{1\over 8}e^{\phi }\sum_n (-1)^{2n}{5-2n\over (2n)!}\slashed{F}_{2n}\cP_n \epsilon,
\end{align}
where $n=1/2,\dots,9/2$ and
\begin{align}
 \cP&=-\sigma^3,\qquad
 \cP_n=\begin{cases}
		  \sigma^1 & \text{$n+1/2$: even}\\
		  i\sigma^2 & \text{$n+1/2$: odd}\\
		 \end{cases}
\end{align}
acts on the doublet index of the gravitino $\psi^{1,2}_\mu $
and dilatino fields $\lambda^{1,2}$, which was suppressed in the expressions above.
We have also introduced the slashed notation 
\begin{align}
  \label{eq:conf1}
\slashed{A}_p &= {1\over p!}A_{m_1\dots m_p}\Gamma^{m_1\dots m_p}\,,
\end{align}
where every form index is contracted with a gamma matrix. The RR field strengths are related to the RR potentials by 
\begin{align}
  \label{eq:conf2}
  H = dB\,,\qquad F_{p} = dC_{p-1} - H \wedge C_{p-3}.
\end{align}
Using the $\Gamma$-matrix algebra and the self-duality relations of the
RR field strengths, one can write the variations \eqref{eq:ferm}
explicitly as
\begin{subequations}
\begin{align}
 \label{typeIIBsusytrfmexpl}
 \delta \psi^1_M&=\left(\nabla_M  -{1\over 8} H_{MNP}\Gamma^{NP}\right)\epsilon^1
 +{1\over 8}e^\phi\left(+\slashed{F}_1+\slashed{F}_3+{1\over 2}\slashed{F}_5\right)\Gamma_M\epsilon^2\,,
 \\
 \delta \psi^2_M&=\left(\nabla_M  +{1\over 8} H_{MNP}\Gamma^{NP}\right)\epsilon^2
 +{1\over 8}e^\phi\left(-\slashed{F}_1+\slashed{F}_3-{1\over 2}\slashed{F}_5\right)\Gamma_M\epsilon^1\,,
 \\
 \delta \lambda^1&=\left(d\phi -{1\over 2}H\right)\epsilon^1
 +{1\over 4}e^\phi(-4\slashed{F}_1 -2\slashed{F}_3)\epsilon^2\,,
 \\
 \delta \lambda^2&=\left(d\phi +{1\over 2}H\right)\epsilon^2
 +{1\over 4}e^\phi(+4\slashed{F}_1 -2\slashed{F}_3)\epsilon^1\,,
 \end{align}
\end{subequations}
where $\nabla_M = \partial_M + {1\over
4}\omega_{M\underline{a}\underline{b}}\Gamma^{\underline{a}\underline{b}}$
with $\omega_M$ the spin connection 1-form. In deriving the above
relations, we also used the fact that $\epsilon^{1,2}$ are Majorana-Weyl
spinors with positive chirality.
\subsection{Spinor conventions}

In the above we mentioned that the supersymmetry variations are
generated by Majorana-Weyl spinors.  The 10-dimensional Weyl condition
is given in \eqref{WeylCond}. If
we take the charge conjugation matrix $\cC$ to be
\begin{align}
\cC = \gu{t},
\end{align}
then the Majorana condition
\begin{align}
 \bar{\epsilon}\equiv \epsilon^\dagger \gu{t} = \epsilon^T\cC
\end{align}
simply means that the spinor is real,
\begin{align}
 \epsilon^*=\epsilon.
\end{align}
Note that for a Majorana spinor one has
\begin{align}\label{conjugation}
\bar{\epsilon}  = \epsilon^T \gu{t}\,.
\end{align}

In most of our analysis we use the coordinates $u$ and $v$ defined in \eqref{sptvw}. It is convenient to express $\gu{t}$ as 
\begin{align}
\gu{t} = \frac1{\sqrt{2}} \left( \gu{u} + \gu{v}\right) \equiv \guh{u} + \guh{v} \,,
\end{align}
where $\guh{u}$ and $\guh{v}$ are defined in \eqref{def_ghat}.  They are
convenient to work with, when acted upon a spinor with a definite
eigenvalue under $\gu{uv}$. They satisfy
\begin{align}
\guh{u}{}^T = \guh{u}{}^\dagger = - \guh{v}\,, \qquad 
\guh{v}{}^T = \guh{v}{}^\dagger = - \guh{u}\,,
\end{align}
where we used the hermiticity properties of the $\Gamma$ matrices,
\begin{align}
{\gu{t}}^\dagger = - \gu{t}, \hspace{10ex} {\gu{i}}^\dagger = \gu{i}, \qquad i \neq t.
\end{align}

As an example, using all these conventions, we find that the transpose of the spinor \eqref{spinorzeta-} used to generate the double variation \eqref{doubleaxionvar1} is given by
\begin{align}\label{transspin}
\left({\epsilon^1}\right)^T = \frac12 e^{i\bigl( {\phit + \psit \over 2} - \frac{v}{\sqrt2 R_y}\bigr)} \left({\zeta_-^{\alpha A}}\right)^T e^{- \frac{\theta}{2} \gu{r\theta}} \left(R_+ + i R_- \guh{ru}\right)\,.
\end{align}

\subsection{Identifying Killing spinors and CFT fermionic generators} \label{sec:aligning}

Here we will give justification for the identification \eqref{STprojtxt},
\eqref{eq:identificationtxt} between the components of the Killing
spinors $\zeta_\pm$ and the CFT fermionic generators
$G_{\pm {1\over 2}}^{\alpha A}$.

As we can see from \eqref{summary1} and \eqref{summary2}, the spinors
$\zeta_\pm$ and $\zetat_\pm$ are related to the $u$- and $v$-dependent
part of the Killing spinors, respectively. So, they are naturally linked
with the left-moving ($G$) and right-moving ($\Gt$) sectors of CFT\@.
Therefore, we will henceforth focus on the identification in the
left-moving sector between $\zeta_\pm$ and $G_{\pm {1\over 2}}^{\alpha A}$.
From the $v$-dependence in~\eqref{eq:chidef}, we have the tentative
identification
\begin{align}
\zeta_\pm  \longleftrightarrow G_{\pm {1\over 2}}^{\alpha A}.
\end{align}

The next issue is to relate the $SU(2)_L\times SU(2)_R$ R-symmetry on
the CFT side and the $SO(4)\cong SU(2)_L\times SU(2)_R$ symmetry of the
sphere~$S^3$ on the supergravity side.  It is clear that these symmetry
groups are to be identified with each other but we would like to
``align'' them by identifying the ``$J^3$'' generators on the two sides.

On the supergravity side, we can write down a set of $SU(2)_L\times
SU(2)_R$ generators acting on spinors as
\begin{align}\label{SU2SU2def1}
J_a = -\frac{i}{4}\left(\gu{r a} + \frac12\epsilon^{\underline{abc}} \gu{bc}\right), \qquad \widetilde J_a = -\frac{i}{4}\left(\gu{r a} - \frac12 \epsilon^{\underline{abc}} \gu{bc}\right),
\end{align}
or, more explicitly,
\begin{subequations}\label{eq:SU(2)Jgen}
\begin{align}
J_\theta = - \frac{i}{4}\left(\gu{r \theta} + \gu{\phit \psit}\right), \qquad J_{\phit} = - \frac{i}{4}\left(\gu{r \phit} + \gu{\psit \theta}\right), \qquad J_{\psit} = - \frac{i}{4}\left(\gu{r \psit} + \gu{\theta \phit}\right)\,,\label{eq:SU(2)Jgen1}\\
\widetilde J_\theta = - \frac{i}{4}\left(\gu{r \theta} - \gu{\phit \psit}\right), \qquad \widetilde J_{\phit} = - \frac{i}{4}\left(\gu{r \phit}- \gu{\psit \theta}\right), \qquad \widetilde J_{\psit} = - \frac{i}{4}\left(\gu{r \psit} - \gu{\theta \phit}\right).
\end{align}
\end{subequations}
These matrices satisfy the commutation relations
\begin{align}
[ J_a , J_b] = i \epsilon_{abc} J_c, \qquad [ \widetilde J_a , \widetilde J_b] = i \epsilon_{abc} \widetilde  J_c, \qquad [ J_a , \widetilde J_b] = 0, 
\end{align}
and have Casimir operators given by
\begin{align}\label{Casimir1}
J^2 &= \left( J_\theta^2 + J_{\phit}^2 + J_{\psit}^2\right) = \frac38\left(1 - \gu{r \theta \phit \psit} \right)\,,\quad 
 \widetilde J^2 = \left( \widetilde J_\theta^2 + \widetilde J_{\phit}^2 + \widetilde J_{\psit}^2\right) = \frac38\left(1 + \gu{r \theta \phit \psit}\right).
\end{align}

We turn our focus to the action of these matrices on the spinors
$\zeta_{\pm}$. Recall that these spinors must satisfy the conditions
\eqref{WeylCond} and \eqref{cond2}. Combining these two implies that 
they must satisfy
\begin{align}
\gu{uvr\theta \phit \psit}\, \zeta_\pm = \zeta_\pm.
\end{align}
From this condition together with \eqref{zetauveigen}, we find that
\begin{align}\label{cond3}
\gu{r\theta \phit \psit}\, \zeta_{\pm} = - \zeta_{\pm}\,.
\end{align}
From the Casimir operators \eqref{Casimir1}, we conclude that
$\zeta_\pm$ transform in the $({\bf 2},{\bf 1})$ representation under
the $SU(2)_L\times SU(2)_R$ generated by $J_a,\widetilde{J}_a$ . This is
as expected, because $\zeta_{\pm}$ is to be identified with
$G_{\pm\frac12}^{\alpha A}$ which transform in the same
representation under the R-symmetry.

We need to identify which of the operators in \eqref{SU2SU2def1}
corresponds to the ``$J^3$'' operator of the $SU(2)_L$ algebra on the
CFT side. This operator will allow us to distinguish between the spinor
components corresponding to $G_{\pm\frac12}^{+ A}$ and
$G_{\pm\frac12}^{- A}$. In order to see that, we look at the spinor
\eqref{summary1}. By setting $\zetat_\pm=0$, we obtain a spinor with
terms that contain the combination $Y_+ \zeta_\pm$. Using the constraint
\eqref{cond3} one can show that the combination appearing in the spinor
can be rewritten as
\begin{align}
Y_+ \zeta_\pm & =  e^{{\theta\over 2}\gu{r\theta}} \Bigl( e^{i {\phit + \psit \over 2}} \cP_S^+ + e^{- i {\phit + \psit \over 2}} \cP_S^-\Bigr) \zeta_\pm\,,\label{ipvz20Dec18}
\end{align}
where $\cP_S^{\pm}$ is the projection operator onto the $J_{\psit}=\pm
\frac{1}{2}$ eigenspace defined in \eqref{STprojtxt}.  Because the
algebra \eqref{n4algc0} says that $\{G_{\frac{1}{2}}^{\pm
A},G_{-\frac{1}{2}}^{\pm B}\}\sim J_0^\pm$, a double variation by
$\zeta_\pm$ should reproduce the bosonic symmetry $J_0^{\pm}$ whose
realisation is given in \eqref{su(2)_gen_NS}.  Since $J_0^{\pm}$ include
a prefactor of $e^{\pm i(\phit + \psit)}$, we conclude that the
$J_{\psit}=\pm \frac{1}{2}$ eigenspaces, multiplied by $e^{\pm i{\phit +
\psit\over 2}}$ in \eqref{ipvz20Dec18}, are precisely the $J_3=\pm
\frac{1}{2}$ eigenspaces. Namely, $J_{\psit}$ can be identified with 
$J^3_0$ on the CFT side.
This leads to a finer identification
\begin{align}
\cP^\alpha_S \zeta_\pm  ~\longleftrightarrow~
 G_{\pm {1\over 2}}^{\alpha A}.
\end{align}

One can repeat the procedure for the $SU(2)_B \times SU(2)_C$ symmetry
on the CFT side, which is to be identified with the symmetry of the
internal $T^4$ on the supergravity side.  We can write down a set of
supergravity generators of $SU(2)_B \times SU(2)_C$ in the spinor representation as follows:
\begin{subequations}
\begin{align}
\begin{aligned}
 B_1 &= - \frac{i}{4}\left(\Gamma^{\underline{78}} - \gu{69}\right), \quad&
 B_2 &= - \frac{i}{4}\left(\Gamma^{\underline{86}} - \gu{79}\right), \quad&
 B_3 &= - \frac{i}{4}\left(\Gamma^{\underline{67}} - \gu{89}\right)\,,\label{SU(2)Bgen}\\
 C_1 &= - \frac{i}{4}\left(\Gamma^{\underline{78}} + \gu{69}\right), &
 C_2 &= - \frac{i}{4}\left(\Gamma^{\underline{86}} + \gu{79}\right), &
 C_3 &= - \frac{i}{4}\left(\Gamma^{\underline{67}} + \gu{89}\right),
\end{aligned}
\end{align}
\end{subequations}
which again satisfy the commutation relations
\begin{align}
\left[B_i, B_j\right] = i\epsilon_{ijk}B_k, \qquad \left[ C_i, C_j\right] = i \epsilon_{ijk} C_k,\qquad \left[B_i, C_j\right] = 0\,,
\end{align}
and have Casimir operators
\begin{align}
B^2  = \frac38\left(1 + \gu{6789}\right)\,,\qquad C^2 = \frac38\left(1 - \gu{6789}\right)\,.
\end{align}
Since our spinors satisfy the condition \eqref{cond2}, we see that
$\zeta_\pm$ both transform in the $({\bf 2},{\bf 1})$ representation 
under $SU(2)_B\times SU(2)_C$ generated by $B_i,C_i$. This is accordance with the fact that
$G_{\pm \frac12}^{\alpha A}$ transform in the same representation under
$SU(2)_B\times SU(2)_C$.

However, unlike the case of $SU(2)_L\times SU(2)_R$, there is no unique
way to align the $SU(2)_B\times SU(2)_C$ groups between supergravity and
CFT, because our ansatz \eqref{ansatzSummary} does not distinguish the
four directions inside the internal $T^4$.  As a result, in the Killing
spinors \eqref{summary1}, no $\Gamma$ matrix with legs in the $T^4$
appear, and all $\gu{k}$ with $k = 6,7,8,9$ are on an equal
footing. Therefore, we can choose the ``$J^3$'' direction of $SU(2)_B$
as we like.  Specifically, we define the projectors $\cP_T^{A}$, $A=1,2$
by \eqref{STprojtxt} and identify its $A$ index with the $A$ index of
$G_{\pm\half}^{\alpha A}$.
This leads to the final identification
\begin{align}
\zeta_\pm^{\alpha A}=\cP^\alpha_S \cP^A_T \zeta_\pm  
 ~\longleftrightarrow~
 G_{\pm {1\over 2}}^{\alpha A}
\end{align}
which is \eqref{eq:identificationtxt} in the main text.

\section{Explicit results in the NS-NS sector}
\label{app:ExplRes}

In section \ref{Sect:KillingSpinors} we have calculated the variations of the dilatino fields generated by the AdS$_3\times S^3 \times T^4$ Killing spinor $\zeta_-^{\alpha A}$. In a similar  manner one can calculate the variations of the components of the gravitino fields. These are presented in this Appendix and one finds some subtleties one does not find in the dilatino variation. Furthermore, we also give the variation of the fermionic fields generated by the spinor $\zeta_+^{\alpha A}$. We expect that these variations vanish, and in fact the dilatino variations do, but the gravitino component variations show the same subtleties as one finds in the variations of generated by the $\zeta_-^{\alpha A}$ spinors. Finally we will present the results of the double variations of the components of the $B$-field in the NS-NS coordinates, which we omitted in the main text.

\subsection{Gravitino variations generated by $\zeta_-^{\alpha A}$}

One finds that the variations of the components of the gravitino field generated by the spinor \eqref{spinorzeta-} are given by
\begin{subequations}\label{gravvar-}
\begin{align}
\delta \psi_{u,b}^1 &= -\frac{b a^k  \sin^{k-1} \theta }{4\sqrt{2}R_y\left(r^2+a^2\right)^{\frac{k+1}{2}}}e^{-i\beta_1}  \biggl[k \left(i R_- + R_+ \guh{vr} \right) \left( \cos\theta \gu{r\theta} - i \gu{r\phit} + \sin\theta \right)  \nonumber\\*
&\quad  - i \frac{\sqrt{r^2+a^2}}{a} \left(  R_+\left(  R_-^4 +  k\right)   -i R_- \guh{vr} \left( R_+^4 +   k\right) \right)\sin\theta \biggr]Y_+ \zeta^{\alpha A}_-\,,
\\
\delta \psi_{v,b}^1 &=  -\frac{b a^k  \sin^{k-1} \theta }{4\sqrt{2}R_y\left(r^2+a^2\right)^{\frac{k+1}{2}}}e^{-i\beta_1}   \biggl[k\left(iR_+^3 + R_-^3\guh{vr}\right) \left( \cos\theta \gu{r\theta}- i \gu{r\phit} + \sin\theta\right)  \nonumber\\*
&\quad  - \left(ik + i\right)  \frac{\sqrt{r^2+a^2}}{a}  \left( R_+  - i R_-  \guh{vr} \right)\sin\theta  \biggr]Y_+ \zeta^{\alpha A}_-\,,
\\
\delta \psi_{r,b}^1 &= -\frac{b a^{k} \sin^{k-1} \theta }{4\left(r^2+a^2\right)^{\frac{k+2}{2}}}  e^{-i\beta_1}   \biggl[k \left(-R_- -iR_+ \guh{vr}\right) \left( \cos\theta \gu{r\theta} - i\gu{r\phit} + \sin\theta\right)   \nonumber\\*
&\quad  + \frac{\sqrt{r^2+a^2}}{a} \left(R_+  \left(1-\frac{kr}{\sqrt{1+r^2}} \right) + i R_-\left(  1 + \frac{kr}{\sqrt{1+r^2}} \right) \guh{vr}   \right)\sin\theta\biggr]Y_+ \zeta^{\alpha A}_-\,,
\\
\delta \psi_{\theta,b}^1 &= - \frac{ba^k \sin^{k-1} \theta }{4\left(r^2+a^2\right)^{\frac{k+1}{2}}}  e^{-i\beta_1}   \biggl[k\left(R_- - i R_+ \guh{vr}\right)\gu{r\theta}  \left( \cos\theta \gu{r\theta} - i \gu{r\phit} + \sin\theta \right) \nonumber\\*
&\quad  + \frac{\sqrt{r^2+a^2}}{a}\left( R_+ - i R_- \guh{vr}\right) \left(k\cos\theta + \sin\theta \gu{r\theta} \right)\biggr]  Y_+ \zeta^{\alpha A}_-\,,
\\
\delta \psi_{\phit,b}^1 &= - \frac{ba^k  \sin^{k} \theta }{4\left(r^2+a^2\right)^{\frac{k+1}{2}}}  e^{-i\beta_1}   \biggl[k \left( R_- -iR_+\guh{vr}\right) \gu{r\phit} \left( \cos\theta \gu{r\theta} - i \gu{r\phit} + \sin\theta \right)  \nonumber\\*
&\quad  +\frac{\sqrt{r^2+a^2}}{a}\left(R_+- i R_- \guh{vr}\right) \left( \sin\theta \gu{r\phit} - ik\right)\biggr]Y_+ \zeta^{\alpha A}_-\,,
\\
\delta \psi_{\psit,b}^1 &= - \frac{ba^k  \sin^{k-1} \theta \cos\theta }{4\left(r^2+a^2\right)^{\frac{k+1}{2}}}  e^{-i\beta_1}  
\biggl[ k \left( R_- -iR_+\, \guh{vr}\right) \gu{r\psit}  \left( \cos\theta \gu{r\theta} - i \gu{r\phit} + \sin\theta \right)  \nonumber\\*
&\quad + \frac{\sqrt{r^2+a^2}}{a}  \left(R_+ - i R_-  \guh{vr} \right)\sin\theta  \gu{r\psit} \biggr]Y_+ \zeta^{\alpha A}_-\,,\\
\delta\psi^1_{k,b} &= 0\,, \qquad k = 6,7,8,9,
\end{align}
\end{subequations}
where
\begin{align}
 \beta_1\equiv k\Bigl({u+v\over \sqrt2 R_y} + \phit\Bigr) + \frac{v	}{\sqrt2 R_y}=\vh_{k,0,\half}.
\end{align}
Note that, for the variations generated by spinor \eqref{spinorzeta-},
we have $\delta \psi_{M, b}^1 = \delta \psi_{M,b}^2$ for all $M$.  We find that the
gravitino variations \eqref{gravvar-} generically have two parts. The
first one is analogous to the dilatino variation, as it contains the
combination $(\cos\theta \gu{r\theta} - i \gu{r\phit} + \sin\theta)$,
which when acting on the combination $(Y_+\zeta_-^{\alpha A})$ projects
out the $\zeta_-^{-A}$ components of the spinor. This part is again
expected from the fact that the perturbed geometry is dual to an
anti-chiral primary state.  The second part of the variations (given in
the second lines of the respective variations) does not distinguish
between the $\zeta_-^{+A}$ and $\zeta_-^{-A}$ components of the spinors.
We believe that these extra factors are related to the residual gauge
freedom that we have in our system and one should be able to
consistently ignore these extra factors. This claim is further
strengthened by the fact that once we use these fermionic variations to
calculate the double variations of bosonic fields, these additional
terms consistently cancel out and thus do not contribute to any
observable fields.

\subsection{Fermionic variations generated by $\zeta_+^{\alpha A}$}

The claim that the extra terms found in \eqref{gravvar-} can be set to zero by gauge transformation is further supported by calculating the variations of fermionic fields generated by the Killing spinor components $\zeta_+^{\alpha A}$. Since these are dual to the modes $G_{+ \frac12}^{\alpha A}$, the variations should vanish. Furthermore, one should not obtain any terms that would distinguish between the 
$\zeta_+^{+ A}$ and $\zeta_+^{- A}$ components as now variations generated by both should vanish. If we set $\zeta_-^{\alpha A} = 0 $ and make   $\zeta_+^{\alpha A}$ arbitrary, the spinor that generates the variation is given by \begin{align}\label{spinorszeta+}
\epsilon^1 = - \epsilon^2  = \frac12\left(iR_- \guh{vr} + R_+ \right)Y_+ \zeta_+^{\alpha A} \,e^{\frac{iv}{\sqrt2 R_y}}\,.
\end{align}
Using this spinor as the generator of the solutions, one obtains that the explicit variations of the fermionic fields are given by  
\begin{subequations}\label{gravvar+}
\begin{align}
\delta \lambda^1_b & =  0,\\
\delta \psi_{u,b}^1 &=  -\frac{b a^{k-1}\sin^{k}\theta}{4	\sqrt2 R_y\left(a^2 + r^2\right)^{k/2}}   e^{-i\beta_2}\biggl[\left( iR_+^4 -ik\right)  i R_-  \guh{vr} + \left( i R_-^4 -ik\right) R_+ \biggr]Y_+ \zeta_+^{\alpha A}
\\
\delta \psi_{v,b}^1 &= - \frac{b a^{k-1} \sin^{k}\theta}{4\sqrt2	 R_y	\left(a^2 + r^2\right)^{k/2}}   e^{-i\beta_2}(i - ik)\left( i R_-  \guh{vr} + R_+ \right)Y_+ \zeta_+^{\alpha A}
\\
\delta \psi_{r,b}^1 &= - \frac{b a^{k-1}\sin^{k}\theta}{4	\left(a^2 + r^2\right)^{\frac{k+1}{2}}}e^{-i\beta_2}\biggl[ \left( -1 -\frac{kr}{\sqrt{a^2 + r^2} }\right) i R_-\guh{vr} 
+ \left(1 - \frac{kr}{\sqrt{a^2 +r^2}}\right)   R_+ \biggr]Y_+ \zeta_+^{\alpha A}
\\
\delta \psi_{\theta,b}^1 &= - \frac{b a^{k-1} \sin^{k-1}\theta}{4	\left(a^2 + r^2\right)^{k/2}}  e^{-i\beta_2} \left(R_+ + i R_- \guh{vr}\right)\left(k\cos\theta + \sin\theta \gu{r\theta} \right) Y_+ \zeta_+^{\alpha A}
\\
\delta \psi_{\phit,b}^1 &= - \frac{b a^{k-1} \sin^k\theta}{4	\left(a^2 + r^2\right)^{k/2}}  e^{-i\beta_2}  \left(R_+ + i R_-  \guh{vr}\right) \left( - ik + \sin\theta \gu{r\phit}\right)Y_+ \zeta_+^{\alpha A}
\\
\delta \psi_{\psit,b}^1 &= - \frac{b a^{k-1} \sin^k \theta \cos\theta }{4	\left(a^2 + r^2\right)^{k/2}} e^{-i\beta_2}\left(  R_+ +  i R_-  \guh{vr} \right)\gu{r\psi} \,Y_+ \zeta_+^{\alpha A}\,,\\
\delta \psi_{k, b}^1 &= 0\,,\qquad k = 6,7,8,9\,,
\end{align}
\end{subequations}
where 
\begin{align}
\beta_2\equiv k\Bigl({u+v\over \sqrt2 R_y}+ \phit\Bigr) - \frac{v	}{\sqrt2 R_y} =\vh_{k,0,-\half},
\end{align}
and $\delta \lambda_b^1 = \delta \lambda_b^2$,  $\delta \psi_{M,b}^1 = \delta \psi_{M,b}^2$. We see that the variations of the dilatino fields vanish, as expected. On the other hand, the variations of the components of the gravitino fields do not vanish. However, note that these variations only contain the terms which we deemed as a consequence of the gauge freedom in our system and lacks the term which would distinguish between the two $SU(2)_L$ components. Since these variations should be trivial, we get another confirmation that these terms appearing in the gravitino variations are not physical.

\subsection{Variations of the $B$-field}

The non-vanishing term in the double $B$-field variation is given by 
\begin{align}
\delta'\delta B_{\mu \nu} = 2 \bar{\epsilon} \,\Gamma_{[\mu} \sigma^3 \delta' \psit_{\nu]} = 2 \left(\epsilon_1\right)^T \gu{t} \left( \Gamma_\mu \delta \psit_\nu^1  - \Gamma_\nu \delta' \psit_\mu^1\right)\,,
\end{align}
where in the last term we again used the fact that the only non-zero spinor components used to generate the variations are $\zeta_-^{+ A}$. The variations of the  individual components of the $B$-field  in the NS-NS sector are then given by 
\begin{subequations}\label{BdvarNS}
\begin{align}
B_{uv}&= -2bk a^{k-\frac52} R_y^{-\frac32} \frac{ r \sin^{k-1}\theta \cos\theta }{\left(r^2+a^2\right)^{\frac{k-1}{2}}}\, e^{-i\beta_3}  \cA\\
 B_{ur}&=  B_{vr} = - \sqrt2 i bk a^{k-\frac12} R_y^{-\frac12} \frac{\sin^{k-1}\theta \cos\theta }{\left(r^2+a^2\right)^{\frac{k+1}{2}}}\, e^{-i\beta_3}  \cA\\ 
B_{u\psit}&= - B_{v\psit} = \sqrt2 b k a^{k-\frac12} R_y^{-\frac12} \frac{\sin^{k-1}\theta \cos\theta }{\left(r^2+a^2\right)^{\frac{k+1}{2}}}\, e^{-i\beta_3}  \cA\\
B_{r\psit}&= -2ibk a^{k-\frac12} R_y^{\frac12} \frac{ \sin^{k-1}\theta\,\cos\theta }{\left(r^2+a^2\right)^{\frac{k+1}{2}}}\, e^{-i\beta_3} \cA\\
B_{\phit\psit}&= 2bk a^{k-\frac12} R_y^{\frac12} \frac{r \sin^{k+1}\theta \cos\theta }{\left(r^2+a^2\right)^{\frac{k+1}{2}}}\, e^{-i\beta_3}   \cA\\
 B_{u\theta}&= -  B_{v\theta}  = \sqrt2 i bk R_y^{-\frac12} a^{k-\frac12} \frac{ r\sin^{k}\theta }{\left(r^2+a^2\right)^{\frac{k+1}{2}}}\, e^{-i\beta_3}  \cA\\
B_{\theta\phit} &= - 2i bk R_y^{\frac12} a^{k-\frac12}  \frac{ r \sin^{k}\theta }{\left(r^2+a^2\right)^{\frac{k+1}{2}}}\, e^{-i\beta_3}   \cA\\
B_{r\theta}& = 2bk a^{k-\frac12} R_y^{\frac12} \frac{ \sin^{k}\theta }{\left(r^2+a^2\right)^{\frac{k+1}{2}}}\, e^{-i\beta_3}  \cA\\
B_{u\phit}&= B_{v\phit}  = B_{\theta\psit} = B_{r\phit} = 0
\end{align}
\end{subequations}
where
\begin{align}
\beta_3&\equiv k\Bigl({u+v \over \sqrt2 R_y } + \phit\Bigr) + \sqrt2 \frac{v}{R_y} - (\phit + \psit),\\
 \cA&=\left[{\left(\zeta_-^{+1}\right)}^T i\gu{r\theta}  \zeta_-^{\prime +2}\right].
\end{align}

\section{Supergravity spectrum around AdS$_3\times S^3$}
\label{app:DegerEtal}

In the main text, we explicitly constructed a new 3-charge supergravity
solution that corresponds to the CFT state \eqref{eq:stateOkmnq} (or
\eqref{eq:RRsmQ}).  Surprisingly, at linear order, the supergravity
solution only involved excitation of $\Theta_4$, which is related to the
NS-NS $B$-field, and not any other fields such as $Z_4$ or the 4D base
metric (even at the non-linear level, the only excited fields are
$\Theta_4,\cF,\omega$).  Actually, this fact can be deduced from the
analysis of the spectrum of 6D supergravity compactified on AdS$_3\times
S^3$, as we will discuss in this Appendix.

The spectrum of 6D supergravity compactified on AdS$_3\times S^3$, which
is relevant for the AdS$_3$/CFT$_2$ correspondence, has been extensively
studied \cite{Deger:1998nm, Maldacena:1998bw,
Larsen:1998xm, deBoer:1998kjm} (see also \cite{Kanitscheider:2006zf}).
Particularly useful for us is Ref.~\cite{Deger:1998nm} which, for $D=6$,
$\cN=(2,0)$ supergravity with $n$ tensor multiplets, worked out the
excitation spectrum around the AdS$_3$/CFT$_2$ background and its
explicit supermultiplet structure. Type IIB supergravity compactified on
K3 corresponds to $n=21$, for which the theory is anomaly free.
On the other hand, $T^4$ compactification can be studied by taking
$n=5$, although this truncates fields that correspond to gravitino
multiplets of the $\cN=(2,0)$ theory (see Ref.~\cite{Larsen:1998xm} which
studied the spectrum in the $T^4$ case including the gravitino
multiplets).

Here we briefly summarise the result of Ref.~\cite{Deger:1998nm} and
relate it to our setup. Note that our convention for the R-charge is
opposite to that in \cite{Deger:1998nm}; so, chiral primaries there
correspond to anti-chiral primaries in our setup, and $G^{\pm A}_n,
\Gt^{\dot{\pm}A}_n$ there correspond to $G^{\mp A}_n,
\Gt^{\dot{\mp}A}_n$ in our setup.

The $D=6$, $\cN=(2,0)$ supergravity theory has the duality group
$SO(5,n)$, and its bosonic fields are the graviton $g_{MN}$, $5$ 2-form
potentials $B_{MN}^i$ with self-dual field strengths, $n$ 2-form
potentials $B_{MN}^r$ with anti-self-dual field strengths, and $5n$
scalars. Here, $M,N,\dots$ are curved 6D indices, $i=1,\dots,5$ is the
$SO(5)$ vector index, and $r=6,\dots,5+n$ is the $SO(n)$ vector index.
The scalars live in the coset space $SO(5,n)/(SO(5)\times SO(n))$ and
can be parametrised by vielbeins $(V_I^i,V_I^r)$ where
$I=(i,r)=1,\dots,5+n$ is the $SO(5,n)$ vector index.  Self-duality
(anti-self-duality) is not imposed on the 3-forms $G^i=dB^i$ and
$G^r=dB^r$ but on $H^i=G^I V_I^i$ and $H^r=G^I V_I^r$; namely, they
satisfy $*_6 H^i=H^i$ and $*_6 H^r=-H^r$.

Small fluctuations around the AdS$_3\times S^3$ background can be
studied by writing the fields as
\begin{align}
\begin{split}
 g_{MN}&=\bar{g}_{MN}+h_{MN},\qquad
 G^I_{MNP}=\bar{G}^I_{MNP}+g^{I}_{MNP},\\
 V^i_I&=\delta^i_I+\phi^{ir}\delta^{r}_I,\qquad
 V^r_I=\delta^r_I+\phi^{ir}\delta^i_I.
\end{split}
\end{align}
Here, $\bar{g}_{MN}$ is the background AdS$_3\times S^3$ metric
and $\bar{G}^I_{MNP}$ is the background 3-form field
strength with components
\begin{align}
 \bar{G}^i_{\mu\nu\rho}=m \epsilon_{\mu\nu\rho}\delta^i_5,\qquad
 \bar{G}^i_{abc}=m \epsilon_{abc}\delta^i_5\qquad
 \bar{G}^r_{MNP}=0,
\end{align}
where $m^{-1}$ is the radius of AdS$_3\times S^3$, $\mu,\nu,\dots$ are
curved AdS$_3$ indices, and $a,b,\dots$ are curved $S^3$ indices.  To
support the AdS$_3\times S^3$ background, we must turn on one of the
self-dual form fields, which we have taken to be the $i=5$ one.  This
breaks the $SO(5)$ R-symmetry of 6D supergravity to
$\widetilde{SO(4)}$.\footnote{In \cite{Deger:1998nm},
$\widetilde{SO(4)}$ is denoted as $SO(4)_R$.} We use
$\underline{i}=1,\dots,4$ for the vector index for this unbroken
$\widetilde{SO(4)}$.  This $\widetilde{SO(4)}=SU(2)_B\times SU(2)_C$
symmetry is not to be confused with the $SO(4)=SU(2)_L\times SU(2)_R$
R-symmetry of the CFT coming from the isometry of the $S^3$. Rather,
$SU(2)_B$ is related to the $SU(2)$ outer automorphism of the
$\cN=(4,4)$ superconformal algebra, while $SU(2)_C$ is related to a
custodial symmetry with respect which the fundamental fields of the CFT
are charged (but not the superconformal generators are).

We can relate the above fields in 6D supergravity to the quantities
appearing in the main text as follows.  First, the self-dual $H^{i=5}$
and the anti-self-dual $H^{r=6}$ are related to the self-dual and
anti-self-dual linear combinations of the RR 3-form $F_3$ and its
6-dimensional dual $*_6 F_3$ \cite{Kanitscheider:2006zf}, while the
scalar $\phi^{56}$ is related to the dilaton which is roughly $Z_1/Z_2$.
To discuss $r=7,\dots,5+n$, let us recall that there are $(n-1)$
$(1,1)$-forms in $T^4$ ($n-1=4$) and K3 ($n-1=20$). Among them is the
K\"ahler form $J\equiv \omega^{(7)}$.  Let us denote the remaining
$(n-2)$ $(1,1)$-forms by $\omega^{(r)}$, $r=8,\dots,5+n$.  The relations are
slightly different for $r=7$ and $r\ge 8$.
The 3-form $H^{r=7}$ is related to the NS-NS 3-form $H_3$ which in
turn is related to the 2-form $\Theta_4$, and the scalar $\phi^{57}$ is
related to $Z_4$.  On the other hand, $H^{r=8,\dots,5+n}$ are related to
the part of the RR 5-form $F_5$ involving $\omega^{(r)}$, and the scalar
$\phi^{5r}$ is related to the part of $H_3$ involving $\omega^{(r)}$
\cite{Bakhshaei:2018vux}.

The fields $h_{MN}$, $g^{I}_{MNP}$, and $\phi^{ir}$ represent small
fluctuations around the background and can be decomposed as
\begin{align}
\begin{split}
  h_{\mu\nu}&=H_{\mu\nu}+\bar{g}_{\mu\nu}M,\qquad \bar{g}^{\mu\nu}H_{\mu\nu}=0,\\
 h_{\mu a}&=K_{\mu a},\\
 h_{ab}&=L_{ab}+\bar{g}_{ab}N,\qquad \bar{g}^{ab}L_{ab}=0
\end{split}
\end{align}
and
\begin{align}
\begin{split}
  g^I_{MNP}&=3\p_{[M}b^I_{NP]},\\
 b^I_{\mu\nu}&=\epsilon_{\mu\nu\rho}X^{I\rho},\qquad
 b^I_{ab}=\epsilon_{abc}U^{Ic},\qquad
 b^I_{\mu a}=Z^I_{\mu a}.
\end{split}
\end{align}

In the de Donder--Lorentz gauge, the fluctuation fields can be expanded
in the harmonic functions in $S^3$
as
\begin{align}
\begin{aligned}
  H_{\mu\nu}&=\sum H_{\mu\nu}^{(\ell\, 0)}Y^{(\ell\, 0)}, &
  M&=\sum M^{(\ell\,0)}Y^{(\ell\,0)}, \\
  K_{\mu a}&=\sum K_{\mu}^{(\ell, \pm 1)}Y_a^{(\ell,\pm 1)}, &\\
  L_{ab}&=\sum L^{(\ell,\pm 2)}Y_{ab}^{(\ell,\pm 2)}, &
  N&=\sum N^{(\ell\,0)}Y^{(\ell\,0)}, &\\
  X^I_\mu &=\sum X_\mu ^{I\,(\ell\,0)}Y^{(\ell\,0)}, &
  Z^I_{\mu a} &=\sum Z_\mu ^{I\,(\ell,\pm 1)}Y_a^{(\ell,\pm 1)}, &
  U^I_a &=\sum U^{I(\ell\,0)}\p_a Y^{(\ell\,0)},\\
 \phi^{ir}&=\sum \phi^{ir(\ell\,0)}Y^{(\ell\,0)}.
\end{aligned}
\label{gzvt6Nov18}
\end{align}
The $SO(4)$ quantum numbers $\ell_1,\ell_2$ of
the $S^3$ harmonic functions $Y^{(\ell_1,\ell_2)}_{(s)}$
are related to the $SU(2)_L\times
SU(2)_R$ quantum numbers $j,\bar{j}$ in the main text as
\begin{align}
 \ell_1=j+\bar{j},\qquad
 \ell_2=j-\bar{j}.\label{mnmk7Nov18}
\end{align}
The subscript $(s)$ denotes the $SO(3)$ content associated with the
tangent space of $S^3$.
By substituting the above expansion into the field equations of the
$D=6$ supergravity, one obtains the spectrum of excitations and their
representation content.  The $SO(2,2)$ representation associated with
the AdS$_3$ can be labeled by the energy $E_0$ and the spin $s_0$, which
are related to the weights $h,\bar{h}$ in the main text as
\begin{align}
 E_0=h+\bar{h},\quad
 s_0=h-\bar{h}.\label{mnmm7Nov18}
\end{align}

This procedure was carried out in \cite{Deger:1998nm}, where they have
explicitly written down which fields are involved in each excitation
mode.  Furthermore, by examining the quantum numbers of the supercharges
associated with the Killing vectors of the AdS$_3\times S^3$ background,
they identified the supermultiplets that these excitation modes are the
members of.  By comparing the representation content of these
supergravity supermultiplets with the representation content of the CFT
supermultiplets obtained by acting with $G^{+A}_{-1/2}$ and
$\Gt^{\dot{+}A}_{-1/2}$ on the anti-chiral primaries\footnote{The notation indicates the $T^4$ or $K^3$ cohomology to which the primary operator $O^{p,q}$ is associated, where $p,q$ can be $(+,0,-)\leftrightarrow (0,1,2)$.} $\ket{\pm\pm}_k$,
$\ket{\pm\mp}_k$ and $\ket{00}_k$, we can find what fields are excited
in the supergravity modes dual to each such state.  In Table
\ref{tbl:excitedfields}, we listed the fields are excited for anti-chiral
primary and superdescendant states (cf.~Table 1 of \cite{Deger:1998nm}).

\newcommand{\susy}[1]{{\color{red}\bm{#1}}}
\begin{table}[htb]
 \begin{center}
  \renewcommand{\arraystretch}{1.5}
 \begin{tabular}{|l||c|c|c|c|c|c|c|}
 \hline
 \multicolumn{1}{|c||}{state}&$h_{\mu\nu}$&$h_{\mu a}$&$h_{ab}$&$B^{I}_{\mu\nu}$&
 $B^{I}_{\mu a}$&$B^{I}_{ab}$&$\phi^{ir}$\\
 \hline 
 $\susy{GG\ket{-+}}$, $\Gt\Gt\ket{+-}$, $\susy{\ket{--}'}$,  $GG\Gt\Gt\ket{--}'$&$H_{\mu\nu}, M$&&$N$&$X_{\mu}^{5}$&&  $U^{5}$& \\
 $G\Gt\ket{--}'$&&&&$X_{\mu}^{\underline{i}}$&& $U^{\underline{i}}$&\\
 $\susy{\ket{00}}$,  $GG\Gt\Gt\ket{00}$, $\susy{\ket{++}'}$, $GG\Gt\Gt\ket{++}'$&&&&$X^{r}_{\mu}$&& $U^{r}$&$\phi^{5r}$\\
 $G\Gt\ket{00}$, $G\Gt\ket{++}'$ &&&&&&&$\phi^{\underline{i}r}$\\
 \hline
 $\susy{\ket{\mp\pm}}$, $GG\Gt\Gt\ket{\mp\pm}$, $\susy{GG\ket{--}'}$, $\Gt\Gt\ket{--}'$&&$K_{\mu}$&&&$Z^{5}_{\mu}$ &&\\
 $G\Gt\ket{\mp\pm}$&&&&& $Z_{\mu}^{\underline{i}}$&&\\
 $\susy{GG\ket{00}}$, $\Gt\Gt\ket{00}$, $\susy{GG\ket{++}'}$, $\Gt\Gt\ket{++}'$  &&&&& $Z_{\mu}^{r}$&&\\
 \hline
 $\susy{GG\ket{+-}}$, $\Gt\Gt\ket{-+}$&&&$L$&&&&\\
 \hline
 \end{tabular}
 \caption{\sl The fields excited in various superdescendant states. The
 fields in each row are described by a coupled system of equations which
 must be diagonalised to find the spectrum~\cite{Deger:1998nm}.  The
 shorthand $GG$ means $G^{+1}_{-1/2}G^{+2}_{-1/2}$, $\Gt\Gt$
 means $\Gt^{\dot{+}1}_{-1/2}\Gt^{\dot{+}2}_{-1/2}$, and $G\Gt$ means the
 four combinations $G^{+A}_{-1/2}\Gt^{\dot{+}B}_{-1/2}$ with
 $A,B=1,2$.  The states in (red) boldface letters are supersymmetric,
 involving only $G$ and not $\Gt$. The meaning of the prime on $\ket{++}'$
 and $\ket{--}'$ is explained in the main text. \label{tbl:excitedfields}}
\end{center}
\end{table}

The anti-chiral primary state $\ket{00}_k$ in Table 1 does not only mean
the one considered in the main text but represents a set of $n-1$ states
corresponding to the $(1,1)$-forms $\omega^{(r)}$, where
$r=7,\dots,5+n$.
In order to furnish a complete vector representation of the $SO(n)$
symmetry that exists in supergravity, we need to add one extra state to
the above $n-1$ states.  The candidate anti-chiral primaries are
$\ket{++}_{k+1}$ and $\ket{--}_{k-1}$, which have the same charges as
$\ket{00}_k$ ($h=-j^3={k\over 2}$).  Because the supergravity modes that
correspond to $\ket{00}_k$ leave the 6D metric unchanged, the extra
state must also leave the 6D metric unchanged.  Indeed, there is a
linear combination of $\ket{++}_{k+1}$ and $\ket{--}_{k-1}$ whose
gravity dual has undeformed 6D metric at linear order; see footnote 9 of
\cite{Shigemori:2013lta}.\footnote{This provides a possible
identification for the linear excitation mode that was
studied in \cite{Mathur:2003hj} in the framework of 6D supergravity.
The non-linear version of this mode is the superstratum with coiffuring
``Style 1'' discussed in \cite{Bena:2016agb}.}  Let us denote by
$\ket{++}'_k$ such a linear combination.  Then the vector of $SO(n)$ is
given by the $n$-vector $(\ket{++}_{k+1}',\ket{00}_k)$.  We also define
$\ket{--}'_k$ to be the linear combination of $\ket{++}_{k+1}$ and
$\ket{--}_{k-1}$ that is orthogonal to $\ket{++}'_k$.

Using the relation between the 6D supergravity in Table
\ref{tbl:excitedfields} and the fields discussed in the main text, we
can figure out what fields (such as $Z_4$ and $\Theta_4$) are excited
for what superdescendants.
From the second last line of Table \ref{tbl:excitedfields}, we see that
the supergravity mode corresponding to the CFT state
$G^{+1}_{-\half}G^{+2}_{-\half}\ket{00}$, which has been the focus of
the current paper, only involves $Z^r_\mu$.
This is related to the
mixed component (between AdS$_3$ and $S^3$) of the anti-self-dual 2-form
$B^r_{MN}$.  In particular, it does not change the 6D metric at the
linear order.  This means that this mode only excites $\Theta_4$ but
none of $Z_{1,2,4}$, which is what we found in the main text.

It is interesting to see that there are other supersymmetric modes that
excite only one field.  First, we see that the state
$G^{+1}_{-\half}G^{+2}_{-\half}\ket{++}'$ do not excite the metric.
The non-linear completion of this would be the $GG$ version of the
``Style 1'' superstratum discussed in \cite{Bena:2016agb}.
Moreover, the state $G^{+1}_{-\half}G^{+2}_{-\half}\ket{+-}$ turns
only on $L$, which is related to the traceless part of the $S^3$
metric. This will probably correspond to some simple deformation of the 4D base
metric with $\beta$ unchanged.  It would be interesting to construct
the non-linear completion of these modes.
Although it must be technically more challenging, some of
non-supersymmetric states, namely $G\Gt\ket{00}$, $G\Gt\ket{++}'$,
$G\Gt\ket{\mp\pm}$, have only one field excited, and constructing their
non-linear completion would be also interesting. See
\cite{Bombini:2017got} for recent work in this direction.

\section{Killing spinors in the RR coordinates}

\label{app:RRspinors}

In the main text, we studied the Killing spinors of AdS$_3\times S^3$ in
the NS-NS coordinates.  Here we derive the expression for the Killing
spinor in the RR coordinates and further show that the two sets of
spinors are related by a local Lorentz transformation and the spectral
flow coordinate transformation.

We summarise some formulas that we make frequent use of in this
Appendix.  

Because of the commutation relations
\begin{align}
 [\Gammau{uv},\Gammau{v}]=2\Gammau{v},\qquad
 [\Gammau{uv},\Gammau{u}]=-2\Gammau{u},
\label{epjw22Nov18}
\end{align}
$\Gammau{v}$ and $\Gammau{u}$ can be regarded as raising and lowering
operators, respectively, with (one-half of) the $\Gammau{uv}$ chirality
being the number operator.

Using the formula 
\begin{align}
 \Gamma_{m_1\dots m_p} \Gamma_{(10)} = 
 (-1)^p\Gamma_{(10)} \Gamma_{m_1\dots m_p}  = 
 {(-1)^{p(p-1)/2}\over (10-p)!} \epsilon_{m_1\dots m_p n_1\dots n_{10-p}}\Gamma^{n_1\dots n_{10-p}}.
 \label{jjkk30Jan18}
\end{align}
one can show   that the following $\Gamma$-matrix
relations hold when they are acting on spinors with
$\Gamma_{(10)}=\Gammau{6789}=1$:
\begin{align}
\begin{aligned}
 \Gammau{i}   &=+{1\over 3!}\epsilonu{ijkl}\,\Gammau{uvjkl},&
 \Gammau{ij}  &=-{1\over 2}\epsilonu{ijkl}\,\Gammau{uvkl},&
 \Gammau{ijk} &=-\epsilonu{ijkl}\,\Gammau{uvl},&
 \Gammau{ijkl}&=+\epsilonu{ijkl}\,\Gammau{uv},\\
 \Gammau{vi}  &=-{1\over 3!}\epsilonu{ijkl}\,\Gammau{vjkl},&
 \Gammau{vij} &=+{1\over 2}\epsilonu{ijkl}\,\Gammau{vkl},&
 \Gammau{vijk}&=+\epsilonu{ijkl}\,\Gammau{vl},\\
 \Gammau{ui}  &=+{1\over 3!}\epsilonu{ijkl}\,\Gammau{ujkl},&
 \Gammau{uij} &=-{1\over 2}\epsilonu{ijkl}\,\Gammau{ukl},&
 \Gammau{uijk}&=-\epsilonu{ijkl}\,\Gammau{ul},\\
\end{aligned}
\label{Gamma_eps_formula1}
\end{align}
where $i,j,k,l$ denote $\bbR^4$ indices and are summed over when
repeated.  In particular, in the convention
$\epsilonu{r\theta\phi\psi}=+1$,
we have
\begin{align}
\begin{aligned}
 \Gammau{\phi \psi}  &=-\Gammau{uv r\theta},&\quad
 \Gammau{\theta\psi} &=+\Gammau{uv r\phi},&\quad
 \Gammau{\theta\phi} &=-\Gammau{uv r\psi},&
 \\
 \Gammau{vr}   &=-\Gammau{v\psi\theta\phi},&
 \Gammau{v\psi}&=+\Gammau{vr\theta \phi},&
 \Gammau{v\phi}&=-\Gammau{vr\theta \psi},&\\
 \Gammau{ur}   &=+\Gammau{u\psi\theta\phi},&
 \Gammau{u\psi}&=-\Gammau{ur\theta \phi},&
 \Gammau{u\phi}&=+\Gammau{ur\theta \psi}
\end{aligned}
\label{Gamma_eps_formula2}
\end{align}
and so on.

\subsection{The RR Killing spinors}

Let us focus on the round supertube solution after the decoupling limit
in the RR coordinates~$(t,r,y,\theta,\phi,\psi)$ (see Eq.~\eqref{stmetric}).  We take the vielbeins
to be
\begin{align}
\label{RRvielbeins}
\begin{split}
 E^{\underline{v}}&={\Sigma\over a R_y}(dv+\beta),\qquad
 E^{\underline{u}}=du+\omega,\qquad
 E^{\underline{r}}=\sqrt{a R_y\over r^2+a^2}\,dr,\qquad
 E^{\underline{\theta}}=\sqrt{a R_y}\,d\theta,\\
 E^{\underline{\phi}}
 &=\sqrt{a R_y(r^2+a^2)\over \Sigma}\,\sin\theta\,d\phi,\qquad
 E^{\underline{\psi}}=\sqrt{a R_y\over \Sigma}\,r\cos\theta\,d\psi,\qquad
 E^{\underline{\alpha}}=\Bigl({Q_1\over Q_5}\Bigr)^{1/4}dx^\alpha,
\end{split}
\end{align}
where $\alpha=6,7,8,9$.  By setting the supersymmetry variations
\eqref{fervar1} to zero, after some manipulations, we can find Killing
spinors that preserve supersymmetry.  The result can be stated in a
simple way in terms of
\begin{align}
 \epsilon_\pm=\epsilon^1\pm \epsilon^2.
\end{align}
The RR Killing spinors are
\begin{align}
 \epsilon_+ &= \kappat^{+} + e^{\Mt u} \kappat^{-},\qquad
 \epsilon_- = \kappa^{-} + e^{M v} \kappa^{+},
 \label{epsilonpm_RR}
\end{align}
where
\begin{subequations} 
\label{coh28Aug18}
 \begin{align}
 \kappat^{+}
 &=
 e^{{\theangle\over 2}\Gammau{r\theta}}
 e^{{1\over 2}(\phi-\psi)\Gammau{\theta\phi}}
 \chit^+
 \cr
 &
 \quad +
 \sqrt{a R_y\over 2}
 {\Gammau{v}\over \Sigma}
 \left(
 -{a^2\sin\theta\cos\theta}\,\Gammau{\phi}\, 
 + r\sqrt{r^2+a^2}\,\Gammau{\psi}
 \right)
 e^{{\theangle\over 2}\Gammau{r\theta}}\,
 e^{{1\over 2}(\phi-\psi)\Gammau{\theta\phi}}\,
 \chit^{-},
\\
 \kappat^{-}
 &=
 e^{-{\theangle\over
 2}\Gammau{r\theta}}e^{{1\over 2}(\phi-\psi)\Gammau{\theta\phi}}\chit^{-},
\\
 \kappa^{-}
 &=
 \sqrt{\tfrac{\Sigma}{a R_y}}\,
 e^{{\theangle\over 2}\Gammau{r\theta}}
 e^{{1\over 2}(\phi+\psi)\Gammau{\theta\phi}}
 \chi^{-}
 \cr
 &
 \quad -
 {\Gammau{u}\over a \sqrt{2\Sigma}}
 \left(
 {a^2\sin\theta\cos\theta}\,\Gammau{\phi}\, 
 + r\sqrt{r^2+a^2}\,\Gammau{\psi}
 \right)
 e^{{\theangle\over 2}\Gammau{r\theta}}\,
 e^{{1\over 2}(\phi+\psi)\Gammau{\theta\phi}}\,
 \chi^{+},
\\
 \kappa^{+}
 &=
 \sqrt{\tfrac{a R_y}{\Sigma}}\,
 e^{-{\theangle\over
 2}\Gammau{r\theta}}e^{{1\over 2}(\phi+\psi)\Gammau{\theta\phi}}\chi^{+},
 \end{align}
\end{subequations}
and
\begin{subequations} 
 \begin{align}
 \Mt&=
 {\Gammau{v} \over \sqrt{a R_y\,}\,\Sigma}
 \left(r\sqrt{r^2+a^2}\,\Gammau{r}-a^2\sin\theta\cos\theta\, \Gammau{\theta}\right)
 \cr
 &\qquad
 -{1\over R_y\sqrt{2\Sigma}}
 \left[
 r\sin\theta(-\Gammau{r\phi}+\Gammau{\theta\psi})
 +\sqrt{r^2+a^2}\cos\theta(+\Gammau{r\psi}+\Gammau{\theta\phi})
 \right]\,\cP_{uv}^-,
 \\
 M&=
 {\Gammau{u} \over \sqrt{a^3R_y^3}}
 \left(r\sqrt{r^2+a^2}\,\Gammau{r}-a^2\sin\theta\cos\theta\, \Gammau{\theta}\right)
 \cr
 &\qquad
 +{1\over R_y\sqrt{2\Sigma}}
 \left[
 r\sin\theta(+\Gammau{r\phi}+\Gammau{\theta\psi})
 +\sqrt{r^2+a^2}\cos\theta(+\Gammau{r\psi}-\Gammau{\theta\phi})
 \right]\,\cP_{uv}^+.
 \end{align}
\end{subequations}
The angle $\theangle$ is defined by
\begin{align}
 \cos \theangle = \sqrt{r^2+a^2\over \Sigma}\,\cos\theta,\qquad
 \sin \theangle = {r\over \sqrt{\Sigma}}\,\sin\theta.
\end{align}
$\cP_{uv}^\pm$ are projection operators onto the $\Gammau{uv}=\pm 1$ subspaces:
\begin{align}
 \cP_{uv}^\pm \equiv {1\over 2}(1\pm\Gammau{uv}).
\end{align}
The spinors $\chit^\pm$, $\chi^\pm$ are constant Majorana-Weyl spinors
with $\Gamma_{(10)}=\Gammau{6789}=1$ on them.  Furthermore, their
superscript indicates the $\Gammau{uv}$ chirality, namely,
\begin{align}
 \Gammau{uv}\,\chit^{\pm}=\pm\chit^{\pm},\qquad
 \Gammau{uv}\,\chi^{\pm}=\pm\chi^{\pm}.
\end{align}
Each of the four spinors $\chit^\pm$, $\chi^\pm$ has $4$ independent
real components.  The total number of unbroken real supercharges is
$4\times 4=16$. The spinors $\kappat^\pm$, $\kappa^\pm$ also have definite
$\Gammau{uv}$ chirality displayed by the superscript:
\begin{align}
 \Gammau{uv}\,\kappat^{\pm}=\pm\kappat^{\pm},\qquad
 \Gammau{uv}\,\kappa^{\pm}=\pm\kappa^{\pm},
\end{align}
while $e^{\Mt u}\kappat^-$,  $e^{M v}\kappa^+$ do not have definite
$\Gammau{uv}$ chirality.

The exponential in  \eqref{epsilonpm_RR} can be written as
\begin{align}
\begin{split}
  e^{\Mt u}&=1
 +R_y^2\biggl(1-\cos{\sqrt{2}u\over R_y}\biggr)\,\Kt
 +{R_y\over\sqrt{2}}\biggl(\sin{\sqrt{2}u\over R_y}\biggr)\,\Mt,
 \\
 e^{M v}&=1
 +R_y^2\biggl(1-\cos{\sqrt{2}v\over R_y}\biggr)\,K
 +{R_y\over\sqrt{2}}\biggl(\sin{\sqrt{2}v\over R_y}\biggr)\,M,
\end{split}
\end{align}
where
\begin{align}
\begin{split}
  \Kt&=
 {\Gammau{v}\over \sqrt{2a R_y^3\Sigma\,}}
 \left[\sqrt{r^2+a^2}\,\sin\theta\,\Gammau{\phi}
 -r\cos\theta\,\Gammau{\psi}\right]-{\cP_{uv}^-\over R_y^2},
 \\
 K&=
 \sqrt{\Sigma\over 2a^3R_y^5}\,\Gammau{u}
 \left[\sqrt{r^2+a^2}\,\sin\theta\,\Gammau{\phi}
 +r\cos\theta\,\Gammau{\psi}\right]-{\cP_{uv}^+\over R_y^2}.
\end{split}
\end{align}

\subsection{Map between the NS-NS and RR spinors}

In the above, we derived the expression for the Killing spinors in the
RR coordinates with the vielbeins \eqref{RRvielbeins}.  On the other
hand, in the main text, we derived the Killing spinors in the NS-NS
coordinate system with the vielbeins \eqref{vielbeine}. The two Killing
spinors are related by a coordinate transformation and a local Lorentz
transformation, which we work out below.

The vielbeins used in the NS-NS coordinates are:
\begin{align}
\begin{aligned}
 \Et^{\underline{v}}&={1\over 2\sqrt{a R_y}}
 [(\sqrt{r^2+a^2}+r)dv+(\sqrt{r^2+a^2}-r)du],\\
 \Et^{\underline{u}}&={1\over 2\sqrt{a R_y}}
 [(\sqrt{r^2+a^2}+r)du+(\sqrt{r^2+a^2}-r)dv],\\
 \Et^{\underline{\phit}}
 &=\sqrt{a R_y}\,\sin\theta\,d\phit,
 \qquad
 \Et^{\underline{\psit}}
 =\sqrt{a R_y}\,r\cos\theta\,d\psit ,
\end{aligned}
\end{align}
where we put tildes on the NS-NS vielbeins, to distinguish them from the
RR ones \eqref{RRvielbeins}.  The components
$E^{\underline{r}},E^{\underline{\theta}},E^{\underline{\alpha} }$ are
the same as the RR ones and we did not write them down.  The RR angles
$(\phi,\psi)$ and the NS-NS ones $(\phit,\psit)$ are related to each
other by the spectral flow transformation \eqref{spectralflowcoord}.

The two sets of vielbeins are related by a local Lorentz transformation
and the coordinate transformation \eqref{spectralflowcoord}.  Let us
focus only on the $(u,v,\phi,\psi)$ and $(u,v,\phit,\psit)$ parts of the
vielbeins because other parts are identical in RR and NS-NS coordinates.
Then the Lorentz transformation can be written as
\begin{align}
 \Et^{\underline{\tilde{a}}}=\Lambda^{\underline{\tilde{a}}}{}_{\underline{b}} E^{\underline{b}},
\end{align}
where
\begin{align}
 \Lambda=\Lambda_5\Lambda_6\Lambda_3\Lambda_4\Lambda_1\Lambda_2,\qquad
 \Lambda_i=e^{a_i g_i}.
\label{pwt23Aug18}
\end{align}
Here $g_i$ are given by
\begin{align}
 g_1= iM^{\underline{uv      }},~~
 g_2=-iM^{\underline{\phi\psi}},~~
 g_3= iM^{\underline{u\phi   }},~~
 g_4= iM^{\underline{u\psi   }},~~
 g_5= iM^{\underline{v\phi   }},~~
 g_6= iM^{\underline{v\psi   }}
\end{align}
with $M^{\underline{ab}}$ the Lorentz generators in the vector representation
given by 
\begin{align}
 (M^{\underline{ab}})^c{}_d &= i(\eta^{ac} \delta^b_d -\eta^{bc} \delta^a_d )
\end{align}
satisfying the Lorentz algebra
\begin{align}
 [M^{\underline{ab}},M^{\underline{cd}}] &= -i(\eta^{ac}M^{\underline{bd}}-\eta^{ad}M^{\underline{bc}}-\eta^{bc}M^{\underline{ad}}+\eta^{bd}M^{\underline{ac}}).
\end{align}
The explicit matrix expressions for $g_i$ are
\begin{align}
&\begin{aligned}
  g_1&=\left(\begin{smallmatrix}      1&0&0&0\\ 0&-1&0&0 \\ 0&0&0&0 \\ 0&0&0&0     \end{smallmatrix}\right),&\quad
 g_2&=\left(\begin{smallmatrix}      0&0&0&0\\ 0&0&0&0 \\ 0&0&0&1 \\ 0&0&-1&0     \end{smallmatrix}\right),&\quad
 g_3&=\left(\begin{smallmatrix}      0&0&1&0\\ 0&0&0&0 \\ 0&1&0&0 \\ 0&0&0&0     \end{smallmatrix}\right),\\
 g_4&=\left(\begin{smallmatrix}      0&0&0&1\\ 0&0&0&0 \\ 0&0&0&0 \\ 0&1&0&0     \end{smallmatrix}\right),&
 g_5&=\left(\begin{smallmatrix}      0&0&0&0\\ 0&0&1&0 \\ 1&0&0&0 \\ 0&0&0&0     \end{smallmatrix}\right),&
 g_4&=\left(\begin{smallmatrix}      0&0&0&0\\ 0&0&0&1 \\ 0&0&0&0 \\ 1&0&0&0     \end{smallmatrix}\right).
\end{aligned} 
\label{mvef22Aug18}
\end{align}
The parameters $a_i$ are given by
\begin{align}
\begin{aligned}
 a_1&=\log{\sqrt{a R_y}\,(\sqrt{r^2+a^2}+r)\over 2\Sigma},&
 a_2&=\theta-\gamma,&
 a_3&=-{a r\sin\theta\over\sqrt{2}\,\Sigma},\\
 a_4&={a\sqrt{r^2+a^2}\,\cos\theta \over \sqrt{2}\,\Sigma },&
 a_5&=-{\sqrt{2}\,a \sin\theta \over \sqrt{r^2+a^2}+r},&
 a_6&=-{\sqrt{2}\,a \cos\theta \over \sqrt{r^2+a^2}+r}.&
\end{aligned}
\end{align}

Spinors transform as scalars under coordinate transformation and
transform in the spinor representation under local Lorentz
transformation.  Therefore, if we act on the RR spinor $\epsilon_-^{\rm
R}$ with the matrix $\Lambda$ in the spinor representation and replace
explicit $\phi,\psi$  appearing in $\epsilon_-^{\rm
R}$ by $\phit,\psit$ using
\eqref{spectralflowcoord}, we must get the NS-NS spinor $\epsilon_-^{\rm
NS}$.  Here, we put ``R'' and ``NS'' on $\epsilon_-$ to clarify the
frame that the spinor is in.  Namely,
\begin{align}
 \epsilon^{\rm NS}= \Lambda^{\rm spinor}\, \epsilon^{\rm R}.
 \label{R2NSspinor}
\end{align}
The local Lorentz generators in the spinor representation are given by
\begin{align}
  M^{\underline{ab}}={i\over 2}\Gammau{ab},
\end{align}
or, more explicitly,
\begin{align}
\begin{aligned}
 g_1&=-{1\over 2}\Gammau{uv},&
 g_2&=+{1\over 2}\Gammau{\phi \psi}&
 g_3&=-{1\over 2}\Gammau{u\phi},\\
 g_4&=-{1\over 2}\Gammau{u\psi},&
 g_5&=-{1\over 2}\Gammau{v\phi},&
 g_6&=-{1\over 2}\Gammau{v\psi}.
\end{aligned}
\label{kkpr22Nov18}
\end{align}
In doing this, we must note that we do not transform the matrices
$\Gammau{\mu}\,$;  they are always constant matrices and the index
$\underline{\mu}$ is only a label. Therefore,
$\Gammau{\phi},\Gammau{\psi}$ are the same constant matrices as
$\Gammau{\phit},\Gammau{\psit}$.

Let us explicitly prove the relation \eqref{R2NSspinor}.  For
simplicity, we consider $\epsilon_-$ in the R sector,
\eqref{epsilonpm_RR}, in the case with $\chi^{+}=0$.  In this case,
\begin{align}
 \epsilon_-^{\rm R}
 =
 \sqrt{\tfrac{\Sigma}{a R_y}} \,
 e^{{\gamma\over 2}\Gammau{r\theta}}\,
 e^{{1\over 2}(\phi + \psi) \Gammau{\theta\phi}}
 \, \chi^{-}.
\label{epsilon-_simple}
\end{align}
Applying the spectral flow coordinate transformation
\eqref{spectralflowcoord} on \eqref{epsilon-_simple}, we have
\begin{align}
 \epsilon_-^{\rm R}
 =
 \sqrt{\tfrac{\Sigma}{a R_y}} \,
 e^{{\gamma\over 2}\Gammau{r\theta}}\,
 e^{{1\over 2}(\phit + \psit+{\sqrt{2}v\over R_y}) \Gammau{\theta\phi}}
 \, \chi^{-}.
\end{align}
Next, let us act on this spinor with the Lorentz transformation matrices
\eqref{pwt23Aug18}, in the spinor representation, one by one.  First, by
the action of
\begin{math}
 \Lambda_2
 =e^{{a_2\over 2}\Gammau{\phi\psi}}
 =e^{-{\theta -\gamma\over 2}\Gammau{uvr\theta}}
 =e^{{\theta -\gamma\over 2}\Gammau{r\theta}}
\end{math}
(for the second equality, see \eqref{Gamma_eps_formula2}; the last equality holds on a
$\Gammau{uv}=-1$ spinor), we have
\begin{align}
 \Lambda_2\epsilon_-^{\rm R}
 =
 \sqrt{\tfrac{\Sigma}{a R_y}} \,
 e^{{\theta\over 2}\Gammau{r\theta}}\,
 e^{{1\over 2}(\phit + \psit+{\sqrt{2}v\over R_y}) \Gammau{\theta\phi}}
 \, \chi^{-}.
\end{align}
Next, acting with $\Lambda_1=e^{-{a_1\over 2}\Gammau{uv}}=e^{{a_1\over
2}}= {(a^3 R_y)^{1/4}\over \sqrt{2\Sigma}}\,R_+$, where $R_\pm$ are defined in
\eqref{eq:Rpmdef}, we get
\begin{align}
 \Lambda_1\Lambda_2\epsilon_-^{\rm R}
 =
 \tfrac{1}{\sqrt{2}}(\tfrac{a}{R_y})^{1/4} R_+
 e^{{\theta\over 2}\Gammau{r\theta}}\,
 e^{{1\over 2}(\phit + \psit+{\sqrt{2}v\over R_y}) \Gammau{\theta\phi}}
 \, \chi^{-}.\label{eeff22Nov18}
\end{align}
Because $\Gammau{v},\Gammau{u}$ square to zero, we find
\begin{align}
 \Lambda_3 \Lambda_4 &= 1-{\Gammau{u}\over 2}(a_3 \Gammau{\phi}+a_4\Gammau{\psi})
 = 1-{a\over 2\sqrt{2\Sigma}}\Gammau{u\psi}\,e^{\gamma \Gammau{\phi\psi}},\label{tla23Aug18}
\\
 \Lambda_5 \Lambda_6 &= 1-{\Gammau{v}\over 2}(a_5 \Gammau{\phi}+a_6\Gammau{\psi})
 =1+{R_-\over \sqrt{2}\,R_+} \Gammau{v\psi}\,e^{-\theta \Gammau{\phi\psi}}.
\end{align}
Because \eqref{eeff22Nov18} has $\Gammau{uv}=-1$, it is killed by
$\Gammau{u}$ (see \eqref{epjw22Nov18}). So, $\Lambda_3\Lambda_4=1$ on
it.  Applying $\Lambda_5\Lambda_6$ on \eqref{eeff22Nov18}, we finally
obtain
\begin{align}
 \Lambda\epsilon_-^{\rm R}
 =
 \tfrac{1}{\sqrt{2}}(\tfrac{a}{R_y})^{1/4}\left(  R_+ e^{{\theta\over 2}\Gammau{r\theta}}
 +
\tfrac{R_-}{\sqrt{2}}
 \Gammau{v\psi}\,e^{-{\theta\over 2} \Gammau{r\theta}}\right)
 e^{{1\over 2}(\phit + \psit+{\sqrt{2}v\over R_y}) \Gammau{\theta\phi}}
 \, \chi^{-}.\label{fhds22Nov18}
\end{align}

This is to be matched with the expression for the NS-NS spinor
\eqref{summary1}, \eqref{summary2}:\footnote{As explained below
\eqref{kkpr22Nov18}, the matrices $\Gammau{\phit},\Gammau{\psit}$ are
identical to the matrices $\Gammau{\phi},\Gammau{\psi}$.  Therefore,
$\Gammau{\theta\phit}=\Gammau{\theta\phi}$,
$\Gammau{r\psit}=\Gammau{r\psi}$.}
\begin{subequations} 
 \begin{align}
 \epsilon_-^{\rm NS}&=R_+ Y_+ \xi + R_- Y_- \eta,\label{hbps22Nov18}\\
 Y_\pm
 &=
 e^{\pm{\theta\over 2}\Gammau{r\theta}}
 e^{{\phit\over 2}\Gammau{\theta\phi}}
 e^{\pm{\psit\over 2}\Gammau{r\psi}}
 ,\\
 \xi &= \zeta_- e^{-\frac{iv}{\sqrt2R_y}}+ \zeta_+  e^{\frac{iv}{\sqrt2R_y}},\qquad 
 \eta = \tfrac{i}{\sqrt{2}}\Gammau{vr} \bigl(- \,\zeta_-e^{-\frac{iv}{\sqrt{2}R_y}} + \, \zeta_+ e^{\frac{iv}{\sqrt{2}R_y}}\bigr).
 \end{align}
\end{subequations}

In order to match two expressions, let us decompose $\chi^{-}$ into the
representation of the $SU(2)_L\times SU(2)_R$ symmetry group, whose
generators are\footnote{These are the same matrices as \eqref{eq:SU(2)Jgen} (also see the comment below \eqref{kkpr22Nov18}).}
\begin{align}
 J^1&=-{i\over 4}(\Gammau{r\theta}+ \Gammau{\phi\psi}),&
 J^2&=-{i\over 4}(\Gammau{r\phi}  + \Gammau{\psi\theta}),&
 J^3&=-{i\over 4}(\Gammau{r\psi}  + \Gammau{\theta\phi}),\\
 \Jt^1&=-{i\over 4}(\Gammau{r\theta}- \Gammau{\phi\psi}),&
 \Jt^2&=-{i\over 4}(\Gammau{r\phi}  - \Gammau{\psi\theta}),&
 \Jt^3&=-{i\over 4}(\Gammau{r\psi}  - \Gammau{\theta\phi}).\label{gmos23Aug18}
\end{align}
Eqs.~\eqref{Gamma_eps_formula2} say that, on the spinor $\chi^-$ with
$\Gamma_{(10)}=\Gammau{6789}=1$ and $\Gammau{uv}=-1$, we have
$(J^1,J^2,J^3)=-{i\over
2}(\Gammau{r\theta},\Gammau{r\phi},\Gammau{\theta \phi})$ and
$\Jt^{1,2,3}=0$, namely, $\chi^-$ is in the $({\bf 2},{\bf 1})$
representation. So, if we decompose $\chi^-$ as
\begin{align}
 \chi^-= \sum_{\alpha=\pm}\chi^{-\alpha},\label{glco22Nov18}
\end{align}
where
\begin{align}
  J^3\chi^{-\alpha }={\alpha\over 2} \chi^{-{}\alpha},
 \qquad\text{or}\qquad
  \Gammau{\theta\phi}\,\chi^{-\alpha}= i \alpha \chi^{-\alpha},
\end{align}
then \eqref{fhds22Nov18} becomes
\begin{align}
 \Lambda\epsilon_-^{\rm R}
 &=
 \tfrac{1}{\sqrt{2}}(\tfrac{a}{R_y})^{1/4}
 \left(  R_+ e^{{\theta\over 2}\Gammau{r\theta}}
 +
 \tfrac{R_-}{\sqrt{2}} \Gammau{v\psi}\,e^{-{\theta\over 2} \Gammau{r\theta}}\right)
 e^{{i\over 2}(\phit + \psit+{\sqrt{2}v\over R_y})}
 \, \chi^{-+}
 \notag\\
 &\qquad
 +
 \tfrac{1}{\sqrt{2}}(\tfrac{a}{R_y})^{1/4}
 \left(R_+ e^{{\theta\over 2}\Gammau{r\theta}}
 +
\tfrac{R_-}{\sqrt2}
 \Gammau{v\psi}\,e^{-{\theta\over 2} \Gammau{r\theta }}\right)
 e^{-{i\over 2}(\phit + \psit+{\sqrt{2}v\over R_y})}
 \, \chi^{--}.\label{jhdl22Nov18}
\end{align}

On the other hand, the NS-NS spinor \eqref{hbps22Nov18} can be written as
\begin{align}
 \epsilon_-^{\rm NS}
 &=
 (R_+ - \tfrac{i}{\sqrt{2}} R_-\Gammau{vr} ) Y_+ e^{-\frac{iv}{\sqrt2R_y}}\zeta_-
+(R_+ + \tfrac{i}{\sqrt{2}} R_- \Gammau{vr}) Y_+ e^{\frac{iv}{\sqrt2R_y}}\zeta_+
\label{jflf22Nov18}
\end{align}
where we used the fact that
$\{\Gammau{r\theta},\Gammau{vr}\}=\{\Gammau{r\psi},\Gammau{vr}\}=0$ and
hence $Y_-\Gammau{vr}=\Gammau{vr}Y_+$.
Because both $\zeta_+$ and $\zeta_-$ have $\Gammau{uv}=-1$, we can
decompose them just as in \eqref{glco22Nov18} as
\begin{align}
 \zeta_+=\sum_{\alpha=\pm}\zeta_+^\alpha,\qquad
 \zeta_-=\sum_{\alpha=\pm}\zeta_-^\alpha.
\end{align}
Using $\Gammau{r\psi}=\Gammau{\theta\phi}=i\alpha$, we see that
$Y_+ = e^{{\theta\over 2}\Gammau{r\theta}} e^{{i\alpha\over
2}(\phit+\psit)}$ on $\zeta_\pm^\alpha$.  So, \eqref{jflf22Nov18} becomes
\begin{align}
 \epsilon_-^{\rm NS}
 &=
 (R_+ + \tfrac{i}{\sqrt{2}} R_- \Gammau{vr}) 
 e^{{\theta\over 2}\Gammau{r\theta}}e^{{i\over 2}(\phit+\psit+\frac{\sqrt{2}v}{R_y})}\zeta_+^+
 +
  (R_+ - \tfrac{i}{\sqrt{2}} R_-\Gammau{vr} ) 
 e^{{\theta\over 2}\Gammau{r\theta}}e^{-{i\over 2}(\phit+\psit+\frac{\sqrt{2}v}{R_y})}\zeta_-^-
 \notag\\
 &\quad
 +(R_+ - \tfrac{i}{\sqrt{2}} R_-\Gammau{vr} ) 
 e^{{\theta\over 2}\Gammau{r\theta}}e^{{i\over 2}(\phit+\psit-{\sqrt{2}v\over R_y})}\zeta_-^+
 +(R_+ + \tfrac{i}{\sqrt{2}} R_- \Gammau{vr}) e^{{\theta\over 2}\Gammau{r\theta}}e^{-{i\over 2}(\phit+\psit-\frac{\sqrt{2} v}{R_y})}\zeta_+^-.
\end{align}
Using the relation $\Gammau{v r}= -\Gammau{v \psi \theta \phi
}=-\Gammau{v \psi}\Gammau{\theta \phi }$, commuting $\Gammau{\theta \phi
}$ through $e^{{\theta\over 2}\Gammau{r\theta}}$, and replacing
$\Gammau{\theta\phi}$ on $\zeta_\pm^\alpha$ by $i\alpha$, we can
rewrite this as
\begin{align}
\epsilon_-^{\rm NS}&=
 (R_+ e^{{\theta\over 2}\Gammau{r\theta}} + \tfrac{1}{\sqrt{2}} R_- \Gammau{v\psi}\, e^{-{\theta\over 2}\Gammau{r\theta}}) 
  e^{{i\over 2}(\phit+\psit+\frac{\sqrt{2}v}{R_y})}\zeta_+^+
 \notag\\
 &\qquad
 +
  (R_+ e^{{\theta\over 2}\Gammau{r\theta}} + \tfrac{1}{\sqrt{2}} R_-\Gammau{v\psi}\, e^{-{\theta\over 2}\Gammau{r\theta}} ) 
 e^{-{i\over 2}(\phit+\psit+\frac{\sqrt{2}v}{R_y})}\zeta_-^-
 ~+~\text{($\zeta_-^+,\zeta_+^-$ terms)}.
\end{align}
This is exactly the same as \eqref{jhdl22Nov18}, with the identification
\begin{align}
 \tfrac{1}{\sqrt{2}}(\tfrac{a}{R_y})^{1/4}\,\chi^{-+}
 ~\leftrightarrow~
 \zeta_+^+,\qquad\qquad
 \tfrac{1}{\sqrt{2}}(\tfrac{a}{R_y})^{1/4}\,\chi^{--}
 ~\leftrightarrow~
 \zeta_-^-.
\end{align}
Similarly, $\zeta_-^+,\zeta_+^-$ must be related to
$\chi^{+}$ which we turned off for simplicity.  

The product representation of $\Lambda$ in \eqref{pwt23Aug18} is
convenient for showing $\Lambda\epsilon_-^{\rm R}=\epsilon_-^{\rm
NS}$. In order to show $\Lambda\epsilon_+^{\rm R}=\epsilon_+^{\rm NS}$,
on the other hand, it is more convenient to use a different product
representation
$\Lambda=\Lambda_3\Lambda_4\Lambda_5\Lambda_6\Lambda_1\Lambda_2$ instead,
where the parameters are now
\begin{align}
 \begin{aligned}
 a_1&=\log{2\sqrt{a R_y}\over \sqrt{r^2+a^2}+r},&
 a_2&=\gamma-\theta,&
 a_3&=-{\sqrt{2}a\sin\theta\over \sqrt{r^2+a^2}+r},\\
 a_4&={\sqrt{2}a\cos\theta\over \sqrt{r^2+a^2}+r},&
 a_5&=-{ar\sin\theta\over \sqrt{2}\Sigma},&
 a_6&=-{a\sqrt{r^2+a^2}\,\cos\theta\over \sqrt{2}\Sigma}.
 \end{aligned}
\end{align}


\newpage

\begin{adjustwidth}{-1mm}{-1mm} 

\providecommand{\href}[2]{#2}\begingroup\raggedright\endgroup

\end{adjustwidth}

\end{document}